\documentclass{emulateapj}
\usepackage{hyperref}
\usepackage{natbib}
\usepackage{color}
\usepackage{soul}
\soulregister\citet7
\soulregister\citep7
\soulregister\citealp7
\soulregister\footnote7
\soulregister\equation7
\soulregister\caption7
\soulregister\subsection7
\soulregister$7
\soulregister\ref7
\soulregister\se7
\soulregister\Fig7
\soulregister\Eq7
\soulregister\Table7
\soulregister\hi7
\soulregister\ha7
\soulregister\magarc7
\soulregister\kms7
\soulregister\solarm7

\usepackage{amsmath}
\usepackage{wasysym}
\usepackage{multirow}
 
\newcommand{\kms}{\ifmmode\,{\rm km}\,{\rm s}^{-1}\else km$\,$s$^{-1}$\fi}
\newcommand{\Rd}{\ifmmode\,R_{\rm d}\else $R_{\rm d}$\fi}
\newcommand{\be}{\begin{equation}}
\newcommand{\ee}{\end{equation}}
\newcommand\ltsima{$\; \buildrel < \over \sim \;$}
\newcommand\ltsim{\lower.5ex\hbox{\ltsima}}
\newcommand\gtsima{$\; \buildrel > \over \sim \;$}
\newcommand\gtsim{\lower.5ex\hbox{\gtsima}}
\newcommand{\solarm}{\ensuremath{M_\odot}}

\newcommand{\magarc}{\ifmmode {{{{\rm mag}~{\rm arcsec}}^{-2}}}
             \else {{{mag}$~${arcsec}$^{-2}$}}
             \fi}

\def\Eq#1{Eq.~(\ref{eq:#1})}

\def\Fig#1{Fig.~\ref{fig:#1}}
\def\Table#1{Table~\ref{tab:#1}}
\def\se#1{\S\ref{sec:#1}}

\def \etal{{et~al.~}} 
\def \ion#1#2{#1{\footnotesize{#2}}\relax}
\newcommand{\ntwo}{[N\textsc{ii}]}
\newcommand{\ha}{H$\alpha$}
\def \hi{\ion{H}\ {I}}

\slugcomment{Submitted to Astrophysical Journal}
\begin{document}

\title{The Spectroscopy and H-band Imaging of Virgo cluster galaxies (SHIVir) Survey: Scaling Relations and the Stellar-to-Total Mass Relation}
\shortauthors{Ouellette et al.}

\author{
Nathalie N.-Q. Ouellette\altaffilmark{1},
St\'ephane Courteau\altaffilmark{1},
Jon A. Holtzman\altaffilmark{2},
Aaron A. Dutton\altaffilmark{3},
Michele Cappellari\altaffilmark{4},
Julianne J. Dalcanton\altaffilmark{5},
Michael McDonald\altaffilmark{6},
Joel C. Roediger\altaffilmark{7},
James E. Taylor\altaffilmark{8},
R. Brent Tully\altaffilmark{9},
Patrick C\^{o}t\'{e}\altaffilmark{7},
%Eric Emsellem\altaffilmark{10,}\altaffilmark{11} X
Laura Ferrarese\altaffilmark{7}
\& Eric W. Peng\altaffilmark{10,}\altaffilmark{11}
}
\altaffiltext{1}{Department of Physics, Engineering Physics and Astronomy, Queen's University, Kingston, ON K7L 3N6, Canada}
\altaffiltext{2}{Department of Physics and Astronomy, New Mexico State University, Las Cruces, NM, 88003-8001, USA}
\altaffiltext{3}{Department of Physics, New York University Abu Dhabi, Abu Dhabi, United Arab Emirates}
\altaffiltext{4}{Sub-department of Astrophysics, Department of Physics, University of Oxford, Denys Wilkinson Building, Keble Road, Oxford, OX1 3RH, UK }
\altaffiltext{5}{Department of Astronomy, University of Washington, Seattle, WA, 98195, USA}
\altaffiltext{6}{MIT Kavli Institute for Astrophysics and Space Research, MIT, Cambridge, MA, 02139, USA}
\altaffiltext{7}{Herzberg Institute of Astrophysics, National Research Council, Victoria, BC, V9E 2E7, Canada}
\altaffiltext{8}{Department of Physics and Astronomy, University of Waterloo, Waterloo, ON, N2L 3G1, Canada}
\altaffiltext{9}{Institute for Astronomy, University of Hawaii, 2680 Woodlawn Drive, Honolulu, HI 96822-1839, USA}
\altaffiltext{10}{Department of Astronomy, Peking University, Beijing 100871, China }
\altaffiltext{11}{Kavli Institute for Astronomy and Astrophysics, Peking University, Beijing 100871, China }
%\altaffiltext{10}{European Southern Observatory, Karl-Schwarschild-Str. 1, Garching, D-85748, Germany}
%\altaffiltext{11}{Universit\'{e} Lyon 1, Observatoire de Lyon, Centre de Recherche Astrophysique de Lyon and \'{E}cole Sup\'{e}rieure de Lyon, Saint-Genis Laval, F-69230, France}

\begin{abstract}
We present parameter distributions and fundamental scaling relations for 190 Virgo cluster galaxies in the SHIVir survey. The distribution of galaxy velocities is bimodal about $V_{\rm circ} \sim 125$ \kms, hinting at the existence of dynamically unstable modes in the inner regions of galaxies. An analysis of the Tully-Fisher relation (TFR) of late-type galaxies (LTGs) and fundamental plane (FP) of early-type galaxies (ETGs) is presented, yielding a compendium of galaxy scaling relations. The slope and zero-point of the Virgo TFR match those of field galaxies, while scatter differences likely reflect distinct evolutionary histories. The velocities minimizing scatter for the TFR and FP are measured at large apertures where the baryonic fraction becomes subdominant. While TFR residuals remain independent of any galaxy parameters, FP residuals (i.e. the FP ``tilt") correlate strongly with the dynamical-to-stellar mass ratio, yielding stringent galaxy formation constraints. We construct a stellar-to-total mass relation (STMR) for ETGs and LTGs and find linear but distinct trends over the range $M_{*} = 10^{8-11} \solarm$. Stellar-to-halo mass relations (SHMRs), which probe the extended dark matter halo, can be scaled down to masses estimated within the optical radius, showing a tight match with the Virgo STMR at low masses; possibly inadequate halo abundance matching prescriptions and broad radial scalings complicate this comparison at all masses. While ETGs appear to be more compact than LTGs of the same stellar mass in projected space, their mass-size relations in physical space are identical. The trends reported here may soon be validated through well-resolved numerical simulations.
\end{abstract}

\section{Introduction}
\label{sec:intro}

A major quest of modern astrophysics is to understand the origin of the broad ensemble of observed galaxy properties. The last decade has heralded a new era of large galaxy surveys designed for this (SDSS, \citealp{Abazajian2003}; SAURON, \citealp{Bacon2001}; ATLAS$^{3{\rm D}}$, \citealp{Cappellari2011a}; CALIFA, \citealp{Sanchez2012}; SAMI, \citealp{Croom2012}; MaNGA, \citealp{Bundy2015}; and others) and will provide for hundreds and thousands of galaxies at a time the high-quality spectrophotometric data required to understand the physical drivers of galaxy formation and evolution in a statistical manner. Scaling relations from parameters extracted for these galaxies, such as the velocity-luminosity or Tully-Fisher relation (hereafter TFR, \citealp{Tully1977}, \citealp{Courteau2007b}), the Faber-Jackson relation \citep{Faber1976}, the fundamental plane of galaxies \citep[hereafter FP,][]{Djorgovski1987,Dressler1987,Bender1992,Bernardi2003,Cappellari2006,LaBarbera2008,Cappellari2013a}, and others have provided empirical evidence of the physical laws governing structure formation in our universe (see \citealp{Cappellari2016a} for a review). 

The scaling relation parameters and their scatter depend on a number of factors: structural parameter definitions \citep{Courteau1996, Courteau1997}, environment \citep{Vogt1995,Mocz2012}, fitting algorithms \citep{Courteau2007b,AvilaReese2008,Hall2012}, redshift and peculiar motions \citep{Willick1997,Willick1998,FernandezLorenzo2011,Miller2011}, projection effects and bandpass \citep{Aaronson1986,Hall2012}, morphology \citep{Courteau2007b,Tollerud2011}, stellar populations \citep{Cappellari2006,FalconBarroso2011b,Cappellari2013b}, and metallicity \citep{Woo2008}, to name a few. Furthermore, despite recent progress, galaxy formation models still struggle with basic relations of galaxies, including color dependencies and structural bimodalities \citep{Dekel2006,McDonald2009b}, angular momentum content \citep{Fall2013,Obreschkow2014}, variations in the stellar initial mass function (IMF) \citep{Dutton2011,Cappellari2012,Smith2014}, central versus satellite distributions \citep{RodriguezPuebla2015}, and more.

Fundamental as they may be, dynamical tracers of structure, such as the circular velocity function and stellar-to-halo mass relations (SHMR), still show acute data-model discrepancies. For instance, the SHMR, which probes the efficiency of star formation processes within certain dark matter halos \citep{Leauthaud2012,Grossauer2015,RodriguezPuebla2015}, is shown to peak for $L^{*}$ galaxies and to decline for both larger and smaller halos as a result of mechanisms such as feedback from supernovae and super-massive black holes, halo strangulation, and ram pressure stripping from the cluster and group environments. However, simulation-based SHMRs are notoriously inaccurate, especially at the low- and high-mass ends, because of erroneous model assumptions such as those involving feedback models and other radiative mechanisms \citep{Sawala2015} as well as problematic data-model comparisons \citep{TrujilloGomez2011,Klypin2015}. Tremendous gains in the calibration and study of the SHMR and other galaxy scaling relations could be made if homogeneous, deep, dynamical compilations of complete (i.e. volume-limited) galaxy samples were available, in particular in the low-mass regime of galaxy building blocks. For instance, the combination of photometrically and spectroscopicallydetermined galaxy metrics has yielded stringent tests of $\Lambda$CDM-motivated galaxy formation models through comparisons with observed velocity-size-luminosity relations of galaxies \citep{Dutton2011,TrujilloGomez2011,Cappellari2013b,Dutton2013,Norris2014,Obreschkow2014,Bekeraite2016}, but these still fail to capture the full range of galaxy properties in a complete, homogeneous manner. We attempt to overcome this predicament with a multi-faceted photometric and spectroscopic survey of the Virgo cluster, and present here the first results of the spectroscopic component of our ongoing ``Spectroscopy and H-band Imaging of the Virgo cluster" (SHIVir) survey.

The Virgo cluster is an ideal laboratory for measuring and characterizing galaxy scaling relations because of its proximity, richness, diverse galaxy population, reliable completeness, and extensive ancillary data. While dynamical tracers, such as \hi\ or \ha, may be truncated in galaxy disks, inner dark matter halos are mostly unaffected by cluster interactions (as verified by comparable field and cluster TFRs). Thus, globally relevant conclusions can be reached by studying Virgo cluster galaxies (hereafter VCGs) and contrasted against similar investigations of field galaxies. Other extensive surveys of the Virgo cluster exist, most notably the Canada-France-Hawaii Telescope's ``Next Generation Virgo cluster Survey'' \citep[NGVS,][]{Ferrarese2012}. SHIVir is however unique for its exploitation of wide-field optical and infrared imaging as well as optical long-slit spectroscopy over a wide areal coverage of the Virgo cluster. Thanks to its complement of deep optical imaging, from the Sloan Digital Sky Survey (SDSS) and near-infrared photometry, collected mostly by ourselves (see \se{data}), for a representative sampling of the Virgo cluster, SHIVir's imaging provides a broad census of galaxy stellar masses, ages, and metallicities\footnote{Short of blue spectra for all VCGs, the NIR photometry alleviates the age-metallicity degeneracy endemic to optical imaging.}. The photometric component of SHIVir has enabled the confirmation and/or discovery of structural bimodalities within the Virgo population and other environments \citep{McDonald2009b}, the ubiquity of stellar disks in early-type galaxies \citep{McDonald2011}, stellar population gradients and their connection with formation models of galaxy bulges and disks \citep{Roediger2011a,Roediger2011b}, stellar radial migrations in Virgo disks \citep{Roediger2012}, and a detailed study of stellar mass-to-light versus color transformations \citep{Roediger2015}. 

The spectroscopic component of the SHIVir survey provides critical data, which naturally complement a photometric survey, for the detailed investigation of galaxy structure and evolution. For instance, as one of the main drivers of galaxy evolution \citep{Courteau2014,Cappellari2016a}, a galaxy's dynamical mass requires that it be measured with well-resolved spectroscopic mapping over its projected surface (or at least major axis). The latter then provides the rotational and dispersion profiles required to link mass and light profiles and assess their dependence on stellar populations and environment. Scaling relations based on a dynamical tracer can then be constructed. Spectra are also most valuable for the determination of cluster membership and peculiar motions; for the Virgo cluster, the former was previously assessed by others \citep{Binggeli1985}\footnote{See also the extensive membership revision by \citet{Ferrarese2016}.}, and we are not concerned with cosmic flow studies in this paper \citep[cf.][]{Tully1984,Lee2014}. Because our program targets a broad range of morphological types for VCGs, we are able to combine dynamical tracers for late- and early-type galaxies (hereafter LTG and ETG, respectively) to enable a direct, unique calibration of stellar-to-total mass relations (hereafter STMR; not to be confused with the SHMR) in a single environment for the first time\footnote{Other SHMRs based on heterogenous databases for varied environments have been presented before \citep[e.g.][]{Dutton2011,TrujilloGomez2011}. We return to them in \se{stmr}.}.  

This paper, which presents the dynamical component of the SHIVir survey, is organized as follows. In \se{data} we introduce the spectroscopic catalog and dataset for the SHIVir survey. The construction of surface brightness (SB) and dynamical distributions/bimodalities and scaling relations (TFR, FP, STMR, stellar mass TFR, mass-size relation, and dark-matter-size relation) is presented in \se{results}. Conclusions and thoughts about future investigations are presented in \se{summary}.
 
\section{Data}
\label{sec:data}

% FIGURE
\begin{figure}
\centering
\includegraphics[width=0.45\textwidth]{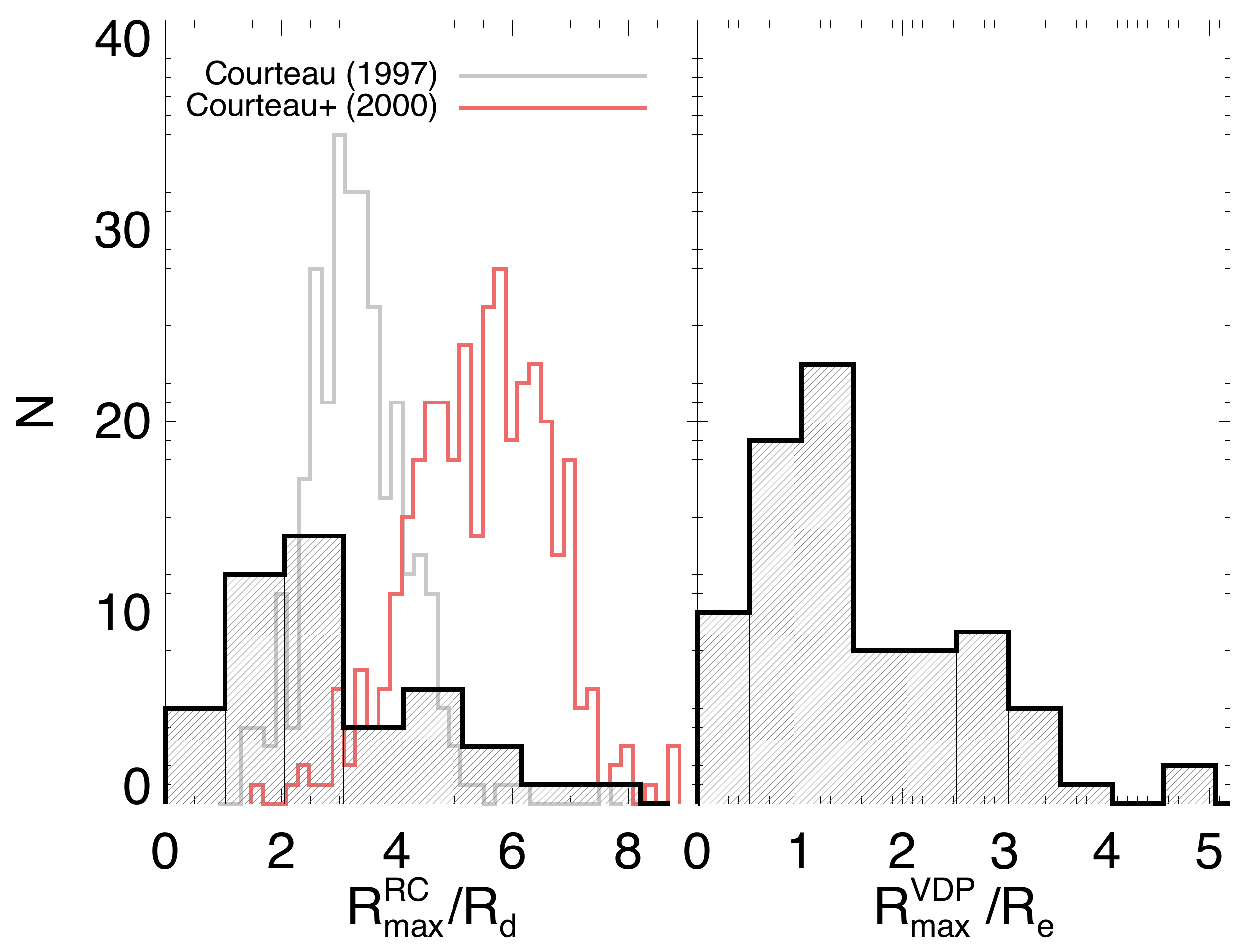}
\caption{(Left) Histogram of the maximum radial extent of the SHIVir \ha\ rotation curves used in our TFR analysis normalized by the $i$-band disk scale length $R_{\rm d}$. Similar data are shown for the field spiral samples of \citet{Courteau1997} and \citet{Courteau2000}. (Right) Histogram of the maximum range of the SHIVir velocity dispersion profiles normalized by the $i$-band effective radius $R_{\rm e}$}.
\label{fig:Radius_Hist}
\end{figure}

The SHIVir survey draws its sample from the magnitude-limited Virgo Cluster Catalog \citep[hereafter VCC]{Binggeli1985} containing 2096 galaxies in a 140 deg$^{2}$ area ($\sim 11.7$ Mpc$^{2}$) around central galaxies M49 and M87. The full SHIVir sample contains 742 VCGs for which $g$-, $r$-, and $i$-band images are available in the SDSS 6th Data Release \citep{AdelmanMcCarthy2008}. A representative subsample of 286 galaxies from the SDSS subsample of VCGs were imaged at $H$-band \citep{McDonald2009b}. The full SHIVir catalog and the $H$-band subsample were both constructed to span the entire range of galaxy morphologies. Each morphological type was also sampled to maintain its proportion within the entire Virgo cluster, see Fig.~2 of \citet{McDonald2009b}. The SHIVir sample is thus meant to be representative of the Virgo cluster. The SHIVir photometric catalog is presented in \citet{McDonald2009b} ($H$-band imaging) and \citet{Roediger2011a,Roediger2011b} (multi-band imaging and stellar population analysis); we refer to these papers for a detailed description of the catalog selection. The SHIVir photometric data are available at http://www.astro.queensu.ca/virgo. 

The construction of unbiased scaling relations relies in part on non-parametric assessments of galaxy structure. Parametric modeling often rests on arbitrary (and internally covariant) multiparameter fitting functions. Our nominal set of non-parametric metrics for galaxies includes local or integrated luminosity $L$, colors, stellar masses $M_{*}$, effective or isophotal SB, radii $R$\footnote{$L_{\rm tot}$ and $R_{\rm e}$ may rely on model-dependent extrapolation of the light profile (growth curve) of a galaxy to infinity or the definition of a galaxy's edge.}, and maximum circular velocity $V_{\rm circ}$ from absorption or emission spectra \citep{Courteau1997,Courteau2007b,Courteau2014}. Photometrically determined values, such as effective or disk scale radius (measured along the semi-major axis) and effective SB, were computed by \citet{McDonald2009b} using isophotal fitting to produce SB profiles from which the quantities were measured. All relevant photometric values are corrected for line-of-sight dust extinction using values from \citet{Schlafly2011}. Throughout this work, $i$-band photometry is favored. This redder band is less affected by dust attenuation, and it allows us to include a larger sample of galaxies in our study, since all 742 SHIVir galaxies have $i$-band photometry available, whereas only 286 VCGs have $H$-band photometry. Radial, luminosity, and inclination parameters are computed via isophotal fitting of galaxy images \citep{Courteau1996}. Radii and SB measurements for LTGs are corrected for inclination-dependent extinction using the method of \citet{Tully1998} and \citet{Hall2012}. All photometric parameters used for the scaling relations presented in this work are from the photometric catalog described here.

In order to directly calibrate the STMR, galaxy stellar masses and other photometrically derived parameters must be contrasted against dynamical masses. While certain direct measurements, such as weak gravitational lensing and satellite kinematics, allow for halo mass measurements out to large radii, they involve complex and approximate methods requiring special circumstances (e.g. the presence of a lensed galaxy or a satellite). On smaller scales, e.g. within the optical radius of a galaxy, the dynamical mass of most galaxies can be estimated quite accurately with circular velocity measurements (rotational or dispersion; see \citealp{Courteau2014} for a review). Providing accurate, well-defined dynamical masses for as many VCGs as possible is one of SHIVir's main goals.

The literature abounds with various, somewhat heterogeneous, mass estimates for VCGs given widely different measurement techniques. Consequently, we embarked in 2008 on a long-term program to acquire our own homogeneous long-slit spectra of VCGs on $4-8$m class telescopes. Deep long-slit spectroscopy was acquired for 138 SHIVir galaxies, including some (40 VCGs) by the ACSVCS \citep{Cote2004} team using both the 2.1 m and the Mayall 4 m telescopes at the Kitt Peak National Observatory (KPNO) using the Ritchey-Chr\'{e}tien Focus Spectrograph. The rest (98 VCGs) were observed by us over the period of 2008--2015 using the following instruments: the Ritchey-Chr\'{e}tien Focus Spectrograph (KPC-007 grating) on the Mayall 4.0 m telescope at the KPNO, the Dual Imaging Spectrograph (B1200/R1200 grating) on the ARC 3.5 m telescope at the Apache Point Observatory (APO), and the Gemini Multi-Object Spectrograph (long-slit mode, B1200 grating) on the Gemini-South 8.2 m telescope. The KPNO spectra covered the wavelength range 3900--5430~\AA, with a spectral resolution of $R \sim 2500$. The APO spectra took advantage of the dual blue and red channels with wavelength coverages of 4160--5420~\AA\ and 6015--7200~\AA, with spectral resolutions of $R \sim 2000$ and $R \sim 4000$, respectively. The Gemini spectra have a wavelength coverage of 4050--5500~\AA\ with a spectral resolution of $R \sim 3750$. The long-slit spectra from KPNO and APO were reduced using a suite of XVISTA routines \footnote{XVISTA is maintained by J. Holtzman, see http://astronomy.nmsu.edu/holtz/xvista/index.html for documentation.}. The Gemini spectra were reduced using the Gemini \texttt{IRAF} package. RC extraction typically required S / N / \AA > 5. The detailed procedures for the extraction of dynamical parameters from our spectra are presented in Ouellette \etal (2017b, in preparation). A brief overview is provided below.

Dynamics for VCG galaxies are measured from red emission features (\ha, \ntwo) for gas-rich systems and blue absorption features for gas-poor systems. The SHIVir emission and absorption spectra are relatively deep, reaching typically beyond $2-4$ $R_{\rm e}$ (see \Fig{Radius_Hist}). The red spectra used $\sim$ 20--30 minute integrations per galaxy on 4 m telescopes while the blue spectra required 1--3 hr per galaxy on 4--8 m telescopes. Because at least half of the Virgo cluster LTGs are stripped by the cluster environment \citep{Koopmann2004}, their \ha\ and \hi\ velocity fields are not as extensive as those of field galaxies (see \Fig{Radius_Hist} for a comparison with \citealp{Courteau1997} and \citealp{Courteau2000}), but they obey the same velocity-luminosity scaling relations (see \se{tfr}). With the added kinematic values taken from reliable literature sources described in \se{ad_data}, the presented SHIVir catalog pertains to a total of 190 VCGs. While the original SHIVir catalog was meant to be representative of the entire VCC catalog, the extensive  integration times of our long-slit spectra prevented the steady observations of dwarf galaxies in absorption (i.e. especially the dwarf ellipticals). Highly inclined LTGs and ETGs with significant emission were generally avoided for cleaner kinematic signatures. The spectroscopic SHIVir sample is fully described in our companion data paper Ouellette \etal (2017b, in preparation).

When available, distances (and their uncertainties) to individual VCGs are taken from \citet{Jerjen2004} and \citet{Blakeslee2009}; otherwise, a distance of 16.5 Mpc \citep{Mei2007} (with an uncertainty of 15\%) is assumed for all other VCGs.

\subsection{Rotational Velocity}
\label{sec:rotvel}

% FIGURE
\begin{figure}
\centering
\includegraphics[width=0.45\textwidth]{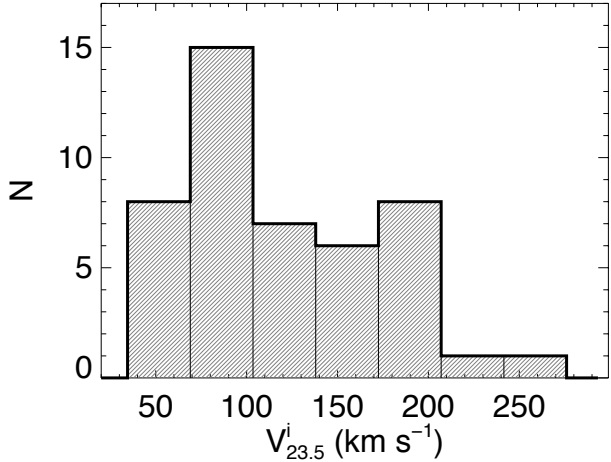}
\caption{Distribution of the extracted inclination-corrected rotational velocities $V^{\rm i}_{23.5}$ used in our TFR analysis.}
\label{fig:Vrot_Hist}
\end{figure}

To construct scaling relations and study the dynamical mass of LTGs, their rotational velocities must be extracted. The wavelength range and resolution of our spectroscopic data allow the creation of \ha\ emission rotation curves (RCs) from which rotational velocities can be measured. 

The SHIVir rotational velocities are extracted by fitting three Gaussian distributions over the \ntwo-\ha-\ntwo\ emission complex. Emission line peaks and uncertainties are computed using an intensity-weighted centroid method \citep{Courteau1997}. The rotational velocities are all corrected for inclination (noted by a superscript `c') using the SHIVir photometric estimates from the reddest, and thus least dust-extincted, images. The typical velocity scatter from duplicate measurements (when literature values were available) is $5-10$ \kms. The constructed RCs are used to select the velocity measure that reduces TFR scatter \citep{Courteau1997} in \se{tfr}. These measurements typically include velocities extracted at isophotal radii, fiducial (e.g. half-light) radii, and metric radii (e.g. in kpc), along the RC. $V^{\rm i}_{2.2}$ is measured at 2.15 disk scale lengths, which corresponds to the peak rotational velocity of a pure exponential disk \citep{Freeman1970,Courteau1997}. $V^{\rm i}_{23.5}$ is measured at the isophotal radius, $R_{23.5}$, corresponding to the $i$-band 23.5 mag arcsec$^{-2}$ isophotal level. The RC can also be collapsed spatially for the measurement of intensity-weighted line widths. $W^{\rm i}_{20}$ is 80\% of the total line width area, as defined by \citet{Courteau1997}. The $V^{\rm i}_{23.5}$ velocity metric yields the smallest TFR scatter (\se{tfr}). $V^{\rm i}_{\rm max}$ is measured along a model fit at the last radial point where the \ha\ tracer is still detected. For that fit, we use the following multiparameter fit function (\citealp[]{Courteau1997}; see also \citealt{Bertola1991}):

\begin{equation}
\label{eq:mpf}
V(R) = V_{\rm crit} \ \frac{1}{(1 + x^{\gamma})^{1/\gamma}},
\end{equation}

\noindent where $x = R_{\rm t}/R$, $\gamma$ controls the degree of sharpness of the RC turnover, $V_{\rm crit}$ is the asymptotic maximum velocity, and $R_{\rm t}$ is the radius at which the transition between the rising and flat parts of the RC occurs. When the RC reaches a flat regime, an average of the observed velocities measured at the appropriate radius is taken. In other cases, nominal velocities are taken from a model fit at the radii described above. 

While the SHIVir compilation of dynamical parameters includes 190 VCGs, visible \ha\ emission was present in only 46 of them. Thirteen complex systems (interacting, highly inclined, or spurious systems) were excluded from our study of RCs and our TFR analysis, resulting in only 33 galaxies with novel, clean extended RCs. We add to our own investigation 6 RCs from \citet{Rubin1997} and 7 RCs from \citet{Chemin2006}. The distribution of rotational velocities for these 46 late-type VCGs with extended RCs is shown in \Fig{Vrot_Hist}. We are also able to build a line width TFR (\se{tfr}). We obtained $W^{\rm i}_{20}$ line widths for 38 of of our 46 SHIVir VCGs with visible \ha\ emission. We augment our line width TFR with \hi\ line widths for 27 VCGs from the ALFALFA $\alpha$.100 catalog \citep{Haynes2011}. Only ALFALFA galaxies with $W^{\rm i}_{50} > 30$~\kms\ are retained because of the instrument's resolution limit.

The maximum radial extent of all the RCs, normalized by disk scale length $R_{\rm d}$, is shown in \Fig{Radius_Hist} (left panel). As previously mentioned, the RCs are slightly truncated because of the cluster environment. This is obvious when comparing with field environments, e.g. the Sb-Sc field sample of \citet{Courteau1997} and \citet{Courteau2000} shown in \Fig{Radius_Hist}. While the majority of our galaxies do not extend beyond a radius of $3R_{\rm d}$, a fair number still extend beyond $4R_{\rm d}$. We do not find a strong correlation ($r = -0.12$) between $R_{\rm max}/R_{\rm d}$ and absolute magnitude $M_{i}$.

\subsection{Velocity Dispersion}
\label{sec:veldisp}

% FIGURE
\begin{figure}
\centering
\includegraphics[width=0.45\textwidth]{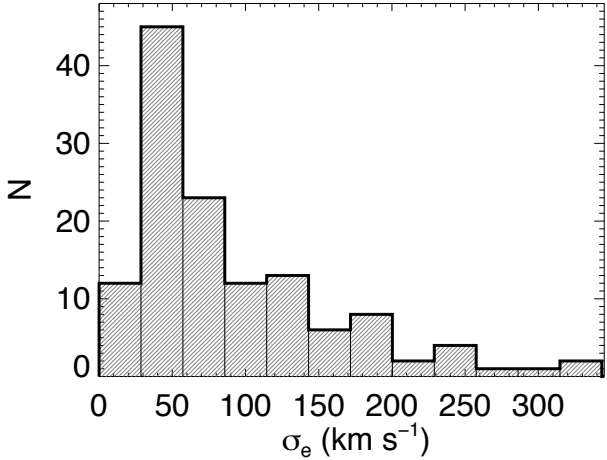}
\caption{Distribution of the SHIVir integrated effective velocity dispersions $\sigma_{\rm e}$.}
\label{fig:Sigma_Hist}
\end{figure}

The velocity dispersion, $\sigma$, is a characteristic kinematic parameter for pressure-supported systems such as ETGs. Stellar kinematics can be extracted from the ETG absorption spectral features using the penalized pixel-fitting method \texttt{pPXF} \citep{Cappellari2004,Cappellari2017}. This algorithm constructs a best-fit linear combination of stellar templates to the observed galaxy spectrum's line-of-sight velocity distribution (LOSVD) parametrized by a Gauss-Hermite function, which is characterized by a mean (rotational) velocity $V$, velocity dispersion $\sigma$, skewness $h_{3}$, and kurtosis $h_{4}$. We used the stellar templates from the MILES library \citep{SanchezBlazquez2006,FalconBarroso2011a} to extract integrated and resolved velocity dispersions. For integrated measurements, spectra collapsed spatially over a range of aperture sizes (e.g. $R_{\rm e}/4$, $1 R_{\rm e}$, $2 R_{\rm e}$, etc.)\footnote{Our use of long-slit spectroscopy restricts our apertures to the stripe of light that is collected along each galaxy major axis. Comparisons with integrated velocity dispersions determined from 2D IFU data are unfortunately not straightforward; this issue will be more closely addressed in Ouellette et.~al. (2017b, in preparation).} are fitted by \texttt{pPXF}. Since it is a summed spectra that is fitted, the resulting velocity dispersion measurement is a $V_{\rm rms} = \sqrt{V_{\rm rot}^{2} + \sigma^{2}}$. These characterize the dynamical mass enclosed within each aperture (see \se{dyn_mass}). They can be used to characterize the FP scatter dependence on the $\sigma$ aperture size (see \se{fp}). No aperture correction is performed. For resolved velocity dispersion measurements, from which dispersion profiles are built, spectral rows are binned radially (along the slit) over three pixels for each radial data point. In the dimmer outer regions of the galaxy, rows are binned until the necessary S / N level of 50 / \AA\ is reached. These dispersion measurements characterize the galaxy kinematics locally; velocity dispersion profiles are constructed from them. The maximum radial extent of these dispersion profiles, normalized by $R_{\rm e}$, is shown in \Fig{Radius_Hist} (right panel). We find only a moderate correlation ($r = -0.33$) between $R_{\rm max}/R_{\rm e}$ and absolute magnitude $M_{i}$.

Various emission lines are masked to enable reliable \texttt{pPXF} fits. While we wish to fit galaxy spectra from as broad a stellar template library as possible, the size of the MILES library (985 templates) and the large number of radial LOSVD measurements required to build a resolved profile would make this effort computationally prohibitive. Instead, we use a spatially collapsed (`mashed') spectrum over an aperture of 2$R_{\rm e}$ for each galaxy for which \texttt{pPXF} selects an optimal stellar template subcatalog from the entire MILES library. Barring any notable variations in the stellar populations of our galaxies along their radii (of which we found none), this yields a subcatalog of approximately $15-20$ stellar templates chosen from the MILES library for each galaxy that produces stable fits in a time-efficient manner. The choice of feature masking and template selection can significantly impact dynamical measurements; this is further explored in Ouellette \etal (2017b, in preparation). 

An integrated central velocity dispersion, $\sigma_{0}$ (taken within an aperture of $R_{\rm e}/8$), was measured for 131 VCGs, whereas an integrated effective velocity dispersion $\sigma_{\rm e}$ could be measured for 128 VCGs (see \Fig{Sigma_Hist}), 88 of which are ETGs. The integrated velocity dispersions for these ETGs are used to extract the tightest FP (\se{fp}).

\subsection{Additional Data}
\label{sec:ad_data}

We supplemented our dynamical catalog with a number of literature sources; all data presented in this paper concern members of the SHIVir catalog (742 galaxies in total; see \se{data}). The morphological classification is taken from the GOLDMine database \citep{Gavazzi2003}, corresponding to numerical Hubble types ranging from $-3$ to $20$. For reference, SHIVir ETGs have a Hubble type between $-3$ and $2$ inclusive, while LTGs range from $3$ to $20$ inclusive.

%Our TFR analysis (\se{tfr}) relies on 31 SHIVir RCs (measured by ourselves) supplemented with 8 RCs from \citet{Rubin1997} and 7 from \citet{Chemin2006}, for which we have fully resolved RC data required for our velocity metric study. The radial extent of those RCs, shown in \Fig{Radius_Hist}, matches that of SHIVir RCs indicating that any truncation of the RCs for Virgo galaxies likely results from the cluster environment rather than any systematic effect or bias. We also augment our $W^{\rm i}_{20}$ line width TFR with 27 VCGs from the ALFALFA $\alpha$.100 catalog \citep{Haynes2011}: these were VCGs for which SHIVir did not measure a line width. 

Our TFR sample is described in \se{rotvel}. Our FP analysis (\se{fp}) only contains SHIVir kinematics in order to study the FP scatter based on different velocity dispersion metrics. For our bimodality (\se{bimo}) and mass relation studies (Sections \ref{se:stmr}--\ref{se:NFW}), multiple dynamical values are used from the following supplementary sources: 72 values from ACSVCS (P. C\^ot\'e 2011, private communication), 43 values from \citet{Fouque1990}, 30 values from \citet{Rubin1997} and \citet{Rubin1999}, 14 values from \citet{Geha2003}, 12 values from \citet{vanZee2004}, 47 values from ATLAS$^{3{\rm D}}$ \citep{Cappellari2011a}, 29 values from SMAKCED \citep{Toloba2011}, 57 values from ALFALFA \citep{Haynes2011}, and 7 values from \citet{Rys2014}. Note that many VCGs in our catalog have values available from multiple sources, which is why the total sum of kinematic values is larger than 190, which is the size of our object catalog. The typical dispersion between multiple estimates of the circular speed is only $10-15$ \%. When multiple entries are available for a galaxy target, we use their statistical average. This procedure ensures that galaxy structural parameters are not counted twice.

% FIGURE
\begin{figure*}
\centering
\includegraphics[width=0.99\textwidth]{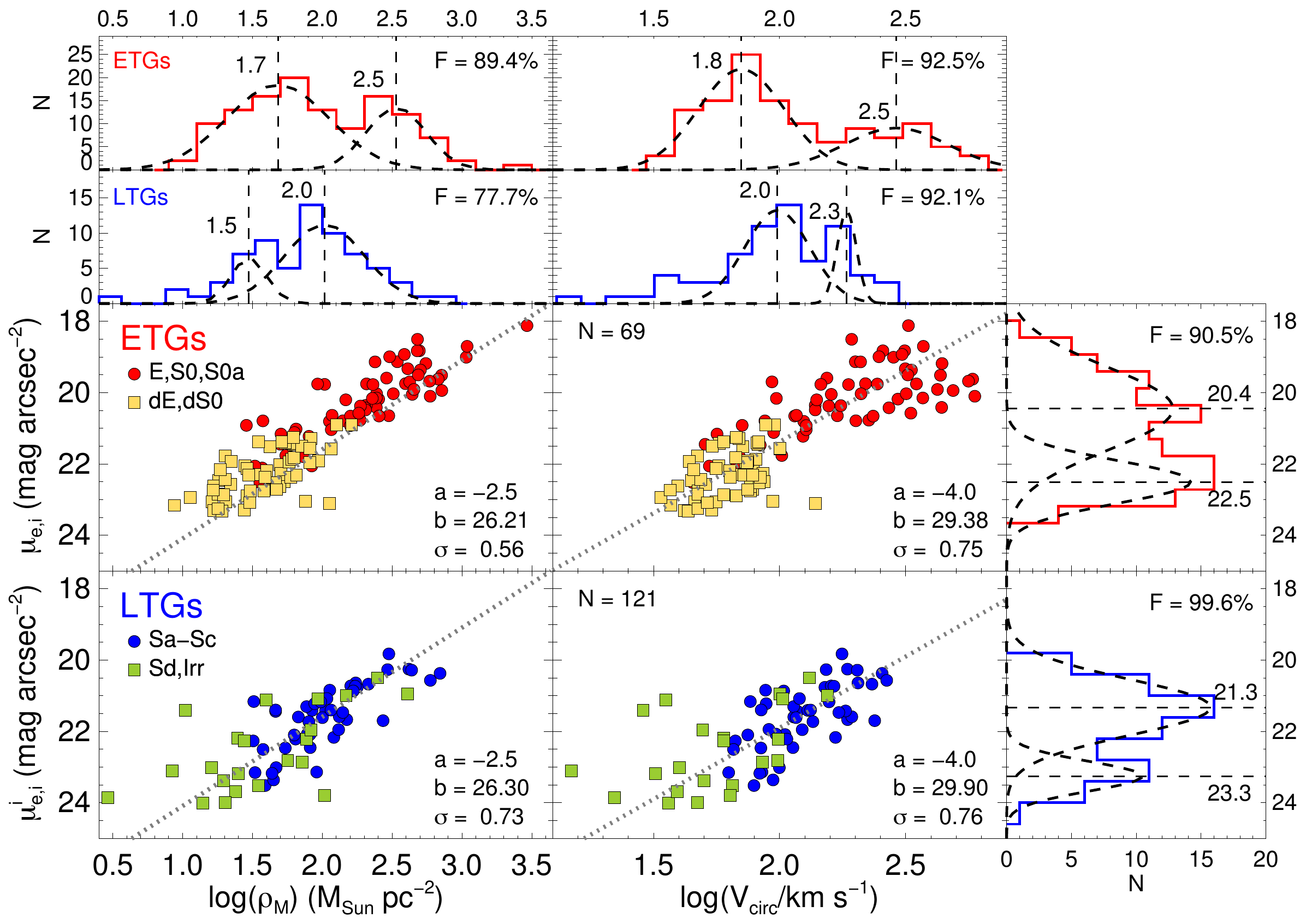}
\caption{Log-scale dynamical mass density and circular velocity versus $i$-band effective SB for ETGs and LTGs. SB is inclination-corrected, $\mu^{\rm i}_{{\rm e},i}$, for LTGs only. Dynamical masses used in $\rho_{M}$ are measured within $r_{1/2}$ (physical radius). The distributions of the mass density, velocity and SB parameters are shown on the periphery as histograms fitted with a double Gaussian function; their peaks are highlighted with a dashed line. {\it F}-test confidence results for a double versus a single Gaussian distribution and sample sizes $N$ are indicated. The $\log \rho_{M}$ vs. $\mu_{\rm e}$ and $\log V_{\rm circ}$ vs. $\mu_{\rm e}$ relations have a fixed slope of $-2.5$ and $-4$ respectively (dotted lines). The scatter $\sigma$ about the best-fit line is also indicated.}
\label{fig:Bimodality_i}
\end{figure*}

\subsection{Stellar and Dynamical Masses}
\label{sec:dyn_mass}

Our computation of stellar mass exploits color transformations such as those presented in \citet{Roediger2015}. The SDSS colors $g-r$, $g-i$, $g-z$, and $g-H$ are used to constrain mass-to-light ratios versus color relations (MLCRs), which allows for optimal modeling and fitting of SEDs using a Chabrier IMF \citep{Chabrier2003}, and then produces stellar mass-to-light ratios from which a stellar mass can be inferred. The colors are computed from the SHIVir photometric catalog described in \se{data}. Our $g$-band and $H$-band SB profiles typically reach a depth of 26 mag arcsec$^{-2}$ and 24 mag arcsec$^{-2}$, respectively \citep{McDonald2011}. The stellar mass errors in this paper account for random uncertainties only. Systematic errors due to the IMF choice may exceed 0.3 dex;  an additional 0.2 dex may contribute to the error budget because of other modeling choices.

Dynamical masses within a fiducial projected radius, $R$, are inferred for rotating disks via the following:

\begin{equation}
\label{eq:mdyn}
M_{\rm dyn}(R) = V_{\rm circ}^{2} R/G,
\end{equation}

\noindent where $G$ is the gravitational constant. We take circular velocity to be $V_{\rm circ} = V^{\rm i}_{\rm rot}=V^{\rm i}_{\rm 23.5} = V_{23.5}/{\rm sin} i$ for LTGs, where $i$ is the inclination of the galaxy disk and the superscript ``c" indicates that the velocity is corrected for inclination. For ETGs, analogous values of $M_{\rm dyn}$ measured inside an effective spherical radius can be computed using

\begin{equation}
\label{eq:mdyn_etgs}
M_{\rm dyn}(r_{1/2}) = c \frac{r_{1/2} \sigma^{2}_{\rm e}}{G},
\end{equation}

\noindent where the structural constant $c$ is computed from the function

$$c = -0.300n + 4.153,$$

\noindent built from the values in Table II of \citet{Courteau2014} and the S\'{e}rsic index $n$ computed from our $i$-band bulge-disk decompositions of the total light profiles described in \citet{McDonald2011}. The constant $c$ is computed for apertures based on a physical spherical radius, namely $r_{1/2}$. It is warranted here to reflect upon the transformation from projected to physical radius, i.e. $r(R)$. The two quantities are comparable for an LTG on the major axis for a pure stellar system. For a spherical ETG, it has been shown that $r_{1/2} \approx (4/3)R_{\rm e}$ for pure stellar systems \citep{Hernquist1990,Ciotti1991}. We have also verified that the assumption of $r_{23.5} \approx (4/3)R_{23.5}$ with a spherical Hernquist profile yields the same enclosed mass within a cylinder of radius $R_{23.5}$ and a sphere of radius $r_{23.5}$ to $\sim 1$\% accuracy\footnote{The ratio $R_{23.5}/R_{\rm e}$ for SHIVir galaxies is $\sim 2$ and ranges from $\sim 1$ to 6. The full range of $R_{23.5}$ can be seen in kpc in \Fig{MassSize}.}. We assume that \Eq{mdyn_etgs} holds true for a spherical radius of $r_{23.5}$. For most of our analysis, we consider masses within $r_{23.5}$, as it matches the radius at which $V_{\rm circ}$ minimizes the TFR scatter (see \se{tfr}) and beyond which surface brightness errors become significant. We find the integrated velocity dispersion for ETGs to not vary greatly enough between ones measured within $r_{1/2}$ and ones measured within $r_{23.5}$ to affect dynamical masses beyond our current level of uncertainty. The variation between $\sigma_{R_{\rm e}}$ and $\sigma_{2R_{\rm e}}$ as defined in \se{fp} is between -15\% and +10\%\footnote{This variation in both directions matches the diversity of resolved velocity dispersion profile shapes (including both rising and falling profiles) in our galaxy sample. While the change in velocity dispersion within $R_{\rm e}$ and $R_{23.5}$ appears small for any given galaxy, the compound effect of these variations on galaxy scaling relations can be significant, as discussed in \se{fp}}. Consequently, we also use $\sigma_{\rm e}$ to measure mass within $r_{23.5}$.

Various attempts to rewrite the Jeans equation as a linear transformation from velocity dispersion to circular velocity have been made (\citealp[][and references therein]{Courteau2007a}; \citealp{Serra2016,Cappellari2017}). \citet{Serra2016} empirically found that $V_{\rm circ}=1.33\sigma_e$ (or $\rho(r) \sim r^{-2.2}$) for ETGs with $\sigma_e > 100$ \kms\ over the broad range $4-6 R_{\rm e}$. However, most of our ETGs have dispersions with $\sigma_e < 100$ \kms, where the density profile is likely shallower ($\rho(r) \sim r^{-1.8}$; \citealp{Cappellari2016a}); a direct calibration between $V_{\rm circ}$ and $\sigma_e$ as in \citet{Serra2016} for this small dispersion range is currently lacking. We chose to use \Eq{mdyn} and \Eq{mdyn_etgs} to set $V_{\rm circ} = \sqrt{c} \times \sigma_{\rm e}$, where $c$ is again the ``virial'' coefficient. We adopt this prescription for the remainder of this work, but note that using the \citet{Serra2016} prescription (extrapolated to small dispersions) would yield similar scaling relations (same slopes), albeit with differences in zero-points on the order of $0.2-0.3$ dex.

\section{Dynamical Distributions and Scaling Relations}
\label{sec:results}

The theoretical basis of galaxy dynamical relations such as the TFR \citep{Tully1977} and the FP of elliptical galaxies remains ill-constrained \citep{Dutton2011,TrujilloGomez2011,Cappellari2016a,Desmond2017}, especially at low masses. Covariances between physical variables such as stellar IMF and baryonic-to-dark matter ratio thwart a conclusive construction of these relations. Still, in addition to expanding theoretical models, the way forward for characterizing the global manifold of galaxy scaling relations is via a comprehensive multiparameter mapping of galaxies that includes dynamics, as we present below. Such an analysis will benefit from understanding the distributions of various key parameters; for instance, the photometric parameters are reviewed in \citet{McDonald2011}, \citet{Roediger2011a}, and \citet{Roediger2011b}. A bimodal distribution in the SBs of disk galaxies is indeed found \citep{Tully1997,McDonald2009a,McDonald2009b,Sorce2013} and our analysis of VCG velocities supports a dynamical connection (see \se{bimo}). As we revisit the TFR and FP to finally build the STMR for VCGs, we wish to tie these various aspects together to unveil new lines of galaxy evolution exploration.

\subsection{A Dynamical Bimodality}
\label{sec:bimo}

Following the discovery by \citet{Tully1997} of a bimodality in the distribution of SBs for UMa cluster (disk) galaxies, \citet{McDonald2009b} used SHIVir optical and infrared imaging to corroborate their finding in the Virgo cluster LTG population. They also extended the notion of SB bimodality to Virgo ETGs. In essence, in each LTG and ETG galaxy class, giant and dwarf galaxies exhibit SB peaks separated by $\sim$2 mag arcsec$^{-2}$. The troughs (or gaps) between these peaks for LTGs and ETGs correspond to a relative paucity of Sc/Sd galaxies and faint ETGs, respectively. The SB peaks for ETGs are also naturally shifted toward brighter systems relative to the LTGs, such that the brightness peak for the fainter ETGs roughly coincides with the trough between the LTG peaks. The current empirical evidence for galaxy SB distributions indicates an environmentally independent structural dichotomy for LTGs, such that high surface brightness (hereafter HSB) galaxies have two distinct classes of high- and low-concentration bulges, likely correlated with low and high central dark matter fractions, whereas low surface brightness (hereafter LSB) galaxies have only low concentration bulges with high central dark matter fractions \citep{McDonald2009b}. There is evidence that the HSB LTG peak may be related to the LSB ETG peak via disk fading on the order of $\sim 1-1.5$ \magarc \citep{Dressler1980,Kent1981}, which partially explains the shift between the peaks of the two galaxy types.

The SB bimodality might emerge from galaxy systems whose baryon and dark matter fractions are comparable within the optical radius, potentially yielding dynamical instabilities; these systems would adjust their equilibrium structure rapidly, thus explaining the observed dearth of intermediate SB systems\footnote{It is noted that the brightness bimodality and the dip in the Virgo cluster luminosity function at $m_i \sim 12.5$ or $M_i \sim -19$ ($M_g \sim -17.5$) \citep{McDonald2009b} are manifestations of two related but different phenomena; the former applies to LTGs and ETGs classes taken separately -- that is, each galaxy class displays its own bimodality -- whereas the latter reflects a transition between the giant and dwarf systems, all classes considered, in the Virgo cluster.}. \citet{Sorce2013} also reinforced the notion of SB bimodality in field LTGs using {\it Spitzer} data\footnote{Bimodality is best measured at infrared wavelengths where dust extinction is minimized.}.

% FIGURE
\begin{figure*}
\centering
\includegraphics[width=0.99\textwidth]{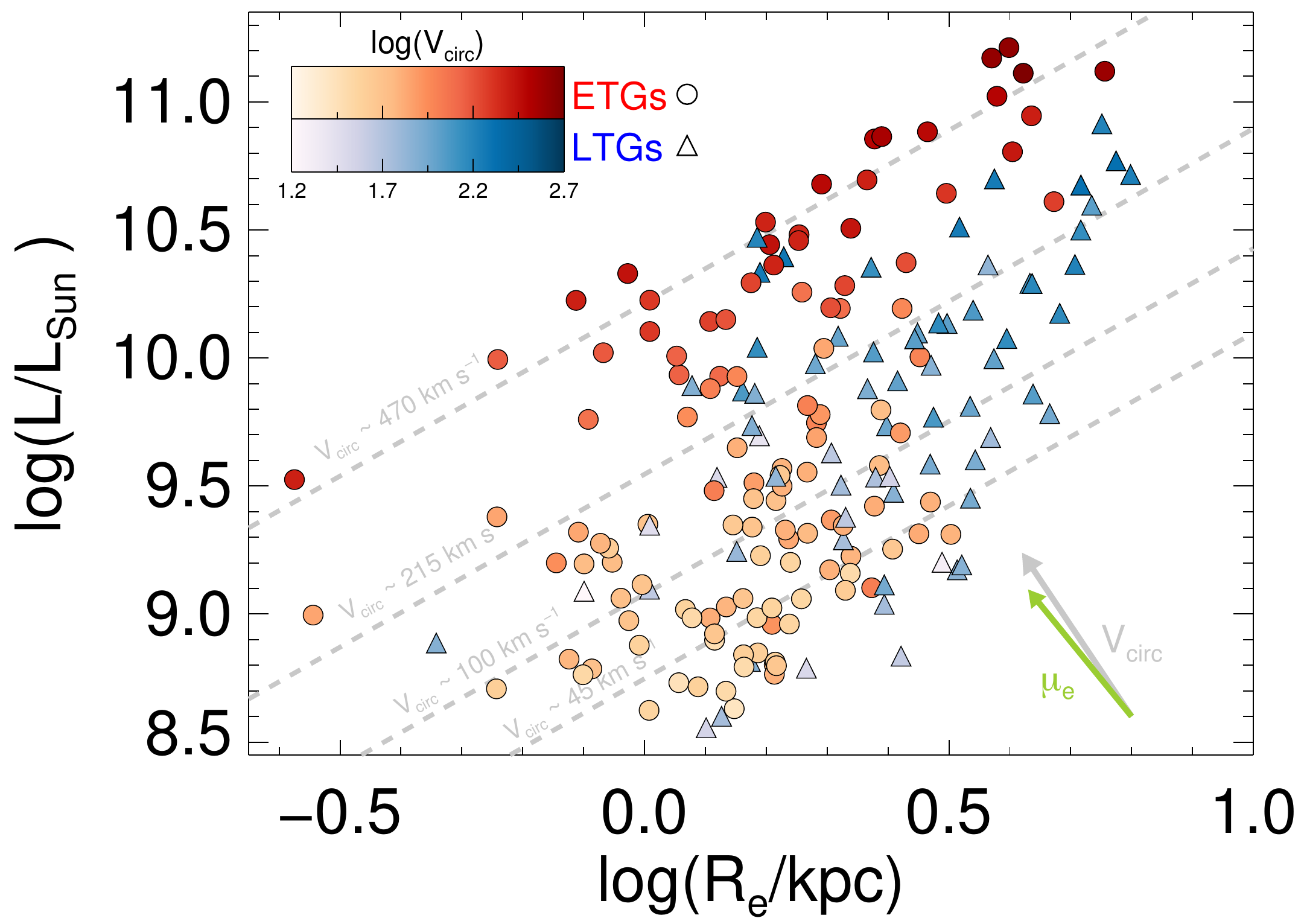}
\caption{Luminosity-size diagram of 190 SHIVir VCGs (the sample is the same as in \Fig{Bimodality_i}). $L$ is the total $i$-band luminosity, and $R_{\rm e}$ is the \emph{projected} effective radius. ETGs are plotted as circles in shades of red, LTGs are plotted as triangles in shades of blue. The saturation of blue and red indicates $V_{\rm circ}$. Dashed lines show lines of constant $V_{\rm circ}$ (approximate best-fit lines for four circular velocity bins from our observational data), and the (nearly parallel) arrows indicate the directions of increasing $V_{\rm circ}$ and $\mu_{\rm e}$. 
}
\label{fig:Lum_Size}
\end{figure*}

If stable configurations are preferred, the SB bimodality ought to be linked to dynamics as proposed by \citet{Tully1997}, \citet{McDonald2009b}, and \citet{Sorce2013}. To assess whether the SB bimodality observed in \citet{McDonald2009b} is dynamically rooted, \Fig{Bimodality_i} compares the distributions of $i$-band effective SB $\mu_{{\rm e},i}$ versus dynamical mass density measured inside the physical radius $r_{1/2}$ (left panels) or circular velocity (right panels) with $V_{\rm circ} = V^{\rm i}_{23.5}$ for all LTGs or $V_{\rm circ} = \sqrt{c} \sigma_{\rm e}$ for all ETGs (as defined in \se{dyn_mass}). The effective SBs are corrected for inclination for the LTGs, $\mu^{\rm i}_{{\rm e},i}$, but not for the ETGs, $\mu_{{\rm e},i}$; that choice of a (somewhat uncertain) correction does not alter the shape of the brightness distribution or affect our conclusion about brightness bimodality for ETGs. Correcting ETG SBs would offset their SB peak locations by $0.3-0.6$ \magarc, but this is not enough to align the ETG SB peaks with the LTG SB peaks\footnote{Four of the LTGs have very low $V^{\rm i}_{23.5}$ measurements that are inconsistent with their intermediate $\mu_{\rm e}$ values, partially because of their extremely truncated RCs (considerably shorter than $R_{\rm d}$). Inspection of their SDSS images confirmed the absence of well-ordered disk-like structure. Velocity dispersion measurements for three of these galaxies are used to compute a $V_{\rm circ}$ value instead of using $V^{\rm i}_{23.5}$ directly. This procedure aligns the galaxies closer to the virial slope of $-4$. The remaining Irr galaxy, VCC 1675, lacked a $\sigma_{\rm e}$ value and is significantly offset from the rest of the LTG sample at $\log V_{\rm circ} \sim 1.18$.}.

The dotted lines drawn in \Fig{Bimodality_i} for $\mu_{\rm e}$ versus $\log V_{\rm circ}$ have a virial slope of $-4$. This results from inserting

$$R_{\rm e} = \frac{G M^{R_{\rm e}}_{\rm dyn}}{V_{\rm circ}^{2}},$$

\noindent taken from \Eq{mdyn}, into the following definitions:

\begin{equation}
\label{eq:intensity}
I = \frac{L}{4 \pi R^{2}},
\end{equation}

\begin{equation}
\label{eq:mu}
\mu = -2.5 {\rm log} I + C_{\lambda},
\end{equation}

\noindent where $I$ is the physical SB, $L$ is the luminosity, $\mu$ is the observed SB, and $C_{\lambda} = M_{\astrosun,\lambda} + 21.572$, a wavelength-dependent constant. We take all values to be measured at or within $R_{\rm e}$. Using these definitions and provided assumptions about dynamical $M/L$ (\citealp{Zwaan1995}, \citealp[see][for caveats about this derivation]{Courteau2007b}), we derive

\begin{equation}
\mu_{\rm e} \propto {\rm log} V_{\rm circ}^{-4}.
\end{equation}

\noindent We also determined the theoretical slope for $\mu_{\rm e}$ versus $\log \rho_{M}$ by using Eqs.~\ref{eq:intensity} and \ref{eq:mu} and defining

$$\rho_{M} = \frac{M}{4 \pi R^{2}}.$$

\noindent This allows us to describe $\mu$ as a function of $\log \rho_{M}$:

$$\mu_{\rm e} = -2.5 \log \rho_{M} + C_{\lambda} + 2.5 \log \bigg{(}\frac{M}{L}\bigg{)}.$$

An {\it F}-test determines the probability, shown next to each histogram, that the observed distributions originate from two distinct Gaussian populations rather than a single one. The {\it F}-test confidence for brightness bimodality exceeds 91\% for both LTG and ETG VCGs; these values differ slightly from those of \citet{McDonald2009b} since the samples also differ slightly. The one presented here reduces the test from the 286 VGC galaxies in \citet{McDonald2009b} to 190 VCGs that must have simultaneous SB and reliable velocity information (sources for kinematics are listed in \se{ad_data}). The version of \citet{McDonald2009b} of this figure relied on the more uncertain and heterogeneous HyperLEDA database for kinematic values \citep{Paturel2003}. The matching bimodality seen in \Fig{Bimodality_i} for $\log \rho_{M}$ and $\log V_{\rm circ}$ also yields F-values higher than $78\%$, or with a significance of $\sim 1.3\sigma$. While this is lower than the ideal $3\sigma$ confidence threshold, the SB bimodality has been observed in a number of environments by a number of other works (\citealp{Tully1997}; M09; \citealp{Sorce2013}), while there have been no studies refuting this bimodality. These statistics again differ slightly from those of M09 since the samples also differ slightly. Henceforth, our discussion of bimodality focuses on $\mu_{\rm e}$ versus $\log V_{\rm circ}$, as $V_{\rm circ}$ is a direct observable, whereas $\rho_{M}$ must be computed from velocity and radius jointly.

A separation of the ($\mu_{\rm e}$, $\log V_{\rm circ}$) relation by Hubble types shows very distinct groupings\footnote{If morphological types are deemed somewhat subjective, a group separation by concentration (see \Eq{concentration}) or $g-i$ color yields the same results and conclusions.}. For the ETGs, the brightness bimodality is clearly delineated by giant (E-S0) versus dwarf (dE-dS0) galaxies. While the morphological dichotomy is trivial, by definition of the types themselves, it is remarkable that the two galaxy groups are equally well delineated in velocity space; the SB gap at $\sim$21 mag arcsec$^{-2}$ lines up with the gap in velocity at $\log V_{\rm circ} \sim 2.1$ ($V_{\rm circ}=125$ \kms\ at $R_{\rm e}$) for the ETGs. The analog for LTGs is slightly muddled by the fact that SHIVir currently suffers from a dearth of Virgo spiral galaxies. While the SHIVir survey was built to be morphologically representative of its parent catalog, the VCC, more recent surveys such as the NGVS (complete down to $M_{*} = 10^{6} \solarm$ and 50\% complete down to $M_{g} = -9.13$ mag) have shown Virgo to contain a larger number of fainter spirals and dwarfs than was previously thought \citep{Ferrarese2016}. A tentative SB gap occurs at $\sim$22.5 mag arcsec$^{-2}$ and again at $\log V_{\rm circ}=2.1$ for the LTGs, despite a significant Hubble type overlap between the two peaks.

In order to understand the dynamical dichotomy, it is speculated that a galaxy undergoes rapid structural readjustments in unstable regimes where baryonic and dark matter are co-dominant. \citet{McDonald2009b} and \citet{Sorce2013} discussed a scenario whereby a rotating system that retains (does not shed) large amounts of angular momentum inhibits the flow of baryons to its center thus delaying the onset of rotational equilibrium at a given radius in the galaxy \citep[see also][]{Dalcanton1997}. The gap between the peaks, whether traced by SB or circular velocity, would then reflect a configuration where baryons and dark matter are equally dominant by mass within the optical radius. We discuss the possible implications of this scenario for the ETG population in \se{lumsize}. While this qualitative picture has physical appeal, the values found in \Fig{Bimodality_i} for the gap velocities ($V^{\rm i}_{23.5} \sim 125$ \kms) seem low compared to those inferred from scaling relation arguments: see Fig.~1 of \citet{Courteau2015}. According to the latter, a 50\% dark matter fraction measured at $\sim 1.3 R_{\rm e}$ ($2.2 R_{\rm d}$) would be found for LTGs rotating at $\sim$ 200 \kms\ at that radius. These two disparate estimates indicate that the problem of dynamical stability in galaxy disks and spheroids requires additional insight, as possibly provided by numerical explorations of galaxy structure with a range of baryons and dark matter at all radii.

\subsection{Size-Luminosity Relation}
\label{sec:lumsize}

The dependence of luminosity/mass on size and velocity has been investigated by the ATLAS$^{3{\rm D}}$ collaboration \citep{Cappellari2013b,Cappellari2016a}. They presented a distribution with a critical mass of $M_{*} \sim 2 \times 10^{11} \solarm$ above which their massive slow-rotator ETGs lie and below which low-mass spirals and fast-rotator ETGs are found. The authors suggested that galaxies evolve across lines of constant $\sigma_{\rm e}$ at low mass and along lines of constant $\sigma_{\rm e}$ at high mass. This notion of a two-stage evolution consistent with our results was introduced in \citet{Faber2007} as a mixed scenario in which blue spirals accrete gas and are eventually quenched into red ellipticals, which then combine via dry mergers to form the most massive ellipticals. \Fig{Lum_Size} revisits this scenario with a distribution of $\log L_{i}$ versus $\log R_{\rm e}$ as a function of circular velocity (for the mass-size relation, see \Fig{MassSize}, top panels). For a given {luminosity,} circular velocity, or fixed dynamical mass, LTGs have noticeably larger effective radii than ETGs. This is expected, since ETGs typically have a more centrally concentrated mass distribution (and higher concentration values) at fixed $V_{\rm circ}$. The difference in mass distribution between the two galaxy classes may indicate an evolutionary sequence between them. To characterize the evolution of different galaxy populations into one another, the luminosity/mass-size distribution of the Virgo cluster can be compared to that of field galaxies and a more evolved cluster such as Coma, as was done in \citet{Cappellari2013c}. The field was found to have a larger fraction of spiral galaxies, whereas Coma has a notable dearth of spirals. LTGs and ETGs have closer proportions within the SHIVir sample, implying that the cluster environment plays a key role in processing spirals into ellipticals. Indeed, the study of \citet{Faber2007} of the evolution of the blue/red galaxy fraction over time showed that the progenitors of the present-day red ETGs must exist in the blue LTG population at $z \geq 1$. 

A scenario whereby red ellipticals formed as blue spirals that are eventually quenched --- via AGN feedback \citep{Granato2004,Springel2005, Dubois2013}, winds \citep{Murray2005}, and other heating mechanisms --- may explain the SB and dynamical bimodality in the ETGs seen in \se{bimo}. Accreted gas turned into stellar mass and bulge growth would increase a galaxy's $V_{\rm circ}$, while star formation shutdown would redden its color. Any bimodality existing in the spiral population as a result of the aforementioned dynamical instability during disk formation could be retained in this blue-to-red evolutionary track. Disk fading \citep{Kent1981} as a possible evolutionary mechanism between spirals and lenticulars would affect the fraction of spirals in the Virgo environment and partially cause the SB peaks' shift between LTGs and ETGs as seen in \Fig{Bimodality_i}. However, staggered quenching --- wherein we find a correlation between halo mass and quenching epochs --- as an evolutionary mechanism in $L^{*}$ galaxies has been shown to increase scatter in growth and star formation histories \citep{Terrazas2016}, potentially muddying any bimodality left over from disk formation instabilities. Minor mergers required to form spheroidal geometries may also add a secondary dynamical instability by which the ETG bimodality is created. Instances where only quenching has occurred may explain the creation of S0s/dS0s for which a disk is still present. The evolution from blue spirals to small red ellipticals would keep much of the disk, and thus the rotational component of the kinematics, intact. Indeed, this is where ATLAS$^{3{\rm D}}$ places their fast-rotator ETGs. Via dry mergers, these fast-rotator ETGs gain stellar mass but also increase in size as they grow larger and more spheroidal \citep{Toomre1977,Kaviraj2014}, therefore moving along lines of constant $\sigma_{\rm e}$ or $V_{\rm circ}$, and along the red sequence. Major mergers almost certainly play an important role in the creation of massive spheroidals ($M_{*} > 10^{10.7} \solarm$), but auxiliary mechanisms such as morphological transformations \citep{Bundy2007} are likely required to drive the observed evolution from intermediate-redshift progenitors to the massive spheroidal population seen in the local universe. Unfortunately, the bimodality observed in the ETG population cannot be attributed to the separation between fast- and slow-rotator ETGs \citep{Emsellem2011}, as the critical mass of this classification is much too high at $M_{*} \sim 2 \times 10^{11} \solarm$.

\subsection{Velocity-Luminosity (Tully-Fisher) Relations}
\label{sec:tfr}

% FIGURE
\begin{figure*}
\centering
\includegraphics[width=0.99\textwidth]{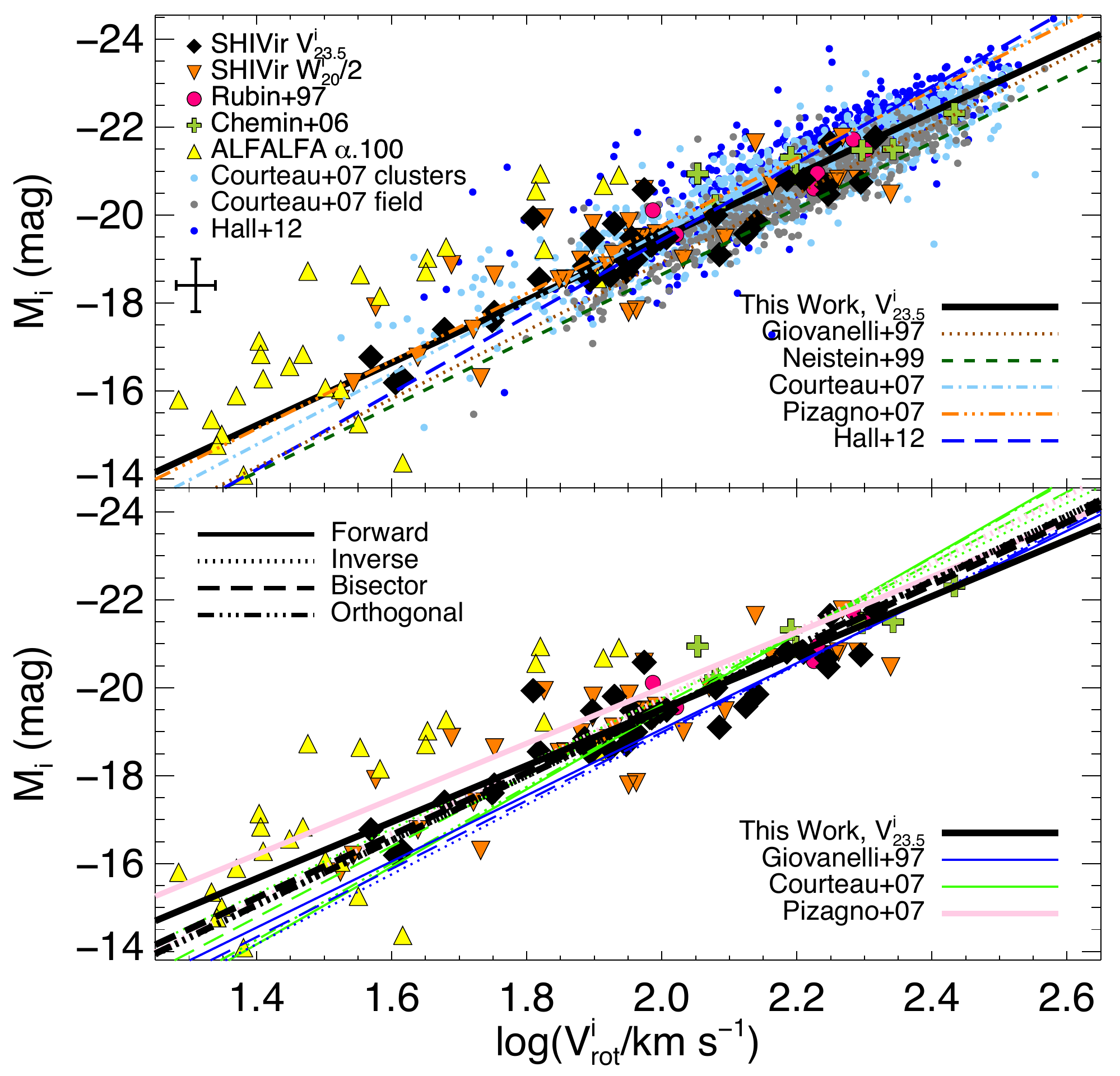}
\caption{(Top) Bisector TFRs for various Virgo databases: SHIVir (both $V_{23.5}$, fitted, and $W_{20}/2$, unfitted), \citet{Rubin1997}, and \citet{Chemin2006}; the ALFALFA \citep{Haynes2011} data are shown but not fitted. The velocity metric $V^{\rm i}_{\rm rot}$ is $V^{\rm i}_{23.5}$ for the first three samples, and the line width $W^{\rm i}_{20}/2$ for SHIVir and $W^{\rm i}_{50}/20$ for ALFALFA. Also shown are cluster and field TF fits for other non-Virgo source catalogs \citep{Giovanelli1997,Neistein1999,Courteau2007b,Pizagno2007,Hall2012}. Our best TFR for VCGs (see Table~1) is shown as a solid black line. The typical uncertainty per SHIVir point is shown on the left. (Bottom) TFRs for various statistical fits of multiple surveys \citep{Giovanelli1997,Courteau2007b,Pizagno2007}, including the present study. The predicted magnitude differences resulting from the choice of fitting method can be comparable to those accounted for by observational scatter.
}
\label{fig:TFR}
\end{figure*}

\begin{table}[t]
\begin{center}
\caption{Bisector Least-squares Tully-Fisher Relations}
{\small
\begin{tabular}{c|cccc}
\hline\hline
$V_{\rm rot}$			& $a$ (mag)		& $b$ (mag)		& $\sigma$ (dex)	& $N$ \\
\hline
$V^{\rm i}_{2.2}$		& $-7.16 \pm 0.35$	& $-21.85 \pm 0.12$	& 0.216					& $46$ \\
$V^{\rm i}_{23.5}$		& $-7.12 \pm 0.28$  	& $-21.63 \pm 0.10$	& 0.197					& $46$ \\
$V^{\rm i}_{\rm max}$	& $-7.15 \pm 0.34$	& $-21.76 \pm 0.12$	& 0.222					& $46$ \\
$W^{\rm i}_{20}/2$		& $-7.68 \pm 0.58$	& $-22.38 \pm 0.33$	& 0.487					& $65$ \\
\hline \hline
\end{tabular} }
\tablecomments{TFRs computed as $M_{i} = a\cdot ({\rm log}V^{\rm i}_{\rm rot} - 2.3) + b$ using different velocity metrics (1st column). $\sigma$ is the forward scatter and $N$ is the number of fitted data points. The catalogs used for the first three velocity metrics are SHIVir, \citet{Rubin1997}, and \citet{Chemin2006}. The catalogs used for the fourth velocity metric are SHIVir and ALFALFA \citep{Haynes2011}.}
\label{tab:tfr}
\end{center}
\end{table}

We now wish to explore the fundamental correlation between galaxian luminosity and velocity, also known as the TFR. While the TFR has been studied extensively \citep[e.g.][to list a few]{Courteau1997,Giovanelli1997,McGaugh2000,Courteau2007b,Pizagno2007,Hall2012,Bekeraite2016,Bradford2016}, differences between field and cluster environments remain ill-constrained especially in the context of the Virgo cluster, where stripping effects have been well documented \citep[e.g.][]{Koopmann2004}. We have compiled a subsample of 46 Virgo LTGs with resolved \ha\ RCs that could successfully be fitted using the multiparameter function in \Eq{mpf}\footnote{The other LTGs have RCs that are either too noisy, still rising at their truncation radius, or have highly irregular shapes. These galaxies all have low $V_{\rm rot}$. Their exclusion from the TFR only results in a dearth of data at the low-velocity end.}. Our TFR analysis takes advantage of $i$-band photometry to mitigate dust extinction effects. \Table{tfr} presents the $i$-band TFRs for LTGs based on a bisector regression and one of the following velocity metrics: $V^{\rm i}_{2.2}$, $V^{\rm i}_{23.5}$, $V^{\rm i}_{\rm max}$, and $W^{\rm i}_{20}/2$ or $W^{\rm i}_{50}/2$ line widths. 

\Fig{TFR} (top panel) shows TFRs for the SHIVir sample and two other VCG data sets \citep{Rubin1997,Chemin2006}, using $V^{\rm i}_{23.5}$. This velocity metric produced the smallest scatter in this TFR analysis, in agreement with \citet{Hall2012}. Integrated line widths, e.g. from 21cm emission profiles, can also be compared to our TFR data. The corrected ALFALFA line widths for VCGs from \citet{Haynes2011} and SHIVir line widths are displayed in the figure as $W^{\rm i}_{50}/2$ and $W^{\rm i}_{20}/2$, respectively. However, given an uncertain transformation from line width to rotational velocity \citep{Papastergis2011}, the first three fits in Table~1 do not include ALFALFA and SHIVir line widths. The TFR slopes using $V^{\rm i}_{\rm max}$, $V^{\rm i}_{2.2}$, and $V^{\rm i}_{23.5}$ are all consistent with each other; $W^{\rm i}_{20}/2$ is the exception with a steeper slope, although it is still consistent within the uncertainties. Its zero-point, however, is noticeably lower than the other three, but this is solely due to the inclusion of the ALFALFA datapoints; the SHIVir $W^{\rm i}_{20}/2$ TFR has a slope of $-7.54 \pm 0.68$ and a zero-point of $-21.85 \pm 0.27$. We confirm that the TFR with the lowest scatter uses $V^{\rm i}_{23.5}$, which is measured far in the outer disk \citep{Courteau1997}.  The TFR scatter with $V^{\rm i}_{23.5}$ is 9\% smaller than for $V^{\rm i}_{2.2}$, 11\% smaller than for $V^{\rm i}_{\rm max}$, and 60\% smaller than for $W^{\rm i}_{20}/2$ (the reported scatter estimates are those of forward fits). In other words, the tightest TFR is achieved for the most extended aperture where the RC is likely at its flattest and where the dark matter is likely dominant. The TFR scatter for our VCGs is somewhat larger than reported for large samples of field and cluster galaxies \citep{Courteau2007b,Hall2012}. This is largely due to our smaller sample size and greater distance errors (as the 3D structure of the Virgo cluster is poorly determined). Typically, the TFR scatter found for samples containing a few hundred galaxies is on the order of 0.2 to 0.5 mag \citep{Courteau1997,Giovanelli1997}. Since our LTG sample is relatively small, the fit uncertainty should be proportionally larger. The scatter for TFRs based on resolved RCs ranges from 0.49 to 0.56 mag, or 0.2 to 0.22 dex. It is even larger if ALFALFA values are included. Regardless of the velocity metric, the TFR scatter always increases at lower velocities ($V_{\rm circ} \leq 90$ \kms, or log($V_{\rm circ}) \leq 1.95$), where baryonic effects (neglect of the gaseous mass, increasing stellar velocity dispersion) steepen and broaden the TFR slope \citep{McGaugh2000,Simons2015,Bekeraite2016}.

% TF scatter
Superimposed on the data in \Fig{TFR} are additional TFRs based on $i$-band photometry for cluster \citep{Giovanelli1997,Neistein1999,Hall2012} and field environments \citep{Courteau2007b,Pizagno2007}. Despite ram pressure stripping and tidal interactions in cluster environments, which affect star formation rates, and thus galaxy luminosities, more effectively than in the field \citep{Koopmann2004}, and while Virgo RCs are truncated relative to field analogs, the VCG data nicely match other cluster and field TF distributions. The environmental independence of the TFR has also been noted by \citet{Vogt1995} and \citet{Mocz2012}, among others. The slope and zero-point of the cluster and field TFRs are statistically the same for the best-fit lines of the included catalogs in \Fig{TFR}. However, the TFR scatters for field and cluster samples may differ, with the field samples showing lower values.  This is likely the result of a quieter mass accretion history and a less perturbed evolution of the dark matter halos for field systems, as well as the presence of kinematically disturbed systems such as tidal dwarf galaxies \citep{Lelli2015}, interacting and stripped galaxies \citep{MendesdeOliveira2003}, and enhanced \citep{MilvangJensen2003} or quenched \citep{Nakamura2006} star formation in galaxy clusters. 

% TFR fits.  TFR zero-point 
\Fig{TFR} (bottom panel) shows the same data as above, but now comparing various fits (forward, inverse, bisector, orthogonal) of the TFR data by \citet{Giovanelli1997}, \citet{Courteau2007b}, \citet{Pizagno2007} and this work. We have tested that the Bayesian formalism of \citet{Kelly2007} yields nearly identical results as the bisector fits (using IDL routines \texttt{sixlin}, \texttt{mpfit} and \texttt{bces}). The range of cluster and field TFR parameters due to the chosen statistical method is as large as the data's own dispersion. Indeed, the study of \citet{Bradford2016} of systematic uncertainties on the baryonic TFR showed a variation of up to 12\% on the slope depending on the fitting algorithm used, which is at the level of both our velocity and magnitude uncertainties. Unfortunately, there is no universal standard for the choice of regression in scaling relation analyses, and the exact statistical method is not always specified. Indeed this ambiguity prevents us from firmly assessing that cluster and field TFRs differ on statistical grounds. Furthermore, the \citet{Courteau2007b} cluster (light blue) and field (gray) points plotted in the background of \Fig{TFR}, top panel, appear to be slightly offset from each other by $M_{i} = 0.5$ at log$(V^{\rm i}_{\rm rot}) = 2.2$. While an environmental dependence is a tempting explanation, one cannot guarantee at this stage that the magnitudes are exactly zero-pointed to the same system \citep{Courteau2007b}. The uniform photometric calibration to the NGVS system should alleviate this concern. Additionally, the \citet{Courteau2007b} data are an amalgamation of different surveys, hence its heterogeneous nature.

% FIGURE
\begin{figure*}
\centering
\includegraphics[width=0.99\textwidth]{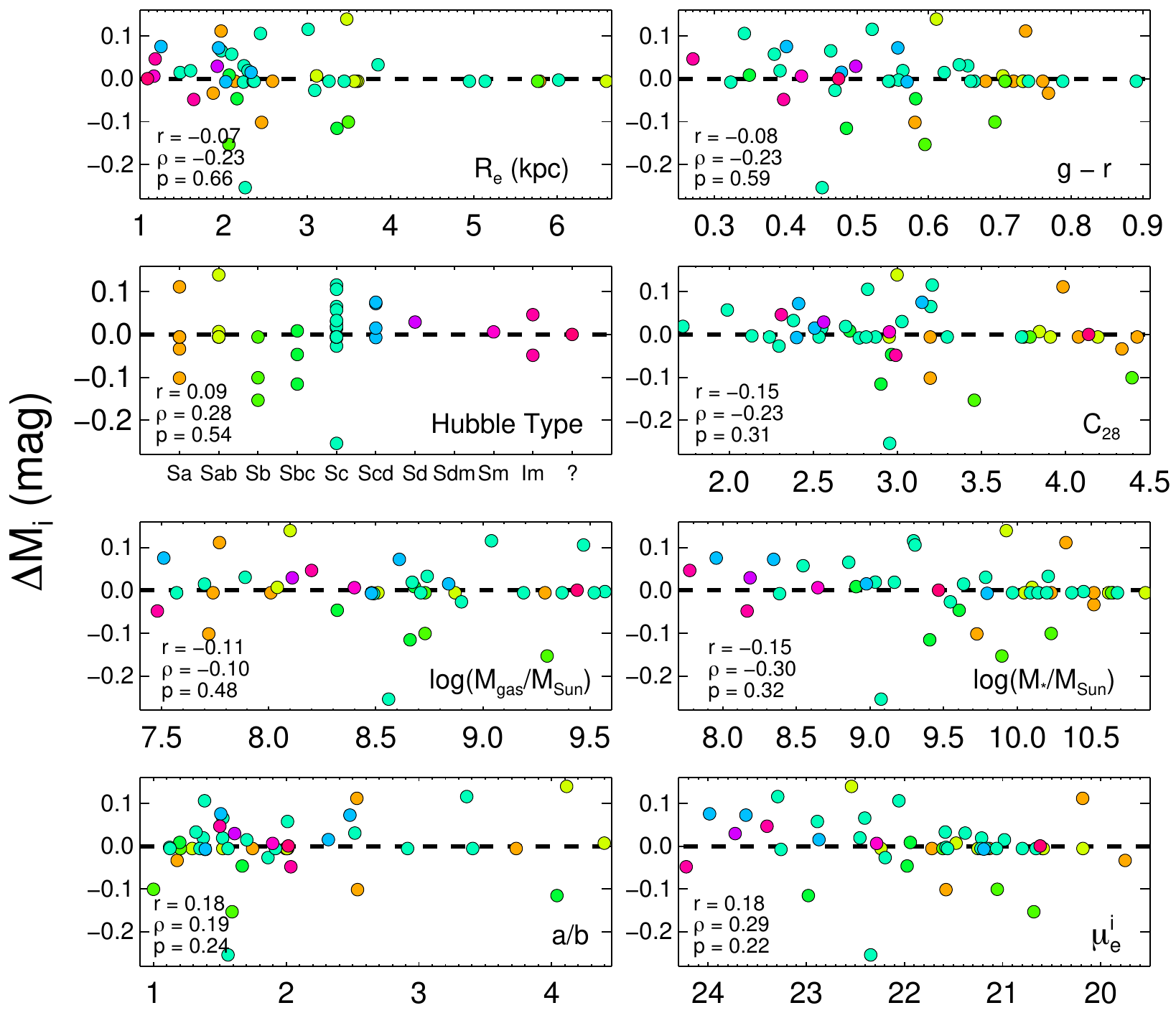}
\caption{Forward TFR residuals $\Delta M_{i}$, versus $R_{\rm e}$, $g - r$, Hubble type, $C_{28}$, gas mass log$(M_{\rm gas})$, stellar mass log$(M_{*})$, $a/b$, and $\mu_{\rm e}$. The multiple windows are arranged in order of increasing Pearson correlation coefficient $r$, shown in the bottom left corner of each window along with the Spearman correlation coefficient $\rho$ and $p$-value. The data are all color-coded by Hubble type.
}
\label{fig:TFR_res}
\end{figure*}

% Distance error
Producing a TFR for the Virgo cluster has its own unique challenges, chief of which is using accurate distances to convert apparent into absolute magnitudes. The size and exact shape of the Virgo cluster remain poorly defined. \citet{Fukugita1993} posited that spiral galaxies in the Virgo cluster may be distributed in an elongated region extending from a distance of 13 to 30 Mpc; \citet{Blakeslee2009} have also measured VCG distances ranging from 10.9 to 30.9 Mpc. The significant distance uncertainty at Virgo makes it challenging to avoid galaxies from contaminating backgrounds, and indeed may be an important source for our larger scatter. This suggests that distance errors contribute to our large scatter. We quantify this contribution here. Based on the SHIVir sample, distance uncertainties average out to 2.27 Mpc, or roughly $\pm$13.8\% in distance, if VCGs lie at a mean distance of 16.5 Mpc. The resulting scatter derived from distance errors alone is 0.3 mag, or over half of the total observed TFR scatter. While it is a common approximation to place all VCGs at 16.5 Mpc, multiple overlapping clouds likely make up the Virgo cluster \citep{Tully2016}. Approximating a VCG distance to be 16.5 Mpc when its true distance is unknown would thus markedly increase the scatter of our scaling relations. Our initial TFR study included only distances with uncertainties on the order of 1 Mpc or less (setting all other distances to 16.5 Mpc). This resulted in a larger scatter than for the TFR presented here, and thus we chose to include all available distances enumerated in \se{ad_data}.

\subsubsection{TFR Residuals}
\label{sec:tfr_res}

The scatter of the TFR is known to be sharply independent of numerous galaxy observables, making this relation a genuine ``fundamental plane" for LTGs \citep{Zwaan1995,Courteau1999,Courteau2007b}. For completeness, we here explore the scatter of this cluster's TFR in \Fig{TFR_res} as a function of a number of parameters: $i$-band effective SB, $g - r$ color, Hubble type, effective radius $R_{\rm e}$, stellar mass, gas mass, semimajor-to-minor axis ratio $a/b$, and $C_{28}$ concentration measured in the $i$-band, defined as

\begin{equation}
\label{eq:concentration}
C_{28} = 5 {\rm log} \bigg{(}\frac{r_{80}}{r_{20}}\bigg{)},
\end{equation}

\noindent where $r_{80}$ and $r_{20}$ are the radii enclosing 80\% and 20\% of the total light. This parameter should be mostly independent of projection effects \citep{McDonald2009b} and is somewhat analogous to morphological class. ETGs with large bulges have high concentrations, while LTGs with smaller bulges have low concentrations. Hubble types have been ordered from 3 (Sa) to 12 (Im), in accordance with the GOLDMine classification \citep{Gavazzi2003}. No included galaxies were found to have Hubble types ranging from 13 (Pec) to 19 (dIm), thus galaxies classified as 20 (?) were placed immediately after 12. Both Spearman and Pearson correlation coefficients for these distributions are quite low, ranging from -0.30 to 0.29 and -0.15 to 0.18, respectively. For a null hypothesis where a given parameter does not correlate with residuals, we find no $p$-value more significant than 0.22; we verify and conclude that the TFR residuals do not strongly depend on any tested galaxy parameters \citep[and references therein]{Courteau2007b,Dutton2007,Hall2012}. 

% FIGURE
\begin{figure*}
\centering
\includegraphics[width=0.99\textwidth]{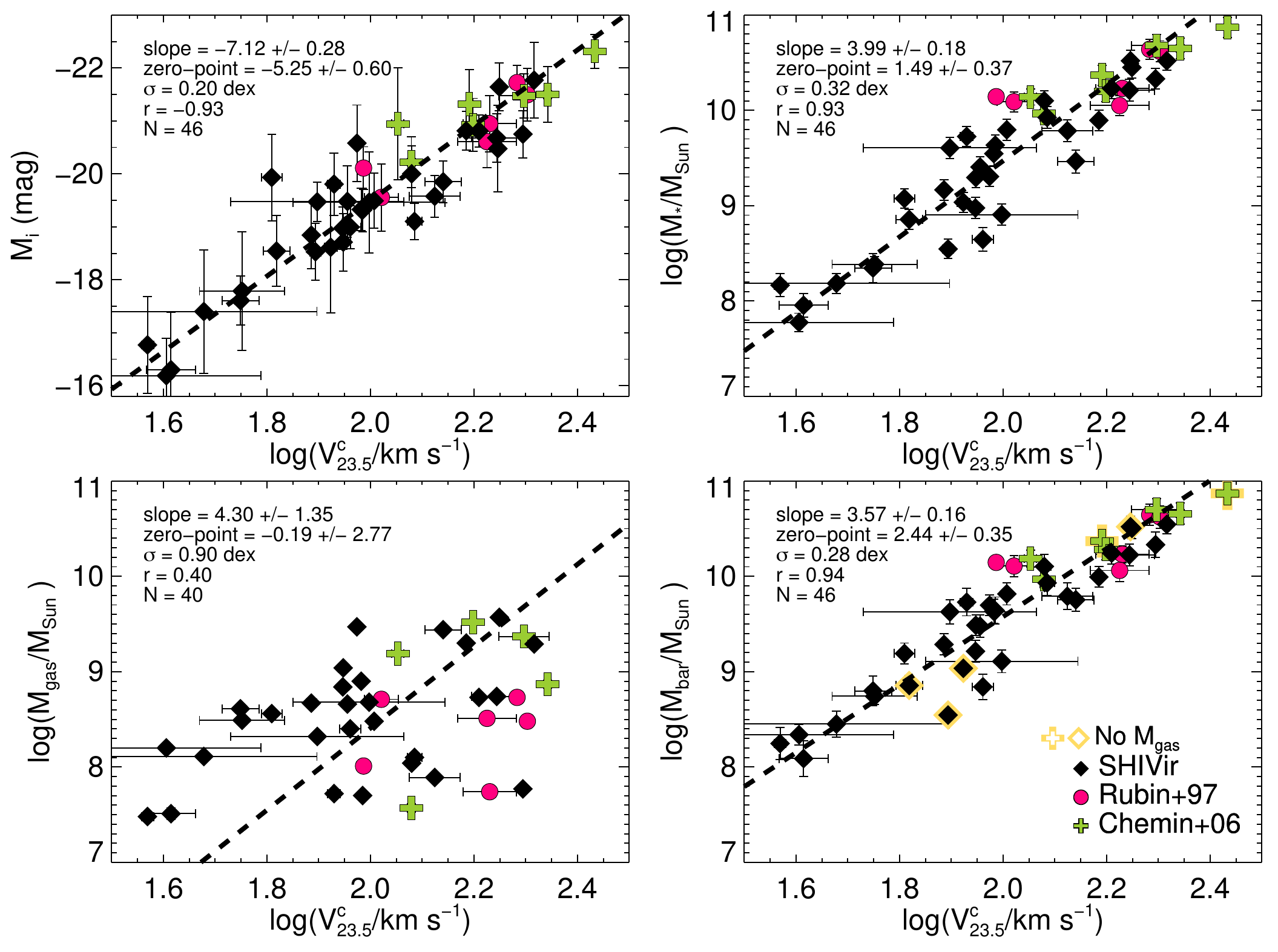}
\caption{Classic (top left), stellar mass (top right), atomic gas (bottom left), and baryonic TFR (bottom right). The velocity metric used is $V^{\rm i}_{23.5}$ for all four relations. \hi\ masses are taken from the ALFALFA $\alpha$.100 catalog \citep{Haynes2011}; only 40 of our 46 VCGs have available gas masses. The baryonic mass is the sum of the stellar and gas masses; the gas mass is set to zero if it was not measured by ALFALFA. The six galaxies with unavailable gas masses are highlighted in yellow in the BTFR plot; their $M_{\rm bar}$ values should be taken as lower limits. TFR slopes, zero-points, scatter $\sigma$, Pearson correlation coefficients $r$, and sample sizes $N$ are shown in each panel. 
}
\label{fig:TFR_all}
\end{figure*}

\subsubsection{Higher Forms of the TFR: Stellar and Baryonic}
\label{sec:all_tfrs}

The best-fit STFR for all SHIVir data plotted in \Fig{TFR_all}, including ALFALFA data, is 

$$\log M_{*} = (4.02 \pm 0.30) \times \log V^{\rm i}_{23.5} + (1.74 \pm 0.53);$$ 

\noindent and without ALFALFA data,

$$\log M_{*} = (3.99 \pm 0.18) \times \log V^{\rm i}_{23.5} + (1.49 \pm 0.37),$$

\noindent as seen in \Fig{TFR_all} (top right panel). \citet{Hall2012} also found $M_{*} \propto V_{\rm rot}^{4}$ using a large SDSS sample ($N=3041$) and radio line widths\footnote{In principle, using line widths for nearly flat rotation curves measured at optical or radio wavelengths ought to yield the same ``maximum" circular velocities, and the respective TFRs should have comparable slopes \citep{Courteau1997}, as they do here.}.  

Our smaller sample size yields a fairly large TFR slope uncertainty, but the slope itself is consistent with a nominal value of 4. The \citet{Bradford2016} STFR for a set of isolated galaxies yielded a slightly larger slope of $4.14 \pm 0.06$, but still matched our relation within the uncertainties. Our best-fit STFR has a scatter of 0.32 dex in $M_{*}$, or 0.08 in $V^{\rm i}_{23.5}$, which exactly matches the values of \citet{Desmond2017}.

Construction of the BTFR took advantage of \hi\ gas masses from the ALFALFA $\alpha$.100 catalog \citep{Haynes2011}. These are only available for 40 of the 46 VCGs used in our TFRs; the six galaxies with missing values are still used in the BTFR; their gas mass is set to zero. An atomic gas TFR is also plotted in \Fig{TFR_all} (bottom left panel). The Virgo cluster BTFR, \Fig{TFR_all} (bottom right panel) yields a flatter relation than the regular TFR, as expected \citep{Gurovich2010}. Our best-fit BTFR for all SHIVir data plotted in \Fig{TFR_all} (bottom right panel), including ALFALFA, is 

$$\log M_{\rm bar} = (3.57 \pm 0.16) \times \log V^{\rm i}_{23.5} + (2.44 \pm 0.35).$$ 

\noindent Our BTFR slope is steeper than that of \citet{Gurovich2010}, $3.2 \pm 0.1$, and \citet{Bradford2016}, $3.24 \pm 0.05$, but matches the slope of \citet{Hall2012} of $3.45 \pm 0.12$. Numerical and semi-analytic galaxy formation simulations based on the $\Lambda$CDM model, which assume $M_{\rm bar} \propto M_{200}$ and $V_{\rm circ} \propto V_{200}$, predict a BTFR slope of 3 \citep{Mo2000,Navarro2000,vandenBosch2000}, but a more realistic picture of disk galaxies that includes the impact of baryons, adiabatic contraction, and angular momentum conservation will likely increase the BTFR slope to lie somewhere between 3 and 4 \citep{Dutton2009,Gurovich2010}. As we increase our sample of gas-rich dwarf galaxies in the future, we may begin to see a shallowing of our BTFR slope at the low-mass end, which has been reflected in dwarf BTFR studies resulting in slopes as low as 2 \citep{McCall2012,Bradford2016,Karachentsev2017}. Complete fit solutions for all TFRs, including scatter and sample size, are found in \Fig{TFR_all}.

\subsection{Fundamental Plane}
\label{sec:fp}

% FIGURE
\begin{figure*}
\centering
\includegraphics[width=0.99\textwidth]{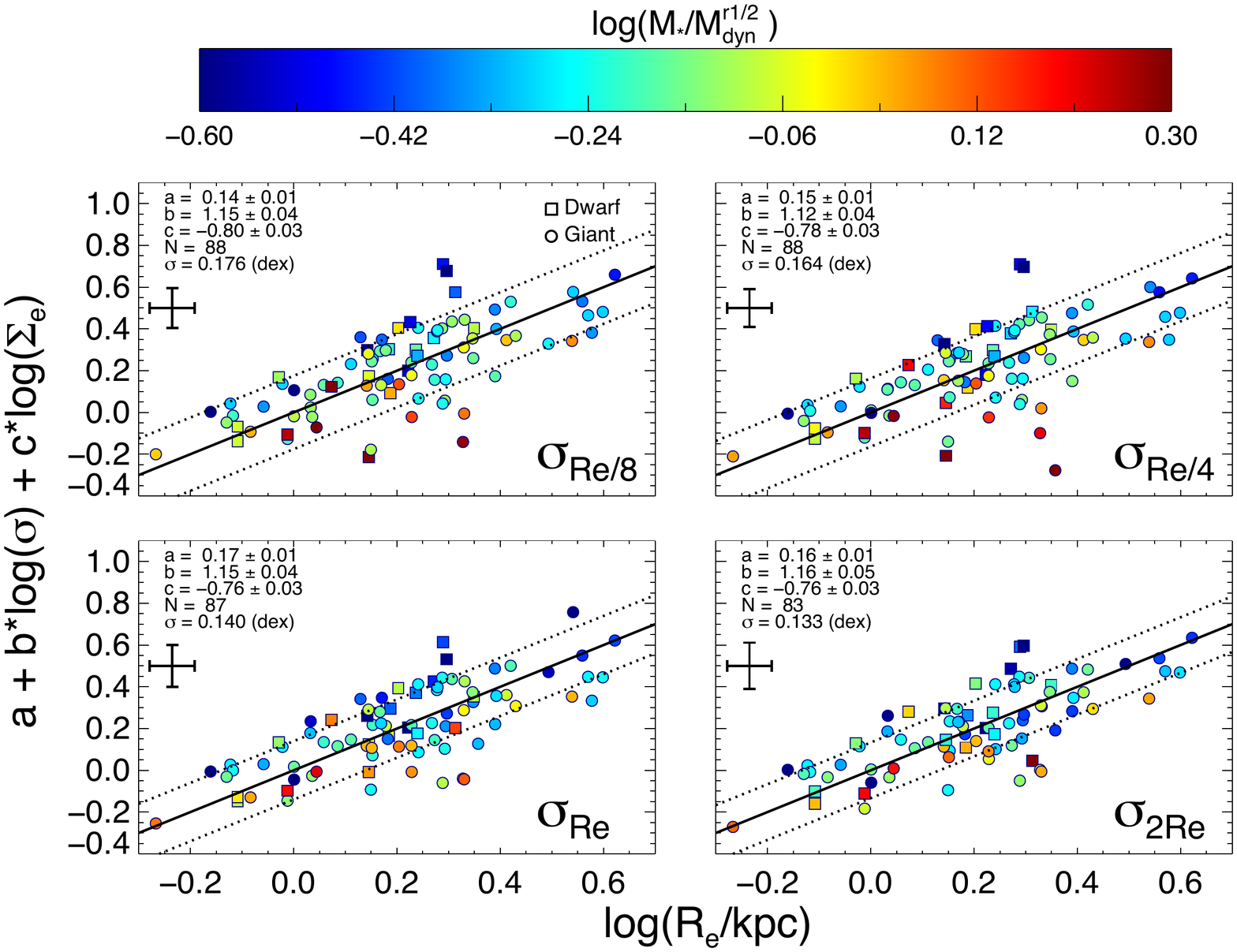}
\caption{FP relations (using a bisector regression) for SHIVir gas-poor galaxies with velocity dispersions measured within a radius of $R_{\rm e}/8$, $R_{\rm e}/4$, $R_{\rm e}$, and $2 R_{\rm e}$. The best-fit line and 1-$\sigma$ uncertainty are shown as solid and dashed black lines, respectively. The best-fit parameters $a,b,c$, sample size $N$, scatter $\sigma$, and typical uncertainty per point are shown in each panel. The data are color-coded by stellar-to-total mass ratio (the most dominant second parameter). Galaxies classified as dwarfs (dS0/dE) by the GOLDMine \citep{Gavazzi2003} database are plotted as squares. The remaining galaxies are plotted as circles.}
\label{fig:FP}
\end{figure*}

We now turn our attention to the FP of the 88 ETGs for which stellar kinematics could be successfully extracted. The FP for E, S0, dE, dS0, and Sa galaxies follows the description: 

$$\log(R_{\rm e}) = a + b~\log(\sigma) + c~\log(\Sigma_{\rm e}),$$ 

\noindent where the effective radius $R_{\rm e}$ and effective SB $\Sigma_{\rm e}$ within $R_{\rm e}$ are determined from $i$-band photometry, and the velocity dispersion $\sigma$ uses multiple definitions described below. Sa galaxies are included in the ETG sample for the FP, despite being classified as LTGs in the rest of this paper, since their velocity dispersions are reliable and dominant (over $V_{\rm rot}$).

Historically, $\sigma$ has been measured for spectra integrated over small radii (typically a fraction of $R_{\rm e}$) or taken over a few central pixels ($\sigma_{0}$). For the SHIVir survey, we wish to probe velocity dispersions well into the transition regime from baryon-to-dark-matter domination, typically in the $2-4$ $R_{\rm e}$ regime. The scatter for FPs based on different velocity metrics is shown to decrease as a function of aperture in this section.

% FIGURE
\begin{figure*}
\centering
\includegraphics[width=0.99\textwidth]{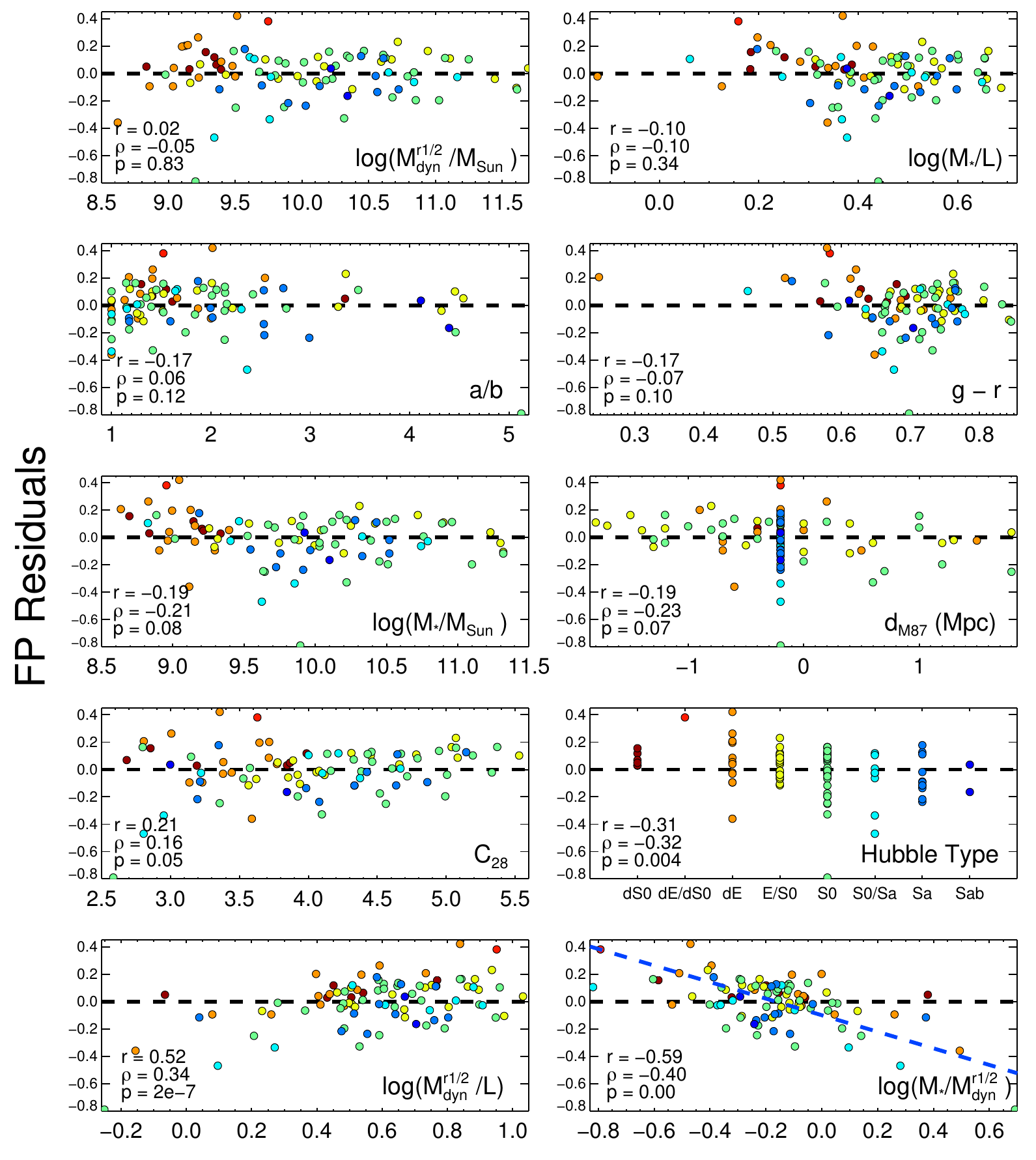}
\caption{FP residuals versus dynamical mass within $r_{1/2}$ log$(M^{r_{1/2}}_{\rm dyn})$, stellar-mass-to-light ratio log$(M_{*}/L)$, semimajor-to-minor axis ratio $a/b$, $g - r$, distance from M87, stellar mass log$(M_{*})$, $C_{28}$, Hubble type, total-mass-to-light ratio log$(M^{r_{1/2}}_{\rm dyn}/L)$ and stellar-to-total mass ratio log$(M_{*}/M^{r_{1/2}}_{\rm dyn})$ (in order of increasing Pearson correlation coefficient $r$, shown on the left corner of each panel along with the Spearman correlation coefficient $\rho$ and $p$-value). $M_{*}$ and $L$ are both measured within $r_{1/2}$. The data are color-coded by Hubble type. A significant correlation exists for the log$(M_{*}/M^{r_{1/2}}_{\rm dyn})$ parameter; the blue dashed line traces the best fit between that mass ratio and the FP residuals.
}
\label{fig:FP_res}
\end{figure*}

\Fig{FP} shows FP fits, computed with the \texttt{lts\_planefit} routine of \citet{Cappellari2013a} for ($R_{\rm e}$,$\sigma$,$\Sigma_{\rm e}$), for four different dispersion measures, with apertures from 0 to $2 R_{\rm e}$. It is encouraging, e.g. for cosmic flow studies or galaxy formation modeling, that the FP scatter is significantly reduced for velocity dispersions measured at large galactocentric radii\footnote{Ultimately, one wishes to find the location where the FP scatter is minimized, likely within the region of dark matter domination.}. Distances can thus be inferred more accurately. This result matches the theoretical predictions of \citet{Dutton2013}, which state that variations in the velocity dispersion profiles that are due to non-homology in anisotropy and structure are decreased when the aperture size is increased and a mass closer to $M_{200}$ is sampled. Integral field spectra of elliptical apertures of up to $R_{\rm e}$ have already revealed hints of this effect \citep{Cappellari2011b,Scott2015}, and we now extend this conclusion to larger radii. \footnote{Signal-to-noise limitations for velocity dispersions measured at larger radii yield smaller samples. However, we have verified that the FP fits remain the same if we restrict the entire analysis to the 83 galaxies with dispersion profiles that extend out to $2 R_{\rm e}$.} 

As mentioned above, we designed the SHIVir survey in order to probe velocity dispersions at large radii. The characteristics of the stellar kinematics are expected to vary with radius. For instance, the classification of ETGs into fast/slow-rotator \citep{Cappellari2011b,Emsellem2011} was defined for kinematics measured within $R_{\rm e}$. The shape of the rotation curve can vary at radii larger than $R_{\rm e}$ however, depending on whether a galaxy is an S0 or a disky elliptical \citep{Arnold2014}. This is because the outer kinematics in fast-rotator ETGs depend on the spatial scale of the stellar disk (see Fig.~3 of \citealp{Cappellari2016a}). Indeed, the FP scatter is reduced when using $\sigma_{\rm e}$ instead of $\sigma_{0}$ \citep{Cappellari2013a,Scott2015}. For the SHIVir sample, the FP scatter drops by 24\% from 0.176 dex at center to 0.133 dex at $2 R_{\rm e}$. This agrees with \citet{Scott2015}, whose FP scatter for an IFU study of 106 galaxies in three nearby clusters dropped from 0.08 dex at center to 0.07 dex at $R_{\rm e}$, a 9\% improvement. Since our integrated velocity dispersions are not aperture corrected, the outer parts of the spectra are possibly overwhelmed by the inner signal. However, the FP scatter dependence on aperture size is clear, indicating that the outer parts still affect the dispersion measurement significantly (and this also despite the rapid loss in slit area coverage at larger galactocentric radii). We have stated in \se{dyn_mass} that the variations in velocity dispersions integrated within different slit lengths are quite minimal, but the combined effects of these small variations for $80-90$ galaxies can contribute a sizeable decrease in FP scatter.

A comparison with literature results is warranted. \citet{Bernardi2003} compiled 11 sets of FP best-fit parameters based largely on central velocity dispersion $\sigma_{0}$ and $r$-band photometry from the literature, the median of which is $b = 1.33 \pm 0.12$ and $c = -0.82 \pm 0.03$. Our own best-fit parameters for the $\sigma_{0}$ case differ slightly: $b = 1.15 \pm 0.04$ and $c = -0.80 \pm 0.03$. Larger apertures may probe a more complex structural pattern; \citet{Cappellari2013a} used velocity dispersions measured within $R_{\rm e}$ and obtained best-fit FP parameters at $i$-band of $b = 1.06 \pm 0.04$ and $c = -0.76 \pm 0.02$. These match our FP solution at $R_{\rm e}$: $b = 1.16 \pm 0.04$ and $c = -0.77 \pm 0.03$. In general, we match ATLAS$^{3{\rm D}}$'s results best. While we may posit that some of those differences are environment-related, \citet{FalconBarroso2011b} showed that kinematic substructure or the environment do not yield a preferred location in the FP for SAURON galaxies. Furthermore, a direct FP comparison might be ill-suited if the FP is a warped surface \citep{Cappellari2013b}. The plane can also change drastically depending on the galaxy sample. The $b$ parameter of \citet{Cappellari2013b} changed by 29\% when only galaxies with $\sigma > 130$ \kms were included. Their $c$ parameter also changed by 14\% depending on the choice of dispersion, whether using $\sigma_{\rm e}$ or integrated over a radius of 1 kpc, $\sigma_{\rm kpc}$. Care is thus called for when comparing different FP planar fits based on different definitions of $\sigma$ and other galaxy structural parameters, especially at large galactocentric radii where the relative fraction of dark matter is non-negligible \citep{Courteau2015}. Nonetheless, while the FP scatter is noticeably reduced for a larger aperture of the $\sigma$ measurement, the best FP fit parameters are rather stable across all definitions.

It is worth noting that the individual FPs for giant and dwarf ellipticals are statistically comparable, thus suggesting that both populations are governed by similar evolutionary processes. This is even more apparent in the FP with $\sigma_{2R_{\rm e}}$ as the velocity dispersion metric. We find, however, that the FP scatter is not only larger for dwarf galaxies ($M_{*} < 10^{9} \solarm$), but their FP residuals are systematically larger. This may imply a continuous but curved FP between dwarfs and giants, as seen for photometric parameters in \citet{Ferrarese2012}. Indeed, galaxies with a higher stellar mass fraction deviate below the FP; we return to this point in our FP residual analysis (see \se{fp_res}). Whether the FP is continuous \citep{Graham2003,Graham2004,Tollerud2011} or has a discontinuity \citep{Kormendy1985,Burstein1997} remains a matter of debate \citep{Graham2005}, and the large scatter in our dwarf FP prevents us from resolving this issue.

\subsubsection{FP Residuals}
\label{sec:fp_res}

As with our TFR analysis, the FP residuals are plotted in \Fig{FP_res} as a function of dynamical mass within $r_{1/2}$, $M^{r_{1/2}}_{\rm dyn}$, stellar-mass-to-light ratio $M_{*}/L$, semi-major-to-minor-axis ratio $a/b$, $g - r$ color, distance from M87 as an analog to environmental density, stellar mass $M_{*}$, Hubble type, $C_{28}$ concentration, total-mass-to-light ratio $M^{r_{1/2}}_{\rm dyn}/L$ and stellar-to-total mass ratio $M_{*}/M^{r_{1/2}}_{\rm dyn}$. We set the distance to M87 at 16.7 Mpc \citep{Mei2007} and subsequently define $d_{\rm M87}$ for each of our VCCs to be $d_{\rm M87} = d_{\rm VCC} - 16.7$ Mpc. We define FP residuals to be the y-axis minus the x-axis in \Fig{FP}. Luminosity $L$ is half the total galaxy light measured in the $i$-band, to compare mass (measured inside $r_{1/2}$) and light within comparable radii. The distance of M87 is taken to be 16.7 Mpc \citep{Blakeslee2009}. Total (dynamical) masses are always calculated using $\sigma_{\rm e}$. This residual analysis benefits from the largest number of VCGs (88) when the FP uses a central velocity dispersion $\sigma_{0}$. Spearman and Pearson correlation coefficients are quite weak (-0.32 to 0.16, and -0.31 to 0.21) for all parameters (whether on a logarithmic or linear scale), save two specific cases: the total-mass-to-light ratio ($r = 0.52$ and $\rho = 0.34$) and the stellar-to-total mass ratio ($r = -0.59$ and $\rho = -0.40$). The nearly null $p$-values for these two parameters show strong statistical significance. These two quantities are closely linked, since our stellar mass measurements exploit luminosity and color values, so we discuss only the strongest of the two correlations with the stellar-to-total mass ratio, $M_{*}/M^{r_{1/2}}_{\rm dyn}$. No correlation between the FP residuals and the stellar mass ($r = -0.21$) or the dynamical mass ($r = 0.00$) is found. We also find $C_{28}$ and Hubble type, both markers of morphology, to have correlations with $p$-values with a significance level of $2\sigma$ or higher, but we believe that both of these correlations trace the $M_{*}/M^{r_{1/2}}_{\rm dyn}$ correlation, since stellar-to-total mass ratio likely correlates with morphology itself. We do note that a large scatter or uncertainty on $\sigma$ would drive the linear correlation we see in $M_{*}/M^{r_{1/2}}_{\rm dyn}$, since an overestimated $\sigma$ would produce a lower $M_{*}/M^{r_{1/2}}_{\rm dyn}$ value and larger FP residual (and vice versa for underestimated $\sigma$ values). We must trust that our VD measurements are reasonably accurate.

% FIGURE
\begin{figure*}
\centering
\includegraphics[width=0.7\textwidth]{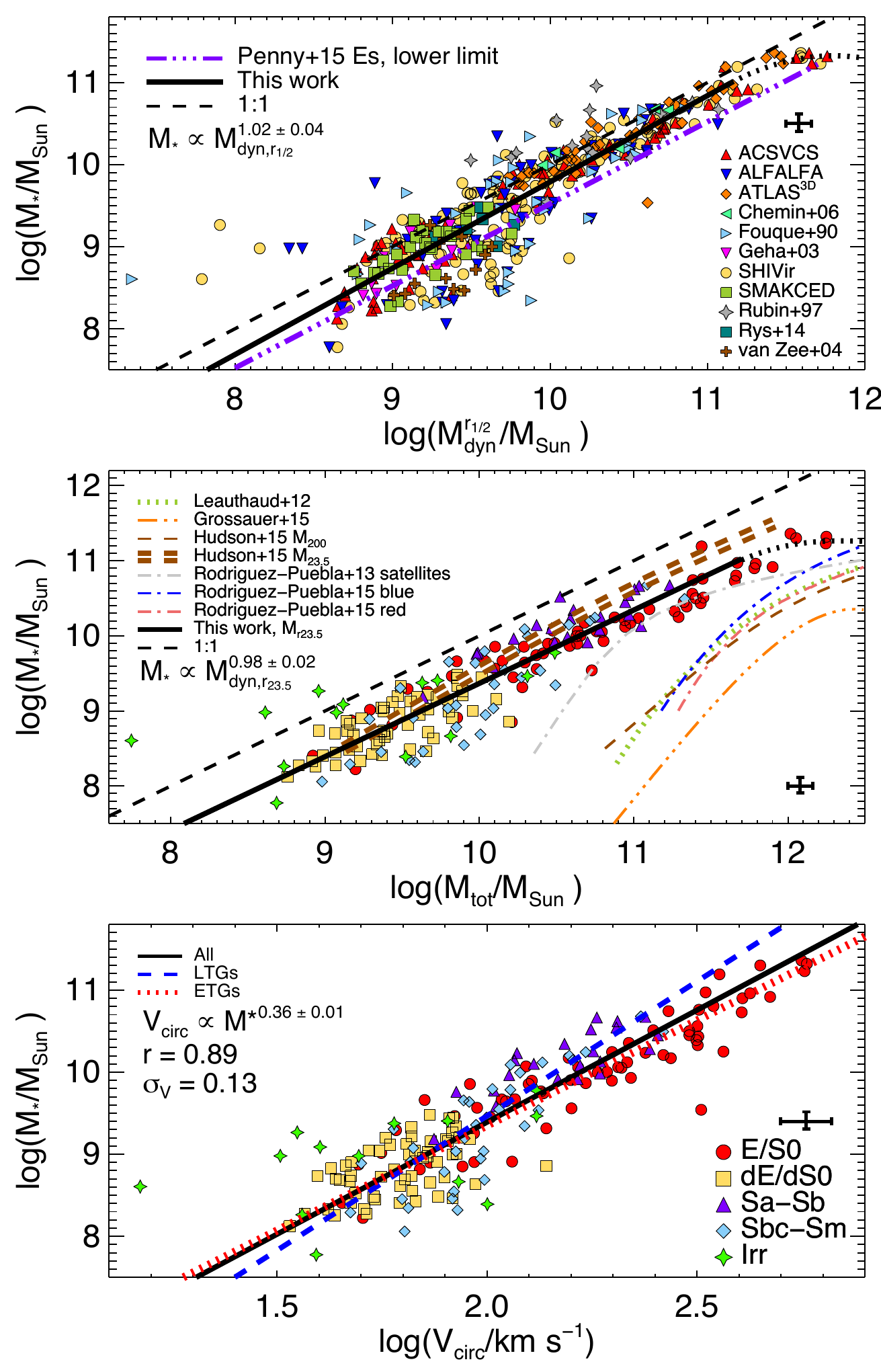}
\caption{(Top) Virgo cluster STMR using a dynamical mass, $M_{\rm dyn}$, measured inside $r_{1/2}$, and a stellar mass $M_{*}$ measured inside $r_{23.5}$ computed from SED fitting. All galaxies are part of the larger SHIVir catalog of 742 VCC galaxies; point colors indicate the source of the dynamical information for $M_{\rm dyn}$. The dashed line is the 1:1 relation and the solid black line is our fit up to $M_{\rm dyn} \sim 10^{11} \solarm$. A curved dashed line accounts for the likely inflection at that transition. The lower limit STMR from \citet{Penny2015} is also shown (purple dash-dotted line). (Middle) STMR, now using the total (dynamical) mass measured inside $r_{23.5}$ for the SHIVir data points, differentiated by morphology. Also shown are SHMRs for central and satellite galaxies that probe dynamical masses at very large radii \citep{Leauthaud2012,Grossauer2015,RodriguezPuebla2013,RodriguezPuebla2015,Hudson2015}, where $M_{\rm tot}$ is $M_{200}$. Predictions for dark matter content of some of these SHMRs are presented in \se{NFW}. \citet{Hudson2015}'s SHMR is interpolated to $M_{\rm tot} = M^{r23.5}_{\rm dyn}$ (see text) for a direct comparison (double dashed brown line) with our data. The match is especially good. (Bottom) Velocity-Stellar Mass relation of VCGs by morphological types, with the Pearson correlation coefficient, $r$, and scatter on log($V_{\rm circ}$), $\sigma_{V}$. $V_{\rm circ}$ is $V_{23.5}$ for LTGs and $\sqrt{c} \times \sigma_{\rm e}$ for ETGs, as defined in \se{dyn_mass}.}
\label{fig:STMR}
\end{figure*}

It is likely the ratio between stellar and total masses that controls the FP tilt \citep{Ciotti1996,Borriello2003}; a higher ratio yields larger negative residuals and is associated with galaxies that move away from the FP best-fit line (also seen in \Fig{FP}). The dark matter fraction within $R_{\rm e}$ increases with total mass, or $\sigma$ used as a proxy, for ETGs in the relevant mass range \citep{Tortora2012} for a Chabrier IMF \citep{Dutton2013}, implying that it is the low-mass population with a higher stellar mass fraction that deviates from the FP. The Hubble type assignments taken from the GOLDMine database \citep{Gavazzi2003} may not accurately reflect the dwarf population in SHIVir, and a strong residual correlation is currently lacking in our analysis for morphology. We still confirm that the galaxies with the largest negative FP residuals (shown to deviate downwards from the FP in \Fig{FP}) are indeed low-total-mass systems. Other important contributors to the tilt of the FP may include variations in the mass-to-light ratio of stellar populations \citep{Faber1987,Prugniel1996,Trujillo2004,Cappellari2013a,Cappellari2016a}, galaxy age \citep{Forbes1998}, dynamical and structural non-homology \citep{Busarello1997,Graham1997,Trujillo2004}, and gas dissipation following mergers \citep{Robertson2006}. \citet{Dutton2013} found that the rate at which the velocity dispersion changes with galaxy size (at any given stellar mass) is a measure of the FP tilt produced by all these combined effects, which ultimately cause the FP to diverge from an idealized virial theorem \citep{Busarello1997}.

For the remainder of our analysis, we now merge the SHIVir spectroscopic and photometric measurements for the LTGs (\se{tfr}) and ETGs into one fundamental STMR of VCGs.

\subsection{Stellar-to-total Mass Relation}
\label{sec:stmr}

While our separate investigations of the TFR and FP have yielded compelling insights about their individual structure and evolution, we now wish to merge our entire catalog, irrespective of morphology, to obtain a most representative picture of a galaxy's evolutionary drivers.  A number of striking relations between ETGs and LTGs emerge as a result. We begin by addressing the relation between stellar and total mass for VCGs. This relates to the ``stellar-to-halo mass relation'' (SHMR) inferred from halo abundance matching (HAM) analyses, whereas we benefit from dynamically determined \emph{total} masses rather than dark matter masses alone. For SHIVir, rotational velocities and velocity dispersions are converted into dynamical masses as described in \se{dyn_mass}. \Fig{STMR} shows the STMRs with $M_{\rm dyn}$ for $r<r_{1/2}$ (top panel) or $r<r_{23.5}$ (middle panel) and stellar masses $M_{*}$ inferred from mean colors within $r_{23.5}$. Only galaxies with $\log V_{\rm circ} > 1.5$ are included in the relations showcased in \S3.5--3.7; a number of Irr galaxies below this range have uncertain circular velocity values (see \se{bimo}).

The scatter of the STMR may indicate different contributors to evolutionary processes depending on mass regime \citep{Gu2016}. Much like the TFR for LTGs with $V_{\rm rot} < 95$ \kms, the increased scatter in the STMR is likely due to the lack of rotational support in low-mass spirals. The STMR is tightest in the regime of massive ellipticals and early-type spirals with $M_{*} \geq 10^{9.5}\solarm$. The latter corresponds to the transition mass where rotational support for the overall dynamical equilibrium becomes significant \citep{Simons2015}. In \Fig{STMR}, middle panel, the scatter for galaxies above and below the transition mass is $\sigma_{\log M_{*}} = 0.18$ dex and $0.33$ dex respectively, in agreement with simulations \citep{RodriguezPuebla2015,Gu2016}, although the concerned halo masses are measured within much larger radii than our own total masses. This also matches typical observational estimates of SHMR scatter ranging from 0.15 to 0.35 dex \citep{Yang2009,More2011,Leauthaud2012,Behroozi2013b,Reddick2013,Tinker2013,Kravtsov2014,Lehmann2017}. Most of these studies used $M_{\rm vir}$ or $M_{200}$ for their SHMRs. Some galaxies have $M_{*} \geq M_{\rm dyn}$, consistent with uncertainties and the fact that dark matter is subdominant in ETGs within $r_{1/2}$ and $r_{23.5}$ \citep{Courteau2015}. All but three data points in \Fig{STMR}, middle panel, fall below the 1:1 line as would be expected for dark-matter-dominated systems.

Various SHMRs are also shown in \Fig{STMR} (middle panel). These are typically computed via HAM or Halo Occupation Distribution (HOD) to estimate the halo mass at cosmological radii (e.g. $R_{200}$). They are reported in \Fig{STMR} as originally published, with no adjustment for the differences in size range. \citet{Leauthaud2012} found a SHMR turnover at $M_{*} = 4.5 \times 10^{10} \solarm$, decreasing slightly with redshift, while \citet{Behroozi2013b} found the same trend for redshifts below $z=2$, but with a reversal at larger redshifts. Thus, if the ratio $M_{\rm halo}/M_{*}$ only changes weakly with redshift, it likely plays a role in regulating quenching and other star formation processes, more so than halo mass alone. The SHIVir STMR (\Fig{STMR}, middle panel) shows a possible inflection (dotted line) above $M_{*} \approx 10^{11} \solarm$, consistent with \citet{Leauthaud2012}, \citet{RodriguezPuebla2013,RodriguezPuebla2015} and to a lesser degree, the low-amplitude curved SHMR for NGVS galaxies with stellar masses in the range $10^{5-10.4} \solarm$ \citep{Grossauer2015}. These comparisons are only qualitative as the metrics for total mass are all slightly different (e.g. a dynamical mass at small radius versus a halo mass at large radius). An {\it F}-test shows a 45\% chance that our STMR is best fitted with a $2^{\rm nd}$ degree polynomial with an inflection rather than a straight line, and a 62\% chance that a $3^{\rm rd}$ degree polynomial fits the plotted inflection best. This is at odds with the linear STMR of \citet{Penny2015}, but their galaxies did not exceed $M_{\rm dyn} = 10^{11.7} \solarm$, the point at which a possible inflection is observed. While this {\it F}-test cannot validate the existence of this turning point in our data set, numerous studies of galaxy groups and clusters have clearly cemented the notion of a maximum efficiency of stellar mass formation at halo masses near $10^{12} \solarm$ \citep{Mandelbaum2006,Behroozi2010,Leauthaud2012,Behroozi2013a}.

% FIGURE
\begin{figure*}
\centering
\includegraphics[width=0.9\textwidth]{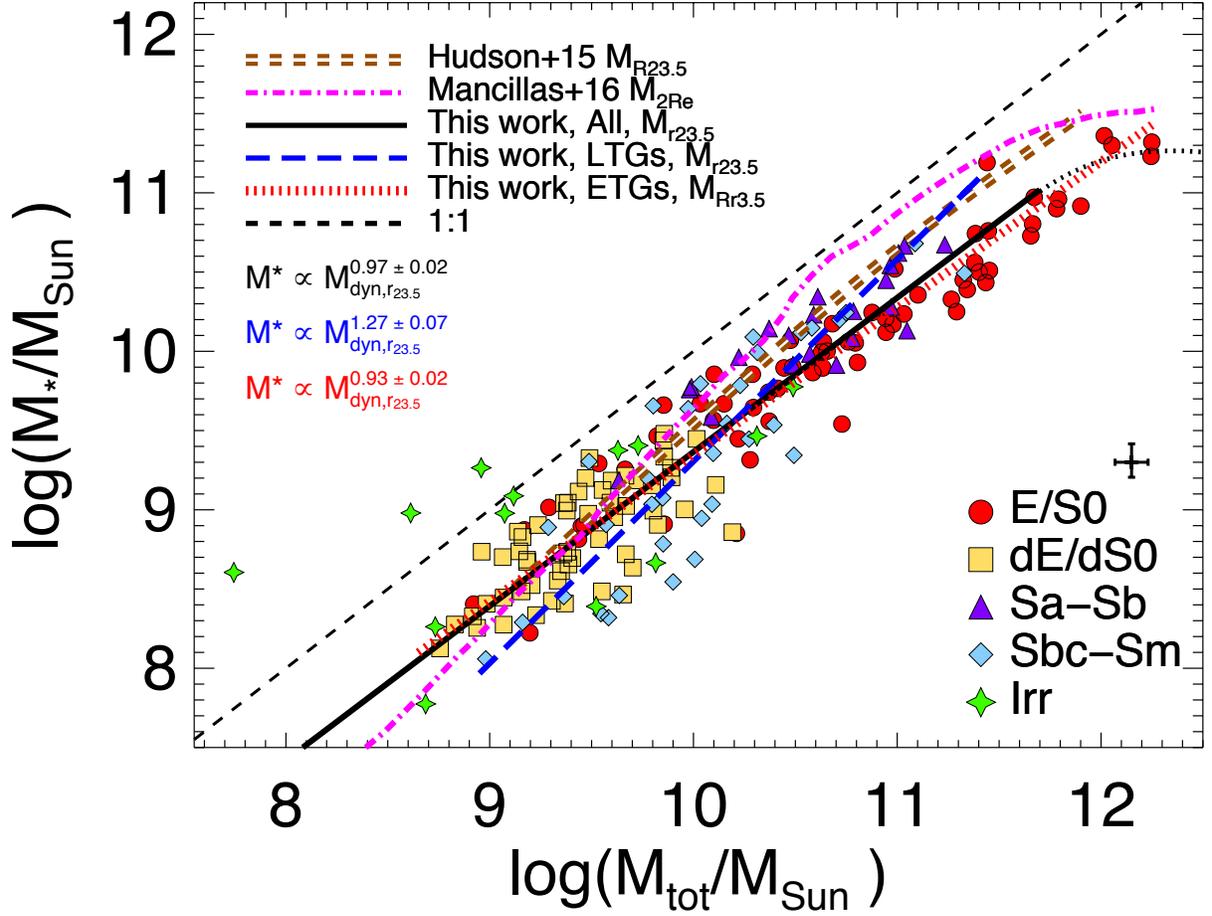}
\caption{As in \Fig{STMR}, middle panel (same data). The STMR from Mancillas et~al. (in preparation) for simulated LTGs, with the stellar mass $M_{*}$ and total mass $M_{\rm tot}$ both measured within $2 R_{\rm e}$, is in pink. The SHIVir STMR is shown in blue for LTGs (Irr galaxies are excluded given their highly truncated and/or uncertain RCs), and in red for ETGs. The dashed line is again the 1:1 line.}
\label{fig:STMR_LTGs}
\end{figure*}

Overall, the broad shapes of SHMRs and our own STMR agree for $M_{*} \ge 10^{10} \solarm$, but the SHMR slopes are much steeper than our STMR's at lower masses. Likewise, \citet{RodriguezPuebla2015} performed a segregated STMR analysis of blue and red central galaxies, concluding that the scatter, shape, and amplitude of their SHMR for the two samples are different. Within their probed stellar mass range of $10^{9-12} \solarm$, $M_{*}/M_{\rm h}$ was found be larger for blue galaxies. The difference in STMR shape between SHIVir and \citet{Leauthaud2012} may well be linked to their investigation of field versus cluster galaxies. It can indeed be argued that most VCGs are satellite rather than central galaxies and our STMR should be compared to that of \citet{RodriguezPuebla2013} for satellite galaxies (gray line); the closest match in shape is indeed found above $M_{*} = 10^{10.2}\ \solarm$. Recall also that our dynamical masses were measured directly using kinematics, rather than inferred through, say, abundance matching. A direct comparison of STMRs measured at $r_{23.5}$ or $R_{200}$ requires linking inner and outer halo profiles. The clear discrepancy between our STMR and the SHMRs of other studies reported above at low masses suggests that feedback effects, adiabatic contraction \citep{Chan2015}, and other baryonic processes are likely responsible for the non-linear scaling between masses computed within $r_{23.5}$ and $R_{200}$ \citep{Dutton2011}. Similar conclusions from a study of the STFR were reached by \citet{Miller2014}. 

In order to compare the SHMRs plotted in \Fig{STMR}, middle panel, with our STMR for masses computed within $r_{23.5}$ (the physical radius), we must interpolate the former to smaller radii. \Fig{MassSize} provides us with an estimate of $r_{23.5}$ for a given $M_{*}$, $V_{\rm circ}$ and morphological type (see \se{sizemass} for a full analysis of these relations). If we assume a NFW dark matter density profile \citep{Navarro1996} with a standard concentration, we can calculate the dark matter mass $M_{\rm halo} (r < r_{23.5})$ expected within this radius (see \se{NFW} for more details). Summing with $M_{*} (r < r_{23.5})$ finally yields $M_{\rm tot} (r < r_{23.5})$, which can be compared with our direct SHIVir measurements in \Fig{STMR}. Using the SHMR of \citet{Hudson2015} --- which describes a mixture of blue (dominant at low-mass) and red (dominant at high-mass) galaxies --- and the concentration-mass relation given by \citet{MunozCuartas2011} --- and converted from $(M_{\rm vir},c_{\rm vir})$ to $(M_{200},c_{200})$ using the method of \citet{Hu2003} --- we can infer one such interpolated STMR (thick brown double-dashed line in \Fig{STMR}, middle panel). Hudson's interpolated STMR is a considerably better match in shape to our STMR, especially  at low masses, than any other original SHMRs reported in \Fig{STMR}.

% FIGURE
\begin{figure}
\centering
\includegraphics[width=0.45\textwidth]{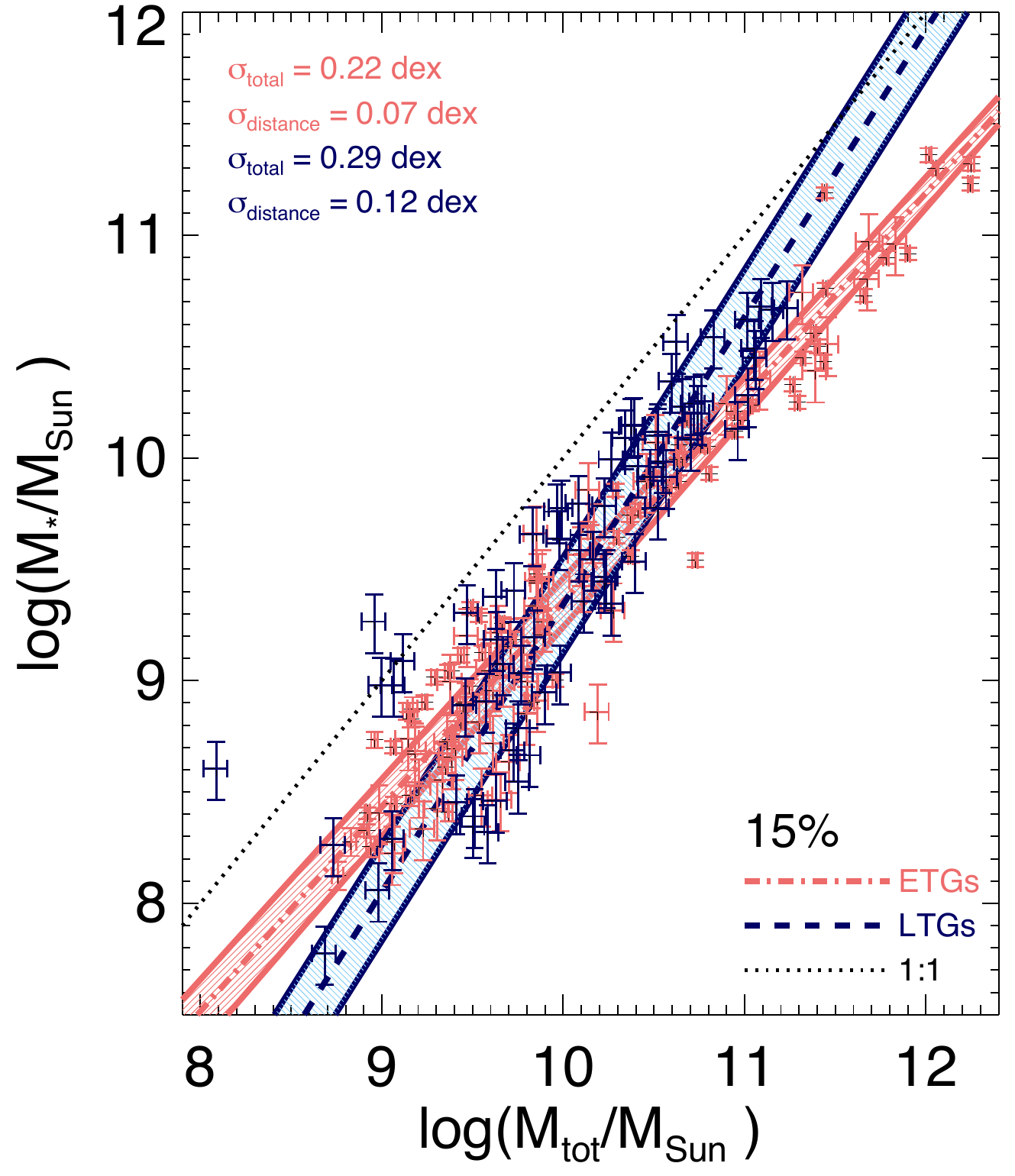}
\caption{As in \Fig{STMR_LTGs} (same data) but data points and STMRs for ETGs/LTGs shown with distance uncertainties only. LTGs and their STMR are shown in navy, and ETGs and their STMR are shown in light red. The black dotted line is the 1:1 line. $1-\sigma$ level total STMR scatters and STMR scatters due to 15\% distance uncertainties (shown about the best-fit STMRs as shaded blue and red regions) are indicated.}
\label{fig:STMR_distance}
\end{figure}

For clarity, we reproduce \Fig{STMR}'s middle panel in \Fig{STMR_LTGs} where we identify the STMR for ETGs and LTGs separately. The Hudson interpolation is shown prominently (double dashed brown line) as well as a similar inward interpolated STMR by Mancillas \& Avila-Reese (in preparation) (single pink line). Mancillas \& Avila-Reese extracted their STMR from mock simulations of LTGs in centrifugal equilibrium within adiabatically-contracted $\Lambda$CDM halos; their stellar and dynamical masses are all inferred within $2 R_{\rm e}$ which is quite appropriate for our comparison --- recall that we find $R_{23.5} \sim 1 - 4 R_{\rm e}$ for our sample, $r_{23.5} = R_{23.5}$ for LTGs, and $r_{23.5} = (4/3)R_{23.5}$ for ETGs. Hudson's STMR provides a generally fair match at low masses. Additionally, Mancillas \& Avila-Reese's STMR is a great match for the SHIVir LTG sample below $M_{\rm tot} < 10^{10} \solarm$, both slopes being near identical. Their mean STMR's slope is also in broad agreement with that of \citet{RodriguezPuebla2015} for blue galaxies (blue solid line in \Fig{STMR}, middle panel). We expect Mancillas \& Avila-Reese's STMR to match our LTG STMR, as it was constructed for LTGs.

The flattening of the STMR (Mancillas \& Avila-Reese) starting at $\log M_{*} \sim 10.8$ appears to be largely the imprint of the $M_{\rm bar}$-$M_{\rm vir}$ relation which turns over at $M_{\rm vir} \sim 10^{12} \solarm$. This well-studied turnover has great significance in galaxy formation models: on the one hand, at smaller virial masses, the gravitational potential is weaker and significant gas outflows due to SN-driven feedback are allowed; on the other, at larger masses, two processes become systematically more relevant:

1) the gas is (virial) shock-heated to very high temperatures in such a way that the cooling timescale becomes very long (in groups and clusters most baryons remain locked in the virial shock-heated gas, inhibiting further galaxy mass growth), and

2) the formation of luminous AGNs is efficient in such a way that the AGN feedback inhibits further stellar mass growth or even promotes strong gas ejection. 

\noindent The above mentioned processes become mostly irrelevant at virial masses near $10^{12} \solarm$. For lower and higher masses, the processes mentioned above systematically reduce the stellar mass growth efficiency \citep{AvilaReese2003, Behroozi2010}. 

\Fig{STMR_LTGs} also shows the STMR relation for our ETG and LTG samples separately. The LTG relation, 

$$ \log M_{*} = (1.27 \pm 0.07) \log M^{r_{23.5}}_{\rm dyn} - (3.44 \pm 0.72),$$ 

\noindent is considerably steeper than for ETGs, 

$$ \log M_{*} = (0.93 \pm 0.02) \log M^{r_{23.5}}_{\rm dyn} + (0.09 \pm 0.20).$$ 

\noindent This may indicate that ETGs live in more massive and/or concentrated halos than LTGs of the same stellar masses, at least above $M_{*} \sim 10^{10} \solarm$ (further discussed in \se{NFW} and \Fig{NFW}) as seen in \citet{More2011} for satellite galaxies. Note that using \citet{Serra2016}'s prescription for $V_{\rm circ}$ for ETGs would change their STMR's zero-point by $\sim 0.2-0.3$ dex. The distinct slope between the LTG and ETG STMRs would not change. The STMR for the full catalog is 

$$ \log M_{*} = (0.97 \pm 0.02) \log M^{r_{23.5}}_{\rm dyn} - (0.36 \pm 0.25).$$

Overall, it can be concluded from comparing abundance-matched SHMRs with our measured STMR that the former are only valid at the highest masses where the stellar-to-halo associations are least affected by stochasticity. \citet{Sawala2015} found HAM to fail at low-masses due to erroneous model assumptions including the assumptions that every halo can host a visible galaxy and that structure formation can be accurately represented by dark-matter only simulations. Usage of HAM has been constrained to galaxies with circular velocities above a threshold value, sometimes as high as 80 \kms\ \citep{TrujilloGomez2011}. Beyond this, the reliability of this technique hinges on including baryons \citep{TrujilloGomez2011} and carefully choosing optimal resolution values for the required simulations \citep{Klypin2015}. When available, high quality observational data such as those presented here are preferable to STMRs extrapolated from SHMRs or predicted by simulations.

In \se{tfr}, we discussed the possible effect of distance uncertainties on the TFR. Due to morphological segregation, assuming a distance of 16.5 Mpc when none other is available, can be more precarious for spiral galaxies who tend to populate cluster outskirts than for ellipticals that are more likely to be concentrated around the cluster center. To aggravate the situation, all 69 SHIVir galaxies morphologically classified as LTGs did not have catalog distances and were consequently assigned a value of 16.5 Mpc. In \Fig{STMR_distance}, we highlight the effects of using a distance uncertainty of 15\% for the 69 LTGs and the 49 ETGs with no formal distance available. When a published distance is available, the quoted uncertainty is readily used. Distance uncertainties affect both the total dynamical mass and total luminosity since physical radius is required. We find that distance errors can contribute $\sim 35$\% of the STMR scatter for both ETGs and LTGs. Nevertheless, despite substantial distance errors, the distinct STMR slopes between ETGs and LTGs remain.

\subsection{The Stellar Mass TFR, Mass-Size and Velocity-Size Relations}
\label{sec:sizemass}

%% FIGURE
%\begin{figure*}
%\centering
%\includegraphics[width=0.99\textwidth]{Fig14.pdf}
%\caption{(Left) Stellar-mass-Size relation, with the Pearson correlation coefficient, $r$, and scatter on log($R_{23.5}$), $\sigma_{R_{23.5}}$. Stellar mass is measured within a cylindrical volume with projected radius $R_{23.5}$. Typical uncertainties are shown for all relations. (Right) Same as in left, but the physical radius $r_{23.5}$ is used, and stellar mass for ETGs is measured inside a spherical volume of radius $r_{23.5}$. }
%\label{fig:MassSize}
%\end{figure*}

% FIGURE
\begin{figure*}
\centering
\includegraphics[width=0.99\textwidth]{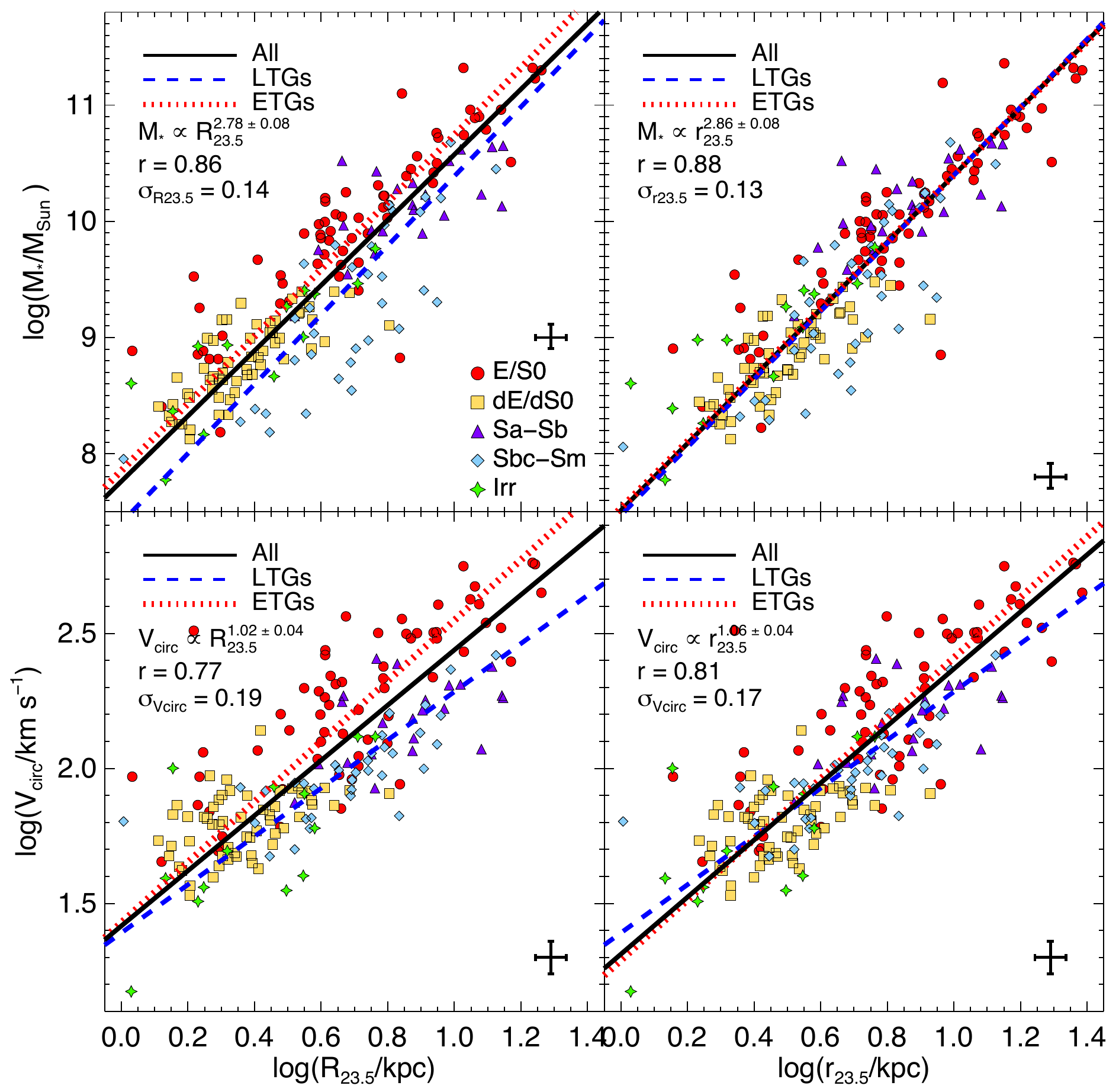}
\caption{(Top Left) Stellar-mass-Size relation, with the Pearson correlation coefficient, $r$, and scatter on log($R_{23.5}$), $\sigma_{R_{23.5}}$. Stellar mass is measured within a cylindrical volume with projected radius $R_{23.5}$. (Top Right) Same as in the top left panel, but the physical radius $r_{23.5}$ is used, and stellar mass for ETGs is measured inside a spherical volume of radius $r_{23.5}$. (Bottom Left) Projected Velocity-Size relation, with the Pearson correlation coefficient, $r$, and scatter on log($V_{\rm circ}$), $\sigma_{V_{\rm circ}}$. Typical uncertainties are shown for all relations. (Bottom Right) Same as in the bottom left panel, but the physical radius $r_{23.5}$ is used. Typical uncertainties are shown for all relations.}
\label{fig:MassSize}
\end{figure*}

% FIGURE
\begin{figure}
\centering
\includegraphics[width=0.45\textwidth]{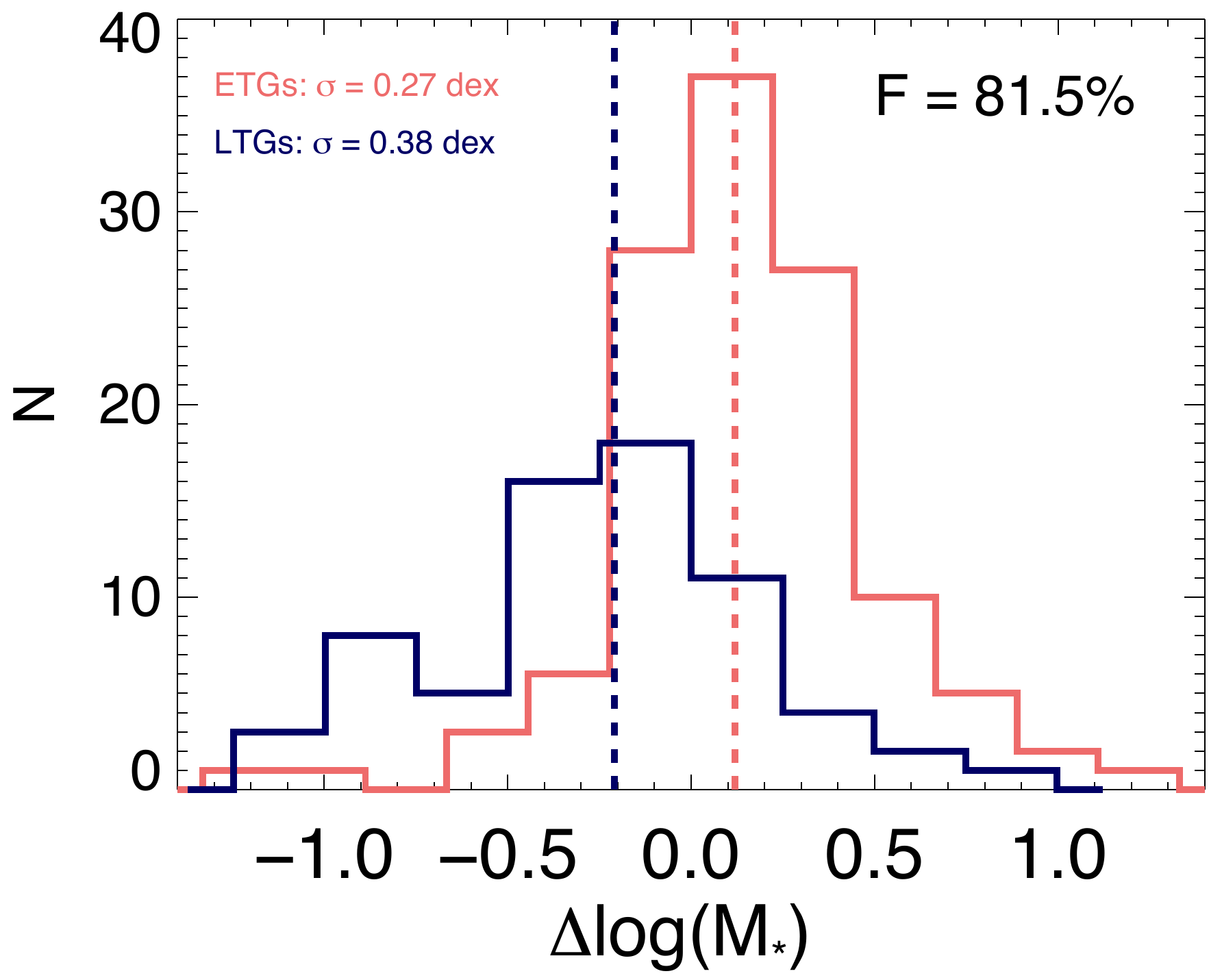}
\caption{Histogram of residuals on log($M_{*}$) in the projected mass-size relation (\Fig{MassSize}, top left panel) for ETGs (light red) and LTGs (navy). The standard deviation $\sigma$ of Gaussians fitted over each individual histogram is posted in the upper right corner. The {\it F}-test confidence result for a double versus a single Gaussian distribution is indicated.}
\label{fig:ms_bimo}
\end{figure}

Figs.~\ref{fig:STMR}-\ref{fig:STMR_LTGs} show two distinct STMRs for ETGs and LTGs, the latter being steeper than the former\footnote{The low mass ends of the two SHMRs for blue and red centrals of \citet{RodriguezPuebla2015} and \citet{Mandelbaum2016} show similar ``bimodal" distributions and distinct slopes, but those concern central rather than satellite galaxies and are thus ill-suited for comparisons with VCGs. This is discussed further in \se{NFW}.}. In \se{tfr}, we computed the STFR for SHIVir spirals for which extended RCs were measured. Including literature sources and all morphologies now yields the following relations: 
$$\log M_{*} = (3.28 \pm 0.26) \times \log V_{\rm circ} + (2.91 \pm 0.57)$$ for a sample of 69 LTGs and 
$$\log M_{*} = (2.56 \pm 0.07) \times \log V_{\rm circ} + (4.23 \pm 0.15)$$ for a sample of 121 ETGs. For all galaxies, this relation becomes 
$$\log M_{*} = (2.73 \pm 0.08) \times \log V_{\rm circ} + (3.93 \pm 0.18)$$ with a scatter of $\sigma_{\log V_{\rm circ}}=0.13$ dex. The mean slope and scatter of the STFR (\Fig{STMR}, bottom panel) matches $\Lambda$CDM expectations \citep{Dutton2011}. In the mass range $M_{*} > 10^{9.5} \solarm$, we find LTGs to contain more stellar mass than ETGs at a given $V_{\rm circ}$ measurement, in agreement with the observational data used by \citet{TrujilloGomez2011}. We also tested the use of different velocity metrics in the STFR, including $S_{0.3}$, $S_{0.5}$ \citep{Kassin2007}, and $V_{\rm rms} = \sqrt{\sigma_{\rm e}^{2} + V_{\rm rot}^{2}}$, and found that $V_{\rm circ}$ produces the smallest STFR scatter.

\Fig{MassSize}, top left panel, shows the mass-size relation for SHIVir VCGs in projected space. At any given projected radius, ETGs contain more stellar mass than LTGs, possibly due to the presence of a relatively larger and more concentrated core; we see a similar result for the trends with circular velocity as a proxy for dynamical mass in \Fig{Lum_Size}. \citet{Cappellari2013b} also found a strong link between bulge mass and global galaxy properties for the ATLAS$^{3{\rm D}}$ sample. As a result, galaxies with higher mean SBs lie above the mean stellar mass-size relation \citep{Hall2012}. We find the relations for the ETGs, 
$$\log M_{*} = (2.89 \pm 0.09) \times \log R_{23.5} + (7.85 \pm 0.06),$$ and LTGs, 
$$\log M_{*} = (2.98 \pm 0.19) \times \log R_{23.5} + (7.41 \pm 0.15),$$ to be closely parallel with a significant offset of $\log M_{*} \sim 0.45$ dex between the samples' zero-points\footnote{For future plans, we aim to add more dwarf galaxies to the SHIVir survey to determine if this offset is maintained across all mass regimes.}. The projected mass-size relation for the entire SHIVir catalog is 
$$\log M_{*} = (2.81 \pm 0.09) \times \log R_{23.5} + (7.76 \pm 0.06).$$ We visualize the bimodal nature of this relation in \Fig{ms_bimo}. Separate histograms of the forward residuals in the mass-size relation for ETGs and LTGs are fitted with Gaussians with a peak offset of 0.32 dex, which roughly matches the intercept offset between the two best-fit lines plotted on \Fig{MassSize}, top left panel. The bimodal nature of the residual distribution is confirmed at the 82\% level.

We considered in \se{lumsize} the possible evolutionary tracks connecting different galaxy populations in luminosity/mass-size relations, to have a more direct probe for comparison with \citet{Faber2007}, \citet{Cappellari2013a}, \citet{Cappellari2013c}, and \citet{Cappellari2016a}. We can, as in \Fig{Lum_Size}, identify the blue spirals that are theorized to evolve into fast-rotator ETGs which may eventually merge to form the massive slow-rotator ETGs. Our projected mass-size relation (\Fig{MassSize}, top left panel) showcases the change in size as the evolution occurs: quenched spirals become more centrally concentrated and turn into ETGs as their bulges grow, and these ETGs become increasingly large and massive via dry mergers.

Corresponding lines of evolution can be drawn in the STFR plot (\Fig{STMR}, bottom panel) as they are in \Fig{Lum_Size} for the luminosity-size relation: quenched LTGs (Sa-Sb) evolve into ETGs (E/S0) with an increase in $V_{\rm circ}$, and these ETGs move upwards in stellar mass and grow into the most massive ETGs along lines of constant $V_{\rm circ}$ via dry mergers. \citet{Cappellari2013a} identified a break at $M_{*} = 3 \times 10^{10} \solarm$ between the two different power laws that delineate the zone of exclusion in this mass-size distribution \citep[see][Fig. 23 for a depiction of this distribution]{Cappellari2016a}. This break could correspond to the slight downward turn in slope we observe in the upper envelope of the E/S0 sample at $\log R_{23.5} \sim 0.6$ and $M_{*} \sim 10^{10} \solarm$ in \Fig{MassSize}, top left panel. \citet{Cappellari2013a} identified this break at $M_{*} \sim 10^{10.5} \solarm$; the difference with our own break location is likely due to different apertures within which stellar mass was measured and systematic uncertainties in stellar masses. Whether this break is a true feature or merely a sample bias will deserve deeper investigation and additional data. A secondary break at $M_{*} \sim 2 \times 10^{9}$ was suggested by \citet{Cappellari2013b}, below which spirals, fast-rotator ETGs and dwarf spheroidals are assumed to follow the same continuous relation. This secondary break coincidentally matches the threshold for quenching of field galaxies found by \citet{Geha2012}: at smaller masses, only environmental effects such as gas stripping \citep{Cappellari2013a} and strangulation \citep{Peng2015} may quench these galaxies. Supernovae and stellar winds should not terminate star formation in these low mass systems \citep{Emerick2016}. Our sample could in principle allow us to observe this second break down to $\log M_{*} = 8.3$, but the increased scatter at low masses (for all the reasons stated above) thwarts any conclusive insight in this regime. \citet{Tinker2016} has shown that different quenching scenarios (involving hitting critical values for parameters such as redshift, halo mass, stellar mass, or stellar-to-halo mass ratio) will yield varying STMR scatter. 

% FIGURE
\begin{figure*}
\centering
\includegraphics[width=0.99\textwidth]{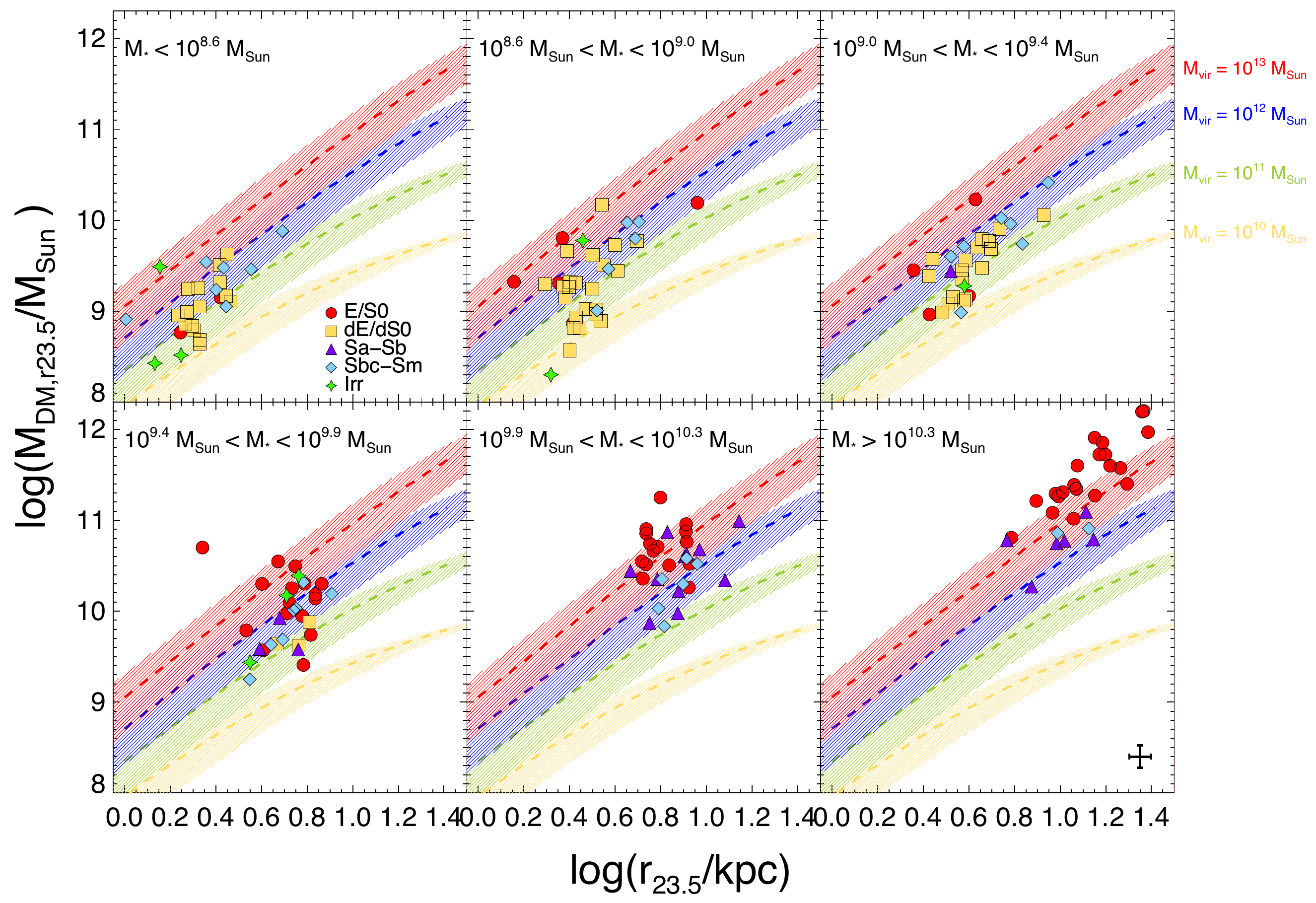}
\caption{Dark halo mass within $r_{23.5}$ as a function of $r_{23.5}$. The SHIVir dataset is divided into 6 stellar mass bins (chosen so that each bin would have roughly the same sample size), and all data points are color-coded by morphological type. NFW predictions based on varying virial halo masses $M_{\rm vir}$ and concentrations $c$ are shown. Typical uncertainties are shown in the bottom right corner.}
\label{fig:NFW}
\end{figure*}

We also inspect the SHIVir mass-size relation in physical space in \Fig{MassSize}, top right panel. Here, we plot the physical radius $r_{23.5}$ versus stellar mass measured within a spherical volume of radius $r_{23.5}$ for ETGs and a cylindrical volume of radius $r_{23.5} = R_{23.5}$ for LTGs. The bimodality seen in projected space is now completely eliminated. We find the relations for the ETGs, 
$$\log M_{*} = (2.87 \pm 0.10) \times \log r_{23.5} + (7.53 \pm 0.07),$$ and LTGs,
$$\log M_{*} = (2.93 \pm 0.18) \times \log r_{23.5} + (7.47 \pm 0.14),$$ to be exactly the same within the uncertainties. The physical mass-size relation for the entire SHIVir catalog is
$$\log M_{*} = (2.89 \pm 0.09) \times \log r_{23.5} + (7.51 \pm 0.07).$$ In this picture, the ``mixed scenario" hypothesis is still compatible with our results, but the first step --- wherein late-type spirals accrete gas to increase in stellar mass and eventually turn into fast-rotator ETGs via quenching (also reducing their size slightly) --- is considerably shorter in the mass-size space. The relative importance of this first step versus that of the dry merger step in galaxies' total mass accretion appears to depend on whether a projected space or physical space mass-size relation is analysed.

As a check, we also plot the velocity-size relation using both projected (\Fig{MassSize}, bottom left panel) and physical (\Fig{MassSize}, bottom right panel) radii. In both cases, we find distinct slopes between the ETG and LTG populations, similar to what was found for the STMR (\Fig{STMR_LTGs}) and the STFR (\Fig{STMR}, bottom panel). Since both the STFR and mass-size relation have the same slope in log-log space, we expect a slope of $\sim 1$ for the velocity-size relation: this is indeed observed within the uncertainties.

A joint analysis of \emph{all} the relations presented in \Fig{STMR} and \Fig{MassSize} should eventually provide the best insight into the physics of galaxy formation.

\subsection{Dark Matter Content and Halo Masses}
\label{sec:NFW}

We can estimate the dark matter mass within $r_{23.5}$ by subtracting the total stellar and gas mass from the dynamical mass estimated at that radius. It is interesting to compare dark matter mass estimates directly to model predictions from cosmological simulations at the same radius. 

Simulations of cosmological structure formation find that halos form with a universal density profile, originally described by \citet{Navarro1996}:

\begin{equation}
\rho(r) = {{\rho_0 r_{\rm s}}\over{r(r+r_{\rm s})^{3}}}. 
\end{equation}

\noindent This form has two free parameters, a characteristic density $\rho_{0}$ and a scale radius $r_{\rm s}$. The scale radius $r_{\rm s}$ is usually specified in terms of the virial radius $r_{\rm vir}$ and the concentration parameter $c = r_{\rm vir}/r_{\rm s}$. There are several common choices for $r_{\rm vir}$; here we adopt the spherical collapse definition, 
$$r_{\rm vir} = \bigg{(}\frac{3M_{\rm vir}}{4\pi\delta_{\rm vir}\rho_{\rm c}}\bigg{)}^{1/3},$$ where $\delta_{\rm vir}$ is an overdensity predicted by the spherical collapse model to be $\delta_{\rm c} \sim 100$ at $z=0$ in $\Lambda$CDM. More recent evidence has shown that the universal profile deviates slightly but systematically from this form, and is better described as a three-parameter Einasto profile \citep{Gao2008,Dutton2014b,Klypin2016}. However, the differences between the two forms in enclosed mass at $R \sim 10$ kpc are fairly minimal for halos in the mass range considered here; for simplicity we will assume the NFW form of the density profile. The corresponding enclosed mass is

\begin{equation}
M(<r) = M_{\rm vir} \frac{f(x)}{f(c)},
\end{equation}

\noindent where $x \equiv r/r_{\rm s}$ and $f(x) \equiv \ln(1+x) - x/(1+x)$.

For the spherical collapse definition, $r_{\rm vir} = 250\ {\rm kpc}\ (M/10^{12}M_\odot)^{1/3}$ at $z=0$. The concentration parameter $c$ has been the subject of many detailed studies; see e.g. \citet{Dutton2014b} or \citet{Klypin2016} for full references. Correcting for the different profiles and definitions assumed, there is reasonable agreement between recent studies in the predicted mean concentration as a function of mass and redshift, at least in our mass range and at low redshift. For a \emph{Planck} cosmology, the expected mean concentration is $c = 7-12$ over the halo mass range $M_{\rm h} \sim 5\times 10^{10}M_\odot - 5\times 10^{12}M_\odot$ \citep{Dutton2014b}. On the other hand, these simulations also predict appreciable halo-to-halo scatter in concentration, with $\log \sigma_c \sim 0.13$, or a factor of 1.35. Thus we will consider a range of representative concentrations, from $c = 5$ to $c = 15$.

\Fig{NFW} shows enclosed dark matter mass versus $r_{23.5}$ for the SHIVir sample (points, colored by morphological type), compared with predictions for NFW profiles of various values of $M_{\rm vir}$ and $c$ (curves, colored by mass and surrounded by a shaded region representing a range of concentrations). Starting from the top left-most panel, the halo masses expected from previous estimates of the SHMR (e.g. \citealp{Grossauer2015}, plotted in the middle panel of \Fig{STMR}) are $\log M_{\rm h} \sim$ 11, 11.2, 11.4, 11.6, 11.8, and 12.2, while the other SHMRs shown in \Fig{STMR}, middle panel, predict values $\sim 0.2-0.3$ dex lower. For the lower mass bins, these predictions seem in good agreement with the data; given observational uncertainties, most individual galaxies are consistent with halos in the predicted mass range and with a reasonable concentration, with only a few points lying $\sim 1-2\sigma$ above the expected relation. At higher stellar masses ($\log M_{*} > 9.9$), we see some evidence for more enclosed mass than expected from the $z=0$ prediction, particularly for the ETGs. The same trend is seen in Figure 12, where the predictions from \citet{Hudson2015} or Mancillas \etal (in preparation) lie $0.2-0.3$ dex to the left of the mean $M_{\rm tot}$ measured for the ETGs, but are roughly consistent with the mean $M_{\rm tot}$ measured for the LTGs.

\begin{table*}
\begin{center}
\caption{Galaxy Scaling Relations for Virgo Cluster}
{\small
\begin{tabular}{ c|c|cc|cccc }
\hline \hline
Relation								& Sample	& y						& x										& Slope			& Zero-point		& Scatter [dex]	& N \\
\hline
TFR									& LTGs	& $M_{i}$					& log($V^{\rm i}_{23.5}$)						& $-7.12 \pm 0.28$	& $-5.25 \pm 0.10$	& 0.216		& 46 \\
\hline
BTFR								& LTGs	& log($M_{\rm bar}$)			& log($V^{\rm i}_{23.5}$)						& $3.57 \pm 0.16$	& $2.44 \pm 0.35$	& 0.280		& 46 \\
\hline
\multirow{3}{*}{STMR}					& LTGs 	& \multirow{3}{*}{log($M_{*}$)}	& \multirow{3}{*}{log($M^{r_{23.5}}_{\rm dyn}$)}	& $1.27 \pm 0.07$	& $-3.44 \pm 0.72$	& 0.342		& 69 \\
 									& ETGs 	&						&										& $0.93 \pm 0.02$	& $0.09 \pm 0.20$	& 0.231		& 121 \\
 									& All 		&						&										& $0.97 \pm 0.02$	& $-0.36 \pm 0.25$	& 0.294		& 190 \\ 
 \hline
\multirow{3}{*}{STFR}					& LTGs 	& \multirow{3}{*}{log($M_{*}$)}	& \multirow{3}{*}{log($V_{\rm circ}$)}				& $3.28 \pm 0.26$	& $2.91 \pm 0.57$	& 0.448		& 69 \\
 									& ETGs 	&						&										& $2.56 \pm 0.07$	& $4.23 \pm 0.15$	& 0.301		& 121 \\
 									& All 		&						&										& $2.73 \pm 0.08$	& $3.93 \pm 0.18$	& 0.367		& 190 \\ 
\hline
\multirow{3}{*}{Projected Mass-Size}			& LTGs 	& \multirow{3}{*}{log($M_{*}$)}	& \multirow{3}{*}{log($R_{23.5}$)}				& $2.98 \pm 0.19$	& $7.41 \pm 0.15$	& 0.448		& 69 \\
 									& ETGs 	&						&										& $2.89 \pm 0.09$	& $7.85 \pm 0.06$	& 0.358		& 121 \\
 									& All 		&						&										& $2.81 \pm 0.09$	& $7.76 \pm 0.06$	& 0.394		& 190 \\ 
\hline
\multirow{3}{*}{Physical Mass-Size}			& LTGs 	& \multirow{3}{*}{log($M_{*}$)}	& \multirow{3}{*}{log($r_{23.5}$)}				& $2.93 \pm 0.18$	& $7.47 \pm 0.14$	& 0.438		& 69 \\
 									& ETGs 	&						&										& $2.87 \pm 0.10$	& $7.53 \pm 0.07$	& 0.350		& 121 \\
 									& All 		&						&										& $2.89 \pm 0.09$	& $7.51 \pm 0.07$	& 0.356		& 190 \\ 
\hline
\multirow{3}{*}{Projected Velocity-Size}		& LTGs 	& \multirow{3}{*}{log($V_{\rm circ}$)}	& \multirow{3}{*}{log($R_{23.5}$)}			& $0.89 \pm 0.08$	& $1.39 \pm 0.06$	& 0.137		& 69 \\
 									& ETGs 	&						&										& $1.12 \pm 0.05$	& $1.43 \pm 0.03$	& 0.182		& 121 \\
 									& All 		&						&										& $1.02 \pm 0.04$	& $1.42 \pm 0.03$	& 0.194		& 190 \\ 
\hline
\multirow{3}{*}{Physical Velocity-Size}		& LTGs 	& \multirow{3}{*}{log($V_{\rm circ}$)}	& \multirow{3}{*}{log($r_{23.5}$)}			& $0.89 \pm 0.08$	& $1.39 \pm 0.06$	& 0.137		& 69 \\
 									& ETGs 	&						&										& $1.12 \pm 0.05$	& $1.29 \pm 0.03$	& 0.182		& 121 \\
 									& All 		&						&										& $1.06 \pm 0.04$	& $1.31 \pm 0.03$	& 0.172		& 190 \\ 
\hline \hline
\end{tabular}}
\tablecomments{LTG sample, ETG sample, and All Morphologies sample are presented separately. $R_{23.5}$ and $r_{23.5}$ are in kpc. $V^{\rm i}_{23.5}$ is in  \kms. All masses are in $\solarm$. Slope, zero-point, scatter $\sigma$ and sample size $N$ are given for each relation.}
\label{tab:SR_sum}
\end{center}
\end{table*}

There are at least four possible explanations for this excess. The first is the effect of formation redshift. Our sample consists of cluster galaxies, some of which formed in the field at higher redshift and fell into the cluster well before $z=0$. The properties of these systems should correspond to those of field halos at higher redshift. Our own calculations indicate that the predictions for $z = 1$ would lie about $0.1-0.2$ dex above those shown, and could explain some of the offset seen in \Fig{NFW} (as well as the possible difference between ETGs and LTGs --- see \citealp{Thomas2009}). Another possibility is that these massive systems have experienced more adiabatic contraction. Furthermore, since $M_{\rm DM} = M_{\rm tot} - M_{*}$, inferred dark matter masses can depend strongly on the IMF assumed for the stellar component, especially at higher masses \citep{Dutton2013,Dutton2014a}. We note however that a different IMF would not affect the total mass estimates discussed earlier, and that these too showed excess mass relative to the predictions at high stellar mass. Finally, we note that \citet{Grossauer2015} found a similar upturn in their Virgo SHMR at large stellar masses as part of the NGVS, relative to field results such as \citet{Behroozi2013b}. If cluster galaxies, which find themselves in relatively dense environments from the outset, formed less efficiently than their field equivalents even before they merged into the cluster, this might explain such an offset. More detailed modeling of Virgo and other clusters will be required to fully test this hypothesis.

We note an offset of $0.3-0.5$ dex in dark matter content between LTGs and ETGs in the two highest stellar mass bins. Note that the use of \citet{Serra2016}'s $V_{\rm circ}$ prescription --- which would subsequently affect both $M_{\rm dyn}$ and $M_{\rm DM}$ measurements --- would likely decrease the size of this offset though not completely erase it. This difference in halo mass between LTGs and ETGs has been observed by \citet{Thomas2009} (in a sample of Coma cluster members), \citet{Dutton2011} and more recently by \citet{Mandelbaum2016} (for a sample of central locally brightest galaxies), whereby ETGs live in more massive halos than LTGs for a given stellar mass at $M_{*} > 10^{9.9} \solarm$; note that \citet{Hudson2015} found the opposite trend. In \citet{Mandelbaum2016}, this difference is further amplified for galaxies with stellar masses larger than $10^{11} \solarm$. It seems likely that this $\sim 2-3 \times$ difference in halo masses is independent of environment and could indicate systematic differences in halo contraction or formation epoch, as a function of galaxy type. The lack of major mergers in the formation history of LTGs may be chief in driving this difference \citep{Dutton2011}.

\section{Summary and Conclusions}
\label{sec:summary}

We have presented the first results from the spectroscopic component of the SHIVir survey. Using carefully-assembled scaling relations, we can paint a global picture of galaxy structural properties within a cluster environment. We find a strong bimodality in both SB and circular velocity for both ETGs and LTGs, possibly indicating dynamically unstable modes of galaxy formation. The TFR for Virgo cluster galaxies shows the same slope and normalization as that of field galaxies, but the TFR scatter for this galaxy cluster is larger than that of the field. Environmentally dependent processes such as tidal stripping and tidal interactions apparently do not influence the dark matter halo which strongly regulates a spiral galaxy's maximal rotational velocity, but may increase the scatter in cluster TFRs. TFR scatter is minimized when $V^{\rm i}_{23.5}$, the deprojected circular velocity measured at the isophotal level of 23.5 mag arcsec$^{-2}$, is used since the likelihood of sampling the flat part of the RC is increased. With slopes of $\sim4$ and $\sim3.5$, respectively, our stellar mass and baryonic TFRs are found to match those of others \citep{Hall2012,Bradford2016}. Following \citet{Hall2012}, we stress that scaling relation parameters are subject to the vagaries of fitting methods. The long-slit spectroscopy used in this study has also enabled the sampling of velocity dispersions beyond current integral field unit maps; as a result, we find that the FP scatter is minimized at (or beyond) $2 R_{\rm e}$ where the transition mid-point between baryon to dark matter domination typically occurs. The FPs for giant and dwarf ellipticals follow a possibly continuous but curved plane, suggesting analogous formation processes. FP residuals, or ``tilt'' of the FP, are found to correlate with galaxies' dark matter fraction (total-to-stellar-mass ratio). This, as well as the independence of the TFR scatter on SB or any other parameter, yield stringent galaxy formation constraints such as the characterization of the stellar IMF, impact of adiabatic contraction, and likely evolutionary paths of galaxies. In the Virgo cluster, we find the contribution of distance errors to the scaling relations' scatters to be substantial: 50\% for the TFR, 25\% for the FP, and 20--35\% for the STMR. For the TFR and STMR, that scatter increases at the low-mass end where the slope of the baryonic TFR is typically steeper. The total typical uncertainties for our stellar and dynamical masses are 0.11 dex and 0.08 dex respectively.

Our presentation of the Virgo cluster STMR and other scaling relations highlights a number of fundamental links between galaxy structural and dynamical parameters involving ill-constrained evolutionary scenarios and self-regulating processes: a possible turnover in the STMR at $M_{*} \approx 10^{11} \solarm$ concordant with other independent results, an increase in the scatter of the STMR and STFR below the transition stellar mass of $\sim 10^{9.5} \solarm$, a discrepancy in the slopes of STMRs/SHMRs based on dynamical masses measured within optical radii ($M_{\rm dyn}^{r_{23.5}}$) or the entire galaxy ($M_{200}$) between ETGs and LTGs, and a gradient in morphology (or SB) in the mass-size relation. 

Most SHMR studies hint at a maximal SF efficiency where $M_{*}/M_{\rm halo}$ is largest. Our data are too sparse in the relevant mass range to confirm this putative turnover, but they are not inconsistent with it either. If this feature is real, it suggests that SF mechanisms are sensitive to the baryon-to-dark-matter ratio at the optical radius even if the dark matter content may be relatively low within that radius. Quenching should thus strongly depend on halo mass rather than stellar mass \citep{Woo2013} for the Virgo cluster, which is mostly dominated by satellite galaxies. In addition to quenching mechanisms affecting central galaxies (gas depletion, heating, feedback), these satellites are subjected to their own unique mechanisms (ram pressure stripping, strangulation). This may explain the consistency of the highest mass range of our data with this turnover, and why the satellite SHMR from \citet{RodriguezPuebla2013} flattens at smaller masses than other SHMRs.

We provide in \Table{SR_sum} a summary of the salient two-parameter scaling relations presented in this paper. Our best-fit three parameter FP, is given by:
\begin{multline} 
$$\log (R_{\rm e}/{\rm kpc}) = 0.17 + 1.15 \log (\sigma_{\rm e}/\kms) \\ \nonumber
- 0.76 \log (\Sigma_{\rm e}/{\rm mag\ }{\rm arcsec}^{-2}),$$ \nonumber
\end{multline}

\noindent The distinct nature of the relations shown in \Fig{STMR}, \Fig{STMR_LTGs}, \Fig{MassSize} (top left, bottom left and right panels) and \Fig{NFW} between ETGs and LTGs solidifies the notion that these two groups are driven by different evolutionary scenarios. The different formation histories of ETGs and LTGs seem to result in a notable offset between their respective galaxy sizes. However, the scatter in our estimates of $\sigma$, and thus $V_{\rm circ}$ and $M_{\rm dyn}$ for ETGs along with the assumptions made in \se{dyn_mass}, complicate the confirmation of genuine, intrinsic physical mechanisms regarding any offset with LTGs. Still, the presence of this offset in the projected mass-size relation, which does not suffer from any velocity transformations, suggests that parameters like bulge mass and concentration may be driving the evolution of galaxies with color, morphological, size, star formation and mass. We do note, however, that this offset disappears in the physical mass-size relation. A number of steps can be taken to reduce the interpretation biases or further uncertainty in the quest towards understanding galaxy formation and evolution based on scaling relations. These include unbiased mass estimates based on robust velocity and mass metrics (free of deprojection uncertainty), bridging the gap in the dynamical mapping of the inner and outer regions of galaxies, augmenting dynamical databases for fainter galaxies into the low-mass regime, and expanding scaling relation manifolds with stellar population parameters such as age and metallicity.

\section*{Acknowledgments}

\noindent
NNQO acknowledges support from the Natural Science and Engineering Research Council (NSERC) of Canada through a PGS D3 scholarship. SC acknowledges support from the NSERC of Canada through a generous Research Discovery Grant. MC acknowledges support from a Royal Society University Research Fellowship. Mike Hudson is thanked warmly for interpolating his SHMR based on our data in order to match the STMR presented in \Fig{STMR_LTGs}, along with Vladimir Avila-Reese and Brisa Mancillas for providing the STMR for their simulated LTGs. We are also grateful to Jakob Walcher, Alexie Leauthaud, and Kristine Spekkens for insightful comments. Finally, our revised manuscript benefited greatly from the referee's thoughtful and constructive suggestions.

\bibliography{SRVirgo}

\begin{thebibliography}{}
\expandafter\ifx\csname natexlab\endcsname\relax\def\natexlab#1{#1}\fi

\bibitem[{{Aaronson} {et~al.}(1986){Aaronson}, {Bothun}, {Mould}, {Huchra},
  {Schommer}, {et~al.}}]{Aaronson1986}
{Aaronson}, M., {Bothun}, G., {Mould}, J., {et~al.} 1986, \apj, 302, 536

\bibitem[{{Abazajian} {et~al.}(2003){Abazajian}, {Adelman-McCarthy},
  {Ag{\"u}eros}, {Allam}, {Anderson}, {et~al.}}]{Abazajian2003}
{Abazajian}, K., {Adelman-McCarthy}, J.~K., {Ag{\"u}eros}, M.~A., {et~al.}
  2003, \aj, 126, 2081

\bibitem[{{Adelman-McCarthy} {et~al.}(2008){Adelman-McCarthy}, {Ag{\"u}eros},
  {Allam}, {Allende Prieto}, {Anderson}, {et~al.}}]{AdelmanMcCarthy2008}
{Adelman-McCarthy}, J.~K., {Ag{\"u}eros}, M.~A., {Allam}, S.~S., {et~al.} 2008,
  \apjs, 175, 297

\bibitem[{{Arnold} {et~al.}(2014){Arnold}, {Romanowsky}, {Brodie}, {Forbes},
  {Strader}, {et~al.}}]{Arnold2014}
{Arnold}, J.~A., {Romanowsky}, A.~J., {Brodie}, J.~P., {et~al.} 2014, \apj,
  791, 80

\bibitem[{{Avila-Reese} {et~al.}(2003){Avila-Reese}, {Firmani}, \&
  {V{\'a}zquez-Semadeni}}]{AvilaReese2003}
{Avila-Reese}, V., {Firmani}, C., \& {V{\'a}zquez-Semadeni}, E. 2003, in
  Revista Mexicana de Astronomia y Astrofisica Conference Series, Vol.~17,
  Revista Mexicana de Astronomia y Astrofisica Conference Series, 66--70

\bibitem[{{Avila-Reese} {et~al.}(2008){Avila-Reese}, {Zavala}, {Firmani}, \&
  {Hern{\'a}ndez-Toledo}}]{AvilaReese2008}
{Avila-Reese}, V., {Zavala}, J., {Firmani}, C., \& {Hern{\'a}ndez-Toledo},
  H.~M. 2008, \aj, 136, 1340

\bibitem[{{Bacon} {et~al.}(2001){Bacon}, {Copin}, {Monnet}, {Miller},
  {Allington-Smith}, {et~al.}}]{Bacon2001}
{Bacon}, R., {Copin}, Y., {Monnet}, G., {et~al.} 2001, \mnras, 326, 23

\bibitem[{{Behroozi} {et~al.}(2010){Behroozi}, {Conroy}, \&
  {Wechsler}}]{Behroozi2010}
{Behroozi}, P.~S., {Conroy}, C., \& {Wechsler}, R.~H. 2010, \apj, 717, 379

\bibitem[{{Behroozi} {et~al.}(2013{\natexlab{a}}){Behroozi}, {Wechsler}, \&
  {Conroy}}]{Behroozi2013a}
{Behroozi}, P.~S., {Wechsler}, R.~H., \& {Conroy}, C. 2013{\natexlab{a}},
  \apjl, 762, L31

\bibitem[{{Behroozi} {et~al.}(2013{\natexlab{b}}){Behroozi}, {Wechsler}, \&
  {Conroy}}]{Behroozi2013b}
---. 2013{\natexlab{b}}, \apj, 770, 57

\bibitem[{{Bekerait{\'e}} {et~al.}(2016){Bekerait{\'e}}, {Walcher},
  {Falc{\'o}n-Barroso}, {Garcia Lorenzo}, {Lyubenova}, {S{\'a}nchez},
  {Spekkens}, {van de Ven}, {Wisotzki}, {Ziegler}, {Aguerri},
  {Barrera-Ballesteros}, {Bland-Hawthorn}, {Catal{\'a}n-Torrecilla}, \&
  {Garc{\'{\i}}a-Benito}}]{Bekeraite2016}
{Bekerait{\'e}}, S., {Walcher}, C.~J., {Falc{\'o}n-Barroso}, J., {et~al.} 2016,
  \aap, 593, A114

\bibitem[{{Bender} {et~al.}(1992){Bender}, {Burstein}, \& {Faber}}]{Bender1992}
{Bender}, R., {Burstein}, D., \& {Faber}, S.~M. 1992, \apj, 399, 462

\bibitem[{{Bernardi} {et~al.}(2003){Bernardi}, {Sheth}, {Annis}, {Burles},
  {Eisenstein}, {et~al.}}]{Bernardi2003}
{Bernardi}, M., {Sheth}, R.~K., {Annis}, J., {et~al.} 2003, \aj, 125, 1866

\bibitem[{{Bertola} {et~al.}(1991){Bertola}, {Bettoni}, {Danziger}, {Sadler},
  {Sparke}, {et~al.}}]{Bertola1991}
{Bertola}, F., {Bettoni}, D., {Danziger}, J., {et~al.} 1991, \apj, 373, 369

\bibitem[{{Binggeli} {et~al.}(1985){Binggeli}, {Sandage}, \&
  {Tammann}}]{Binggeli1985}
{Binggeli}, B., {Sandage}, A., \& {Tammann}, G.~A. 1985, \aj, 90, 1681

\bibitem[{{Blakeslee} {et~al.}(2009){Blakeslee}, {Jord{\'a}n}, {Mei},
  {C{\^o}t{\'e}}, {Ferrarese}, {et~al.}}]{Blakeslee2009}
{Blakeslee}, J.~P., {Jord{\'a}n}, A., {Mei}, S., {et~al.} 2009, \apj, 694, 556

\bibitem[{{Borriello} {et~al.}(2003){Borriello}, {Salucci}, \&
  {Danese}}]{Borriello2003}
{Borriello}, A., {Salucci}, P., \& {Danese}, L. 2003, \mnras, 341, 1109

\bibitem[{{Bradford} {et~al.}(2016){Bradford}, {Geha}, \& {van den
  Bosch}}]{Bradford2016}
{Bradford}, J.~D., {Geha}, M.~C., \& {van den Bosch}, F.~C. 2016, \apj, 832, 11

\bibitem[{{Bundy} {et~al.}(2015){Bundy}, {Bershady}, {Law}, {Yan}, {Drory},
  {et~al.}}]{Bundy2015}
{Bundy}, K., {Bershady}, M.~A., {Law}, D.~R., {et~al.} 2015, \apj, 798, 7

\bibitem[{{Bundy} {et~al.}(2007){Bundy}, {Treu}, \& {Ellis}}]{Bundy2007}
{Bundy}, K., {Treu}, T., \& {Ellis}, R.~S. 2007, \apjl, 665, L5

\bibitem[{{Burstein} {et~al.}(1997){Burstein}, {Bender}, {Faber}, \&
  {Nolthenius}}]{Burstein1997}
{Burstein}, D., {Bender}, R., {Faber}, S., \& {Nolthenius}, R. 1997, \aj, 114,
  1365

\bibitem[{{Busarello} {et~al.}(1997){Busarello}, {Capaccioli}, {Capozziello},
  {Longo}, \& {Puddu}}]{Busarello1997}
{Busarello}, G., {Capaccioli}, M., {Capozziello}, S., {Longo}, G., \& {Puddu},
  E. 1997, \aap, 320, 415

\bibitem[{{Cappellari}(2013)}]{Cappellari2013c}
{Cappellari}, M. 2013, \apjl, 778, L2

\bibitem[{{Cappellari}(2016)}]{Cappellari2016a}
---. 2016, \araa, 54, 597

\bibitem[{{Cappellari}(2017)}]{Cappellari2017}
---. 2017, \mnras, 466, 798

\bibitem[{{Cappellari} {et~al.}(2006){Cappellari}, {Bacon}, {Bureau}, {Damen},
  {Davies}, {et~al.}}]{Cappellari2006}
{Cappellari}, M., {Bacon}, R., {Bureau}, M., {et~al.} 2006, \mnras, 366, 1126

\bibitem[{{Cappellari} \& {Emsellem}(2004)}]{Cappellari2004}
{Cappellari}, M., \& {Emsellem}, E. 2004, \pasp, 116, 138

\bibitem[{{Cappellari} {et~al.}(2011{\natexlab{a}}){Cappellari}, {Emsellem},
  {Krajnovi{\'c}}, {McDermid}, {Scott}, {et~al.}}]{Cappellari2011a}
{Cappellari}, M., {Emsellem}, E., {Krajnovi{\'c}}, D., {et~al.}
  2011{\natexlab{a}}, \mnras, 413, 813

\bibitem[{{Cappellari} {et~al.}(2011{\natexlab{b}}){Cappellari}, {Emsellem},
  {Krajnovi{\'c}}, {McDermid}, {Serra}, {et~al.}}]{Cappellari2011b}
---. 2011{\natexlab{b}}, \mnras, 416, 1680

\bibitem[{{Cappellari} {et~al.}(2012){Cappellari}, {McDermid}, {Alatalo},
  {Blitz}, {Bois}, {et~al.}}]{Cappellari2012}
{Cappellari}, M., {McDermid}, R.~M., {Alatalo}, K., {et~al.} 2012, \nat, 484,
  485

\bibitem[{{Cappellari} {et~al.}(2013{\natexlab{a}}){Cappellari}, {McDermid},
  {Alatalo}, {Blitz}, {Bois}, {et~al.}}]{Cappellari2013b}
---. 2013{\natexlab{a}}, \mnras, 432, 1862

\bibitem[{{Cappellari} {et~al.}(2013{\natexlab{b}}){Cappellari}, {Scott},
  {Alatalo}, {Blitz}, {Bois}, {et~al.}}]{Cappellari2013a}
{Cappellari}, M., {Scott}, N., {Alatalo}, K., {et~al.} 2013{\natexlab{b}},
  \mnras, 432, 1709

\bibitem[{{Chabrier}(2003)}]{Chabrier2003}
{Chabrier}, G. 2003, \pasp, 115, 763

\bibitem[{{Chan} {et~al.}(2015){Chan}, {Kere{\v s}}, {O{\~n}orbe}, {Hopkins},
  {Muratov}, {Faucher-Gigu{\`e}re}, \& {Quataert}}]{Chan2015}
{Chan}, T.~K., {Kere{\v s}}, D., {O{\~n}orbe}, J., {et~al.} 2015, \mnras, 454,
  2981

\bibitem[{{Chemin} {et~al.}(2006){Chemin}, {Balkowski}, {Cayatte}, {Carignan},
  {Amram}, {et~al.}}]{Chemin2006}
{Chemin}, L., {Balkowski}, C., {Cayatte}, V., {et~al.} 2006, \mnras, 366, 812

\bibitem[{{Ciotti}(1991)}]{Ciotti1991}
{Ciotti}, L. 1991, \aap, 249, 99

\bibitem[{{Ciotti} {et~al.}(1996){Ciotti}, {Lanzoni}, \&
  {Renzini}}]{Ciotti1996}
{Ciotti}, L., {Lanzoni}, B., \& {Renzini}, A. 1996, \mnras, 282, 1

\bibitem[{{C{\^o}t{\'e}} {et~al.}(2004){C{\^o}t{\'e}}, {Blakeslee},
  {Ferrarese}, {Jord{\'a}n}, {Mei}, {et~al.}}]{Cote2004}
{C{\^o}t{\'e}}, P., {Blakeslee}, J.~P., {Ferrarese}, L., {et~al.} 2004, \apjs,
  153, 223

\bibitem[{{Courteau}(1996)}]{Courteau1996}
{Courteau}, S. 1996, \apjs, 103, 363

\bibitem[{{Courteau}(1997)}]{Courteau1997}
---. 1997, \aj, 114, 2402

\bibitem[{{Courteau} {et~al.}(2014){Courteau}, {Cappellari}, {de Jong},
  {Dutton}, {Emsellem}, {et~al.}}]{Courteau2014}
{Courteau}, S., {Cappellari}, M., {de Jong}, R.~S., {et~al.} 2014, Reviews of
  Modern Physics, 86, 47

\bibitem[{{Courteau} \& {Dutton}(2015)}]{Courteau2015}
{Courteau}, S., \& {Dutton}, A.~A. 2015, \apjl, 801, L20

\bibitem[{{Courteau} {et~al.}(2007{\natexlab{a}}){Courteau}, {Dutton}, {van den
  Bosch}, {MacArthur}, {Dekel}, {et~al.}}]{Courteau2007b}
{Courteau}, S., {Dutton}, A.~A., {van den Bosch}, F.~C., {et~al.}
  2007{\natexlab{a}}, \apj, 671, 203

\bibitem[{{Courteau} {et~al.}(2007{\natexlab{b}}){Courteau}, {McDonald},
  {Widrow}, \& {Holtzman}}]{Courteau2007a}
{Courteau}, S., {McDonald}, M., {Widrow}, L.~M., \& {Holtzman}, J.
  2007{\natexlab{b}}, \apjl, 655, L21

\bibitem[{{Courteau} \& {Rix}(1999)}]{Courteau1999}
{Courteau}, S., \& {Rix}, H.-W. 1999, \apj, 513, 561

\bibitem[{{Courteau} {et~al.}(2000){Courteau}, {Willick}, {Strauss},
  {Schlegel}, \& {Postman}}]{Courteau2000}
{Courteau}, S., {Willick}, J.~A., {Strauss}, M.~A., {Schlegel}, D., \&
  {Postman}, M. 2000, \apj, 544, 636

\bibitem[{{Croom} {et~al.}(2012){Croom}, {Lawrence}, {Bland-Hawthorn},
  {Bryant}, {Fogarty}, {et~al.}}]{Croom2012}
{Croom}, S.~M., {Lawrence}, J.~S., {Bland-Hawthorn}, J., {et~al.} 2012, \mnras,
  421, 872

\bibitem[{{Dalcanton} {et~al.}(1997){Dalcanton}, {Spergel}, {Gunn}, {Schmidt},
  \& {Schneider}}]{Dalcanton1997}
{Dalcanton}, J.~J., {Spergel}, D.~N., {Gunn}, J.~E., {Schmidt}, M., \&
  {Schneider}, D.~P. 1997, \aj, 114, 635

\bibitem[{{Dekel} \& {Birnboim}(2006)}]{Dekel2006}
{Dekel}, A., \& {Birnboim}, Y. 2006, \mnras, 368, 2

\bibitem[{{Desmond} \& {Wechsler}(2017)}]{Desmond2017}
{Desmond}, H., \& {Wechsler}, R.~H. 2017, \mnras, 465, 820

\bibitem[{{Djorgovski} \& {Davis}(1987)}]{Djorgovski1987}
{Djorgovski}, S., \& {Davis}, M. 1987, \apj, 313, 59

\bibitem[{{Dressler}(1980)}]{Dressler1980}
{Dressler}, A. 1980, \apj, 236, 351

\bibitem[{{Dressler} {et~al.}(1987){Dressler}, {Lynden-Bell}, {Burstein},
  {Davies}, {Faber}, {et~al.}}]{Dressler1987}
{Dressler}, A., {Lynden-Bell}, D., {Burstein}, D., {et~al.} 1987, \apj, 313, 42

\bibitem[{{Dubois} {et~al.}(2013){Dubois}, {Gavazzi}, {Peirani}, \&
  {Silk}}]{Dubois2013}
{Dubois}, Y., {Gavazzi}, R., {Peirani}, S., \& {Silk}, J. 2013, \mnras, 433,
  3297

\bibitem[{{Dutton} {et~al.}(2011){Dutton}, {Conroy}, {van den Bosch}, {Simard},
  {Mendel}, {et~al.}}]{Dutton2011}
{Dutton}, A.~A., {Conroy}, C., {van den Bosch}, F.~C., {et~al.} 2011, \mnras,
  416, 322

\bibitem[{{Dutton} \& {Macci{\`o}}(2014)}]{Dutton2014b}
{Dutton}, A.~A., \& {Macci{\`o}}, A.~V. 2014, \mnras, 441, 3359

\bibitem[{{Dutton} {et~al.}(2013){Dutton}, {Macci{\`o}}, {Mendel}, \&
  {Simard}}]{Dutton2013}
{Dutton}, A.~A., {Macci{\`o}}, A.~V., {Mendel}, J.~T., \& {Simard}, L. 2013,
  \mnras, 432, 2496

\bibitem[{{Dutton} \& {Treu}(2014)}]{Dutton2014a}
{Dutton}, A.~A., \& {Treu}, T. 2014, \mnras, 438, 3594

\bibitem[{{Dutton} \& {van den Bosch}(2009)}]{Dutton2009}
{Dutton}, A.~A., \& {van den Bosch}, F.~C. 2009, \mnras, 396, 141

\bibitem[{{Dutton} {et~al.}(2007){Dutton}, {van den Bosch}, {Dekel}, \&
  {Courteau}}]{Dutton2007}
{Dutton}, A.~A., {van den Bosch}, F.~C., {Dekel}, A., \& {Courteau}, S. 2007,
  \apj, 654, 27

\bibitem[{{Emerick} {et~al.}(2016){Emerick}, {Mac Low}, {Grcevich}, \&
  {Gatto}}]{Emerick2016}
{Emerick}, A., {Mac Low}, M.-M., {Grcevich}, J., \& {Gatto}, A. 2016, \apj,
  826, 148

\bibitem[{{Emsellem} {et~al.}(2011){Emsellem}, {Cappellari}, {Krajnovi{\'c}},
  {Alatalo}, {Blitz}, {et~al.}}]{Emsellem2011}
{Emsellem}, E., {Cappellari}, M., {Krajnovi{\'c}}, D., {et~al.} 2011, \mnras,
  414, 888

\bibitem[{{Faber} {et~al.}(1987){Faber}, {Dressler}, {Davies}, {Burstein}, \&
  {Lynden-Bell}}]{Faber1987}
{Faber}, S.~M., {Dressler}, A., {Davies}, R.~L., {Burstein}, D., \&
  {Lynden-Bell}, D. 1987, in Nearly Normal Galaxies. From the Planck Time to
  the Present, ed. S.~M. {Faber}, 175--183

\bibitem[{{Faber} \& {Jackson}(1976)}]{Faber1976}
{Faber}, S.~M., \& {Jackson}, R.~E. 1976, \apj, 204, 668

\bibitem[{{Faber} {et~al.}(2007){Faber}, {Willmer}, {Wolf}, {Koo}, {Weiner},
  {et~al.}}]{Faber2007}
{Faber}, S.~M., {Willmer}, C.~N.~A., {Wolf}, C., {et~al.} 2007, \apj, 665, 265

\bibitem[{{Falc{\'o}n-Barroso}
  {et~al.}(2011{\natexlab{a}}){Falc{\'o}n-Barroso}, {S{\'a}nchez-Bl{\'a}zquez},
  {Vazdekis}, {Ricciardelli}, {Cardiel}, {et~al.}}]{FalconBarroso2011a}
{Falc{\'o}n-Barroso}, J., {S{\'a}nchez-Bl{\'a}zquez}, P., {Vazdekis}, A.,
  {et~al.} 2011{\natexlab{a}}, \aap, 532, A95

\bibitem[{{Falc{\'o}n-Barroso}
  {et~al.}(2011{\natexlab{b}}){Falc{\'o}n-Barroso}, {van de Ven}, {Peletier},
  {Bureau}, {Jeong}, {et~al.}}]{FalconBarroso2011b}
{Falc{\'o}n-Barroso}, J., {van de Ven}, G., {Peletier}, R.~F., {et~al.}
  2011{\natexlab{b}}, \mnras, 417, 1787

\bibitem[{{Fall} \& {Romanowsky}(2013)}]{Fall2013}
{Fall}, S.~M., \& {Romanowsky}, A.~J. 2013, \apjl, 769, L26

\bibitem[{{Fern{\'a}ndez Lorenzo} {et~al.}(2011){Fern{\'a}ndez Lorenzo},
  {Cepa}, {Bongiovanni}, {P{\'e}rez Garc{\'{\i}}a}, {Ederoclite},
  {et~al.}}]{FernandezLorenzo2011}
{Fern{\'a}ndez Lorenzo}, M., {Cepa}, J., {Bongiovanni}, A., {et~al.} 2011,
  \aap, 526, A72

\bibitem[{{Ferrarese} {et~al.}(2012){Ferrarese}, {C{\^o}t{\'e}}, {Cuillandre},
  {Gwyn}, {Peng}, {et~al.}}]{Ferrarese2012}
{Ferrarese}, L., {C{\^o}t{\'e}}, P., {Cuillandre}, J.-C., {et~al.} 2012, \apjs,
  200, 4

\bibitem[{{Ferrarese} {et~al.}(2016){Ferrarese}, {C{\^o}t{\'e}},
  {S{\'a}nchez-Janssen}, {Roediger}, {McConnachie}, {et~al.}}]{Ferrarese2016}
{Ferrarese}, L., {C{\^o}t{\'e}}, P., {S{\'a}nchez-Janssen}, R., {et~al.} 2016,
  \apj, 824, 10

\bibitem[{{Forbes} {et~al.}(1998){Forbes}, {Ponman}, \& {Brown}}]{Forbes1998}
{Forbes}, D.~A., {Ponman}, T.~J., \& {Brown}, R.~J.~N. 1998, \apjl, 508, L43

\bibitem[{{Fouqu{\'e}} {et~al.}(1990){Fouqu{\'e}}, {Bottinelli}, {Gouguenheim},
  \& {Paturel}}]{Fouque1990}
{Fouqu{\'e}}, P., {Bottinelli}, L., {Gouguenheim}, L., \& {Paturel}, G. 1990,
  \apj, 349, 1

\bibitem[{{Freeman}(1970)}]{Freeman1970}
{Freeman}, K.~C. 1970, \apj, 160, 811

\bibitem[{{Fukugita} {et~al.}(1993){Fukugita}, {Okamura}, \&
  {Yasuda}}]{Fukugita1993}
{Fukugita}, M., {Okamura}, S., \& {Yasuda}, N. 1993, \apjl, 412, L13

\bibitem[{{Gao} {et~al.}(2008){Gao}, {Navarro}, {Cole}, {Frenk}, {White},
  {et~al.}}]{Gao2008}
{Gao}, L., {Navarro}, J.~F., {Cole}, S., {et~al.} 2008, \mnras, 387, 536

\bibitem[{{Gavazzi} {et~al.}(2003){Gavazzi}, {Boselli}, {Donati}, {Franzetti},
  \& {Scodeggio}}]{Gavazzi2003}
{Gavazzi}, G., {Boselli}, A., {Donati}, A., {Franzetti}, P., \& {Scodeggio}, M.
  2003, \aap, 400, 451

\bibitem[{{Geha} {et~al.}(2012){Geha}, {Blanton}, {Yan}, \&
  {Tinker}}]{Geha2012}
{Geha}, M., {Blanton}, M.~R., {Yan}, R., \& {Tinker}, J.~L. 2012, \apj, 757, 85

\bibitem[{{Geha} {et~al.}(2003){Geha}, {Guhathakurta}, \& {van der
  Marel}}]{Geha2003}
{Geha}, M., {Guhathakurta}, P., \& {van der Marel}, R.~P. 2003, \aj, 126, 1794

\bibitem[{{Giovanelli} {et~al.}(1997){Giovanelli}, {Haynes}, {Herter}, {Vogt},
  {da Costa}, {Freudling}, {Salzer}, \& {Wegner}}]{Giovanelli1997}
{Giovanelli}, R., {Haynes}, M.~P., {Herter}, T., {et~al.} 1997, \aj, 113, 53

\bibitem[{{Graham} \& {Colless}(1997)}]{Graham1997}
{Graham}, A., \& {Colless}, M. 1997, \mnras, 287, 221

\bibitem[{{Graham}(2005)}]{Graham2005}
{Graham}, A.~W. 2005, in IAU Colloq. 198: Near-fields cosmology with dwarf
  elliptical galaxies, ed. H.~{Jerjen} \& B.~{Binggeli}, 303--310

\bibitem[{{Graham} \& {Guzm{\'a}n}(2003)}]{Graham2003}
{Graham}, A.~W., \& {Guzm{\'a}n}, R. 2003, \aj, 125, 2936

\bibitem[{{Graham} \& {Guzman}(2004)}]{Graham2004}
{Graham}, A.~W., \& {Guzman}, R. 2004, in Astrophysics and Space Science
  Library, Vol. 319, Penetrating Bars Through Masks of Cosmic Dust, ed. D.~L.
  {Block}, I.~{Puerari}, K.~C. {Freeman}, R.~{Groess}, \& E.~K. {Block}, 723

\bibitem[{{Granato} {et~al.}(2004){Granato}, {De Zotti}, {Silva}, {Bressan}, \&
  {Danese}}]{Granato2004}
{Granato}, G.~L., {De Zotti}, G., {Silva}, L., {Bressan}, A., \& {Danese}, L.
  2004, \apj, 600, 580

\bibitem[{{Grossauer} {et~al.}(2015){Grossauer}, {Taylor}, {Ferrarese},
  {MacArthur}, {C{\^o}t{\'e}}, {et~al.}}]{Grossauer2015}
{Grossauer}, J., {Taylor}, J.~E., {Ferrarese}, L., {et~al.} 2015, \apj, 807, 88

\bibitem[{{Gu} {et~al.}(2016){Gu}, {Conroy}, \& {Behroozi}}]{Gu2016}
{Gu}, M., {Conroy}, C., \& {Behroozi}, P. 2016, \apj, 833, 2

\bibitem[{{Gurovich} {et~al.}(2010){Gurovich}, {Freeman}, {Jerjen},
  {Staveley-Smith}, \& {Puerari}}]{Gurovich2010}
{Gurovich}, S., {Freeman}, K., {Jerjen}, H., {Staveley-Smith}, L., \&
  {Puerari}, I. 2010, \aj, 140, 663

\bibitem[{{Hall} {et~al.}(2012){Hall}, {Courteau}, {Dutton}, {McDonald}, \&
  {Zhu}}]{Hall2012}
{Hall}, M., {Courteau}, S., {Dutton}, A.~A., {McDonald}, M., \& {Zhu}, Y. 2012,
  \mnras, 425, 2741

\bibitem[{{Haynes} {et~al.}(2011){Haynes}, {Giovanelli}, {Martin}, {Hess},
  {Saintonge}, {et~al.}}]{Haynes2011}
{Haynes}, M.~P., {Giovanelli}, R., {Martin}, A.~M., {et~al.} 2011, \aj, 142,
  170

\bibitem[{{Hernquist}(1990)}]{Hernquist1990}
{Hernquist}, L. 1990, \apj, 356, 359

\bibitem[{{Hu} \& {Kravtsov}(2003)}]{Hu2003}
{Hu}, W., \& {Kravtsov}, A.~V. 2003, \apj, 584, 702

\bibitem[{{Hudson} {et~al.}(2015){Hudson}, {Gillis}, {Coupon}, {Hildebrandt},
  {Erben}, {et~al.}}]{Hudson2015}
{Hudson}, M.~J., {Gillis}, B.~R., {Coupon}, J., {et~al.} 2015, \mnras, 447, 298

\bibitem[{{Jerjen} {et~al.}(2004){Jerjen}, {Binggeli}, \&
  {Barazza}}]{Jerjen2004}
{Jerjen}, H., {Binggeli}, B., \& {Barazza}, F.~D. 2004, \aj, 127, 771

\bibitem[{{Karachentsev} {et~al.}(2017){Karachentsev}, {Kaisina}, \&
  {Kashibadze (Nasonova}}]{Karachentsev2017}
{Karachentsev}, I.~D., {Kaisina}, E.~I., \& {Kashibadze (Nasonova}, O.~G. 2017,
  \aj, 153, 6

\bibitem[{{Kassin} {et~al.}(2007){Kassin}, {Weiner}, {Faber}, {Koo}, {Lotz},
  {Diemand}, {Harker}, {Bundy}, {Metevier}, {Phillips}, {Cooper}, {Croton},
  {Konidaris}, {Noeske}, \& {Willmer}}]{Kassin2007}
{Kassin}, S.~A., {Weiner}, B.~J., {Faber}, S.~M., {et~al.} 2007, \apjl, 660,
  L35

\bibitem[{{Kaviraj} {et~al.}(2014){Kaviraj}, {Huertas-Company}, {Cohen},
  {Peirani}, {Windhorst}, {et~al.}}]{Kaviraj2014}
{Kaviraj}, S., {Huertas-Company}, M., {Cohen}, S., {et~al.} 2014, \mnras, 443,
  1861

\bibitem[{{Kelly}(2007)}]{Kelly2007}
{Kelly}, B.~C. 2007, \apj, 665, 1489

\bibitem[{{Kent}(1981)}]{Kent1981}
{Kent}, S.~M. 1981, \apj, 245, 805

\bibitem[{{Klypin} {et~al.}(2015){Klypin}, {Prada}, {Yepes}, {He{\ss}}, \&
  {Gottl{\"o}ber}}]{Klypin2015}
{Klypin}, A., {Prada}, F., {Yepes}, G., {He{\ss}}, S., \& {Gottl{\"o}ber}, S.
  2015, \mnras, 447, 3693

\bibitem[{{Klypin} {et~al.}(2016){Klypin}, {Yepes}, {Gottl{\"o}ber}, {Prada},
  \& {He{\ss}}}]{Klypin2016}
{Klypin}, A., {Yepes}, G., {Gottl{\"o}ber}, S., {Prada}, F., \& {He{\ss}}, S.
  2016, \mnras, 457, 4340

\bibitem[{{Koopmann} \& {Kenney}(2004)}]{Koopmann2004}
{Koopmann}, R.~A., \& {Kenney}, J.~D.~P. 2004, \apj, 613, 866

\bibitem[{{Kormendy}(1985)}]{Kormendy1985}
{Kormendy}, J. 1985, \apj, 295, 73

\bibitem[{{Kravtsov} {et~al.}(2014){Kravtsov}, {Vikhlinin}, \&
  {Meshscheryakov}}]{Kravtsov2014}
{Kravtsov}, A., {Vikhlinin}, A., \& {Meshscheryakov}, A. 2014, ArXiv e-prints,
  arXiv:1401.7329

\bibitem[{{La Barbera} {et~al.}(2008){La Barbera}, {Busarello}, {Merluzzi}, {de
  la Rosa}, {Coppola}, {et~al.}}]{LaBarbera2008}
{La Barbera}, F., {Busarello}, G., {Merluzzi}, P., {et~al.} 2008, \apj, 689,
  913

\bibitem[{{Leauthaud} {et~al.}(2012){Leauthaud}, {Tinker}, {Bundy}, {Behroozi},
  {Massey}, {et~al.}}]{Leauthaud2012}
{Leauthaud}, A., {Tinker}, J., {Bundy}, K., {et~al.} 2012, \apj, 744, 159

\bibitem[{{Lee} {et~al.}(2014){Lee}, {Rey}, \& {Kim}}]{Lee2014}
{Lee}, J., {Rey}, S.~C., \& {Kim}, S. 2014, \apj, 791, 15

\bibitem[{{Lehmann} {et~al.}(2017){Lehmann}, {Mao}, {Becker}, {Skillman}, \&
  {Wechsler}}]{Lehmann2017}
{Lehmann}, B.~V., {Mao}, Y.-Y., {Becker}, M.~R., {Skillman}, S.~W., \&
  {Wechsler}, R.~H. 2017, \apj, 834, 37

\bibitem[{{Lelli} {et~al.}(2015){Lelli}, {Duc}, {Brinks}, {Bournaud},
  {McGaugh}, {et~al.}}]{Lelli2015}
{Lelli}, F., {Duc}, P.-A., {Brinks}, E., {et~al.} 2015, \aap, 584, A113

\bibitem[{{Mandelbaum} {et~al.}(2006){Mandelbaum}, {Seljak}, {Kauffmann},
  {Hirata}, \& {Brinkmann}}]{Mandelbaum2006}
{Mandelbaum}, R., {Seljak}, U., {Kauffmann}, G., {Hirata}, C.~M., \&
  {Brinkmann}, J. 2006, \mnras, 368, 715

\bibitem[{{Mandelbaum} {et~al.}(2016){Mandelbaum}, {Wang}, {Zu}, {White},
  {Henriques}, \& {More}}]{Mandelbaum2016}
{Mandelbaum}, R., {Wang}, W., {Zu}, Y., {et~al.} 2016, \mnras, 457, 3200

\bibitem[{{McCall} {et~al.}(2012){McCall}, {Vaduvescu}, {Pozo Nunez}, {Barr
  Dominguez}, {Fingerhut}, {et~al.}}]{McCall2012}
{McCall}, M.~L., {Vaduvescu}, O., {Pozo Nunez}, F., {et~al.} 2012, \aap, 540,
  A49

\bibitem[{{McDonald} {et~al.}(2009{\natexlab{a}}){McDonald}, {Courteau}, \&
  {Tully}}]{McDonald2009a}
{McDonald}, M., {Courteau}, S., \& {Tully}, R.~B. 2009{\natexlab{a}}, \mnras,
  393, 628

\bibitem[{{McDonald} {et~al.}(2009{\natexlab{b}}){McDonald}, {Courteau}, \&
  {Tully}}]{McDonald2009b}
---. 2009{\natexlab{b}}, \mnras, 394, 2022

\bibitem[{{McDonald} {et~al.}(2011){McDonald}, {Courteau}, {Tully}, \&
  {Roediger}}]{McDonald2011}
{McDonald}, M., {Courteau}, S., {Tully}, R.~B., \& {Roediger}, J. 2011, \mnras,
  414, 2055

\bibitem[{{McGaugh} {et~al.}(2000){McGaugh}, {Schombert}, {Bothun}, \& {de
  Blok}}]{McGaugh2000}
{McGaugh}, S.~S., {Schombert}, J.~M., {Bothun}, G.~D., \& {de Blok}, W.~J.~G.
  2000, \apjl, 533, L99

\bibitem[{{Mei} {et~al.}(2007){Mei}, {Blakeslee}, {C{\^o}t{\'e}}, {Tonry},
  {West}, {et~al.}}]{Mei2007}
{Mei}, S., {Blakeslee}, J.~P., {C{\^o}t{\'e}}, P., {et~al.} 2007, \apj, 655,
  144

\bibitem[{{Mendes de Oliveira} {et~al.}(2003){Mendes de Oliveira}, {Amram},
  {Plana}, \& {Balkowski}}]{MendesdeOliveira2003}
{Mendes de Oliveira}, C., {Amram}, P., {Plana}, H., \& {Balkowski}, C. 2003,
  \aj, 126, 2635

\bibitem[{{Miller} {et~al.}(2011){Miller}, {Bundy}, {Sullivan}, {Ellis}, \&
  {Treu}}]{Miller2011}
{Miller}, S.~H., {Bundy}, K., {Sullivan}, M., {Ellis}, R.~S., \& {Treu}, T.
  2011, \apj, 741, 115

\bibitem[{{Miller} {et~al.}(2014){Miller}, {Ellis}, {Newman}, \&
  {Benson}}]{Miller2014}
{Miller}, S.~H., {Ellis}, R.~S., {Newman}, A.~B., \& {Benson}, A. 2014, \apj,
  782, 115

\bibitem[{{Milvang-Jensen} {et~al.}(2003){Milvang-Jensen},
  {Arag{\'o}n-Salamanca}, {Hau}, {J{\o}rgensen}, \&
  {Hjorth}}]{MilvangJensen2003}
{Milvang-Jensen}, B., {Arag{\'o}n-Salamanca}, A., {Hau}, G.~K.~T.,
  {J{\o}rgensen}, I., \& {Hjorth}, J. 2003, \mnras, 339, L1

\bibitem[{{Mo} \& {Mao}(2000)}]{Mo2000}
{Mo}, H.~J., \& {Mao}, S. 2000, \mnras, 318, 163

\bibitem[{{Mocz} {et~al.}(2012){Mocz}, {Green}, {Malacari}, \&
  {Glazebrook}}]{Mocz2012}
{Mocz}, P., {Green}, A., {Malacari}, M., \& {Glazebrook}, K. 2012, \mnras, 425,
  296

\bibitem[{{More} {et~al.}(2011){More}, {van den Bosch}, {Cacciato}, {Skibba},
  {Mo}, {et~al.}}]{More2011}
{More}, S., {van den Bosch}, F.~C., {Cacciato}, M., {et~al.} 2011, \mnras, 410,
  210

\bibitem[{{Mu{\~n}oz-Cuartas} {et~al.}(2011){Mu{\~n}oz-Cuartas}, {Macci{\`o}},
  {Gottl{\"o}ber}, \& {Dutton}}]{MunozCuartas2011}
{Mu{\~n}oz-Cuartas}, J.~C., {Macci{\`o}}, A.~V., {Gottl{\"o}ber}, S., \&
  {Dutton}, A.~A. 2011, \mnras, 411, 584

\bibitem[{{Murray} {et~al.}(2005){Murray}, {Quataert}, \&
  {Thompson}}]{Murray2005}
{Murray}, N., {Quataert}, E., \& {Thompson}, T.~A. 2005, \apj, 618, 569

\bibitem[{{Nakamura} {et~al.}(2006){Nakamura}, {Arag{\'o}n-Salamanca},
  {Milvang-Jensen}, {Arimoto}, {Ikuta}, {et~al.}}]{Nakamura2006}
{Nakamura}, O., {Arag{\'o}n-Salamanca}, A., {Milvang-Jensen}, B., {et~al.}
  2006, \mnras, 366, 144

\bibitem[{{Navarro} {et~al.}(1996){Navarro}, {Frenk}, \& {White}}]{Navarro1996}
{Navarro}, J.~F., {Frenk}, C.~S., \& {White}, S.~D.~M. 1996, \apj, 462, 563

\bibitem[{{Navarro} \& {Steinmetz}(2000)}]{Navarro2000}
{Navarro}, J.~F., \& {Steinmetz}, M. 2000, \apj, 538, 477

\bibitem[{{Neistein} {et~al.}(1999){Neistein}, {Maoz}, {Rix}, \&
  {Tonry}}]{Neistein1999}
{Neistein}, E., {Maoz}, D., {Rix}, H.-W., \& {Tonry}, J.~L. 1999, \aj, 117,
  2666

\bibitem[{{Norris} {et~al.}(2014){Norris}, {Kannappan}, {Forbes}, {Romanowsky},
  {Brodie}, {et~al.}}]{Norris2014}
{Norris}, M.~A., {Kannappan}, S.~J., {Forbes}, D.~A., {et~al.} 2014, \mnras,
  443, 1151

\bibitem[{{Obreschkow} \& {Glazebrook}(2014)}]{Obreschkow2014}
{Obreschkow}, D., \& {Glazebrook}, K. 2014, \apj, 784, 26

\bibitem[{{Papastergis} {et~al.}(2011){Papastergis}, {Martin}, {Giovanelli}, \&
  {Haynes}}]{Papastergis2011}
{Papastergis}, E., {Martin}, A.~M., {Giovanelli}, R., \& {Haynes}, M.~P. 2011,
  \apj, 739, 38

\bibitem[{{Paturel} {et~al.}(2003){Paturel}, {Petit}, {Prugniel}, {Theureau},
  {Rousseau}, {et~al.}}]{Paturel2003}
{Paturel}, G., {Petit}, C., {Prugniel}, P., {et~al.} 2003, \aap, 412, 45

\bibitem[{{Peng} {et~al.}(2015){Peng}, {Maiolino}, \& {Cochrane}}]{Peng2015}
{Peng}, Y., {Maiolino}, R., \& {Cochrane}, R. 2015, \nat, 521, 192

\bibitem[{{Penny} {et~al.}(2015){Penny}, {Janz}, {Forbes}, {Benson}, \&
  {Mould}}]{Penny2015}
{Penny}, S.~J., {Janz}, J., {Forbes}, D.~A., {Benson}, A.~J., \& {Mould}, J.
  2015, \mnras, 453, 3635

\bibitem[{{Pizagno} {et~al.}(2007){Pizagno}, {Prada}, {Weinberg}, {Rix},
  {Pogge}, {et~al.}}]{Pizagno2007}
{Pizagno}, J., {Prada}, F., {Weinberg}, D.~H., {et~al.} 2007, \aj, 134, 945

\bibitem[{{Prugniel} \& {Simien}(1996)}]{Prugniel1996}
{Prugniel}, P., \& {Simien}, F. 1996, \aap, 309, 749

\bibitem[{{Reddick} {et~al.}(2013){Reddick}, {Wechsler}, {Tinker}, \&
  {Behroozi}}]{Reddick2013}
{Reddick}, R.~M., {Wechsler}, R.~H., {Tinker}, J.~L., \& {Behroozi}, P.~S.
  2013, \apj, 771, 30

\bibitem[{{Robertson} {et~al.}(2006){Robertson}, {Cox}, {Hernquist}, {Franx},
  {Hopkins}, {et~al.}}]{Robertson2006}
{Robertson}, B., {Cox}, T.~J., {Hernquist}, L., {et~al.} 2006, \apj, 641, 21

\bibitem[{{Rodr{\'{\i}}guez-Puebla} {et~al.}(2013){Rodr{\'{\i}}guez-Puebla},
  {Avila-Reese}, \& {Drory}}]{RodriguezPuebla2013}
{Rodr{\'{\i}}guez-Puebla}, A., {Avila-Reese}, V., \& {Drory}, N. 2013, \apj,
  767, 92

\bibitem[{{Rodr{\'{\i}}guez-Puebla} {et~al.}(2015){Rodr{\'{\i}}guez-Puebla},
  {Avila-Reese}, {Yang}, {Foucaud}, {Drory}, {et~al.}}]{RodriguezPuebla2015}
{Rodr{\'{\i}}guez-Puebla}, A., {Avila-Reese}, V., {Yang}, X., {et~al.} 2015,
  \apj, 799, 130

\bibitem[{{Roediger} \& {Courteau}(2015)}]{Roediger2015}
{Roediger}, J.~C., \& {Courteau}, S. 2015, \mnras, 452, 3209

\bibitem[{{Roediger} {et~al.}(2011{\natexlab{a}}){Roediger}, {Courteau},
  {MacArthur}, \& {McDonald}}]{Roediger2011a}
{Roediger}, J.~C., {Courteau}, S., {MacArthur}, L.~A., \& {McDonald}, M.
  2011{\natexlab{a}}, \mnras, 416, 1996

\bibitem[{{Roediger} {et~al.}(2011{\natexlab{b}}){Roediger}, {Courteau},
  {McDonald}, \& {MacArthur}}]{Roediger2011b}
{Roediger}, J.~C., {Courteau}, S., {McDonald}, M., \& {MacArthur}, L.~A.
  2011{\natexlab{b}}, \mnras, 416, 1983

\bibitem[{{Roediger} {et~al.}(2012){Roediger}, {Courteau},
  {S{\'a}nchez-Bl{\'a}zquez}, \& {McDonald}}]{Roediger2012}
{Roediger}, J.~C., {Courteau}, S., {S{\'a}nchez-Bl{\'a}zquez}, P., \&
  {McDonald}, M. 2012, \apj, 758, 41

\bibitem[{{Rubin} {et~al.}(1997){Rubin}, {Waterman}, \& {Kenney}}]{Rubin1997}
{Rubin}, V.~C., {Waterman}, A.~H., \& {Kenney}, J.~D.~P. 1997, in Bulletin of
  the American Astronomical Society, Vol.~29, American Astronomical Society
  Meeting Abstracts, 105.15

\bibitem[{{Rubin} {et~al.}(1999){Rubin}, {Waterman}, \& {Kenney}}]{Rubin1999}
{Rubin}, V.~C., {Waterman}, A.~H., \& {Kenney}, J.~D.~P. 1999, \aj, 118, 236

\bibitem[{{Ry{\'s}} {et~al.}(2014){Ry{\'s}}, {van de Ven}, \&
  {Falc{\'o}n-Barroso}}]{Rys2014}
{Ry{\'s}}, A., {van de Ven}, G., \& {Falc{\'o}n-Barroso}, J. 2014, \mnras, 439,
  284

\bibitem[{{S{\'a}nchez} {et~al.}(2012){S{\'a}nchez}, {Kennicutt}, {Gil de Paz},
  {van de Ven}, {V{\'{\i}}lchez}, {et~al.}}]{Sanchez2012}
{S{\'a}nchez}, S.~F., {Kennicutt}, R.~C., {Gil de Paz}, A., {et~al.} 2012,
  \aap, 538, A8

\bibitem[{{S{\'a}nchez-Bl{\'a}zquez} {et~al.}(2006){S{\'a}nchez-Bl{\'a}zquez},
  {Peletier}, {Jim{\'e}nez-Vicente}, {Cardiel}, {Cenarro},
  {et~al.}}]{SanchezBlazquez2006}
{S{\'a}nchez-Bl{\'a}zquez}, P., {Peletier}, R.~F., {Jim{\'e}nez-Vicente}, J.,
  {et~al.} 2006, \mnras, 371, 703

\bibitem[{{Sawala} {et~al.}(2015){Sawala}, {Frenk}, {Fattahi}, {Navarro},
  {Bower}, {et~al.}}]{Sawala2015}
{Sawala}, T., {Frenk}, C.~S., {Fattahi}, A., {et~al.} 2015, \mnras, 448, 2941

\bibitem[{{Schlafly} \& {Finkbeiner}(2011)}]{Schlafly2011}
{Schlafly}, E.~F., \& {Finkbeiner}, D.~P. 2011, \apj, 737, 103

\bibitem[{{Scott} {et~al.}(2015){Scott}, {Fogarty}, {Owers}, {Croom},
  {Colless}, {et~al.}}]{Scott2015}
{Scott}, N., {Fogarty}, L.~M.~R., {Owers}, M.~S., {et~al.} 2015, \mnras, 451,
  2723

\bibitem[{{Serra} {et~al.}(2016){Serra}, {Oosterloo}, {Cappellari}, {den
  Heijer}, \& {J{\'o}zsa}}]{Serra2016}
{Serra}, P., {Oosterloo}, T., {Cappellari}, M., {den Heijer}, M., \&
  {J{\'o}zsa}, G.~I.~G. 2016, \mnras, 460, 1382

\bibitem[{{Simons} {et~al.}(2015){Simons}, {Kassin}, {Weiner}, {Heckman},
  {Lee}, {et~al.}}]{Simons2015}
{Simons}, R.~C., {Kassin}, S.~A., {Weiner}, B.~J., {et~al.} 2015, \mnras, 452,
  986

\bibitem[{{Smith}(2014)}]{Smith2014}
{Smith}, R.~J. 2014, \mnras, 443, L69

\bibitem[{{Sorce} {et~al.}(2013){Sorce}, {Courtois}, {Sheth}, \&
  {Tully}}]{Sorce2013}
{Sorce}, J.~G., {Courtois}, H.~M., {Sheth}, K., \& {Tully}, R.~B. 2013, \mnras,
  433, 751

\bibitem[{{Springel} {et~al.}(2005){Springel}, {Di Matteo}, \&
  {Hernquist}}]{Springel2005}
{Springel}, V., {Di Matteo}, T., \& {Hernquist}, L. 2005, \mnras, 361, 776

\bibitem[{{Terrazas} {et~al.}(2016){Terrazas}, {Bell}, {Henriques}, \&
  {White}}]{Terrazas2016}
{Terrazas}, B.~A., {Bell}, E.~F., {Henriques}, B.~M.~B., \& {White}, S.~D.~M.
  2016, \mnras, 459, 1929

\bibitem[{{Thomas} {et~al.}(2009){Thomas}, {Saglia}, {Bender}, {Thomas},
  {Gebhardt}, {Magorrian}, {Corsini}, \& {Wegner}}]{Thomas2009}
{Thomas}, J., {Saglia}, R.~P., {Bender}, R., {et~al.} 2009, \apj, 691, 770

\bibitem[{{Tinker} {et~al.}(2016){Tinker}, {Wetzel}, {Conroy}, \&
  {Mao}}]{Tinker2016}
{Tinker}, J., {Wetzel}, A., {Conroy}, C., \& {Mao}, Y.-Y. 2016, ArXiv e-prints,
  arXiv:1609.03388

\bibitem[{{Tinker} {et~al.}(2013){Tinker}, {Leauthaud}, {Bundy}, {George},
  {Behroozi}, {et~al.}}]{Tinker2013}
{Tinker}, J.~L., {Leauthaud}, A., {Bundy}, K., {et~al.} 2013, \apj, 778, 93

\bibitem[{{Tollerud} {et~al.}(2011){Tollerud}, {Bullock}, {Graves}, \&
  {Wolf}}]{Tollerud2011}
{Tollerud}, E.~J., {Bullock}, J.~S., {Graves}, G.~J., \& {Wolf}, J. 2011, \apj,
  726, 108

\bibitem[{{Toloba} {et~al.}(2011){Toloba}, {Boselli}, {Cenarro}, {Peletier},
  {Gorgas}, {et~al.}}]{Toloba2011}
{Toloba}, E., {Boselli}, A., {Cenarro}, A.~J., {et~al.} 2011, \aap, 526, A114

\bibitem[{{Toomre}(1977)}]{Toomre1977}
{Toomre}, A. 1977, in Evolution of Galaxies and Stellar Populations, 401

\bibitem[{{Tortora} {et~al.}(2012){Tortora}, {La Barbera}, {Napolitano}, {de
  Carvalho}, \& {Romanowsky}}]{Tortora2012}
{Tortora}, C., {La Barbera}, F., {Napolitano}, N.~R., {de Carvalho}, R.~R., \&
  {Romanowsky}, A.~J. 2012, \mnras, 425, 577

\bibitem[{{Trujillo} {et~al.}(2004){Trujillo}, {Burkert}, \&
  {Bell}}]{Trujillo2004}
{Trujillo}, I., {Burkert}, A., \& {Bell}, E.~F. 2004, \apjl, 600, L39

\bibitem[{{Trujillo-Gomez} {et~al.}(2011){Trujillo-Gomez}, {Klypin}, {Primack},
  \& {Romanowsky}}]{TrujilloGomez2011}
{Trujillo-Gomez}, S., {Klypin}, A., {Primack}, J., \& {Romanowsky}, A.~J. 2011,
  \apj, 742, 16

\bibitem[{{Tully} {et~al.}(2016){Tully}, {Courtois}, \& {Sorce}}]{Tully2016}
{Tully}, R.~B., {Courtois}, H.~M., \& {Sorce}, J.~G. 2016, \aj, 152, 50

\bibitem[{{Tully} \& {Fisher}(1977)}]{Tully1977}
{Tully}, R.~B., \& {Fisher}, J.~R. 1977, \aap, 54, 661

\bibitem[{{Tully} {et~al.}(1998){Tully}, {Pierce}, {Huang}, {Saunders},
  {Verheijen}, {et~al.}}]{Tully1998}
{Tully}, R.~B., {Pierce}, M.~J., {Huang}, J.-S., {et~al.} 1998, \aj, 115, 2264

\bibitem[{{Tully} \& {Shaya}(1984)}]{Tully1984}
{Tully}, R.~B., \& {Shaya}, E.~J. 1984, \apj, 281, 31

\bibitem[{{Tully} \& {Verheijen}(1997)}]{Tully1997}
{Tully}, R.~B., \& {Verheijen}, M.~A.~W. 1997, \apj, 484, 145

\bibitem[{{van den Bosch}(2000)}]{vandenBosch2000}
{van den Bosch}, F.~C. 2000, \apj, 530, 177

\bibitem[{{van Zee} {et~al.}(2004){van Zee}, {Skillman}, \&
  {Haynes}}]{vanZee2004}
{van Zee}, L., {Skillman}, E.~D., \& {Haynes}, M.~P. 2004, \aj, 128, 121

\bibitem[{{Vogt}(1995)}]{Vogt1995}
{Vogt}, N.~P. 1995, PhD thesis, CORNELL UNIVERSITY.

\bibitem[{{Willick} {et~al.}(1997){Willick}, {Courteau}, {Faber}, {Burstein},
  {Dekel}, {et~al.}}]{Willick1997}
{Willick}, J.~A., {Courteau}, S., {Faber}, S.~M., {et~al.} 1997, \apjs, 109,
  333

\bibitem[{{Willick} \& {Strauss}(1998)}]{Willick1998}
{Willick}, J.~A., \& {Strauss}, M.~A. 1998, \apj, 507, 64

\bibitem[{{Woo} {et~al.}(2008){Woo}, {Courteau}, \& {Dekel}}]{Woo2008}
{Woo}, J., {Courteau}, S., \& {Dekel}, A. 2008, \mnras, 390, 1453

\bibitem[{{Woo} {et~al.}(2013){Woo}, {Dekel}, {Faber}, {Noeske}, {Koo},
  {et~al.}}]{Woo2013}
{Woo}, J., {Dekel}, A., {Faber}, S.~M., {et~al.} 2013, \mnras, 428, 3306

\bibitem[{{Yang} {et~al.}(2009){Yang}, {Mo}, \& {van den Bosch}}]{Yang2009}
{Yang}, X., {Mo}, H.~J., \& {van den Bosch}, F.~C. 2009, \apj, 695, 900

\bibitem[{{Zwaan} {et~al.}(1995){Zwaan}, {van der Hulst}, {de Blok}, \&
  {McGaugh}}]{Zwaan1995}
{Zwaan}, M.~A., {van der Hulst}, J.~M., {de Blok}, W.~J.~G., \& {McGaugh},
  S.~S. 1995, \mnras, 273, L35

\end{thebibliography}

\end{document}